\documentclass[fleqn,usenatbib]{mnras}

\usepackage{newtxtext,newtxmath}

\usepackage[T1]{fontenc}

\DeclareRobustCommand{\VAN}[3]{#2}
\let\VANthebibliography\thebibliography
\def\thebibliography{\DeclareRobustCommand{\VAN}[3]{##3}\VANthebibliography}

\usepackage{graphicx}	
\usepackage{amsmath}	
\usepackage{multirow}
\usepackage{natbib}
\usepackage{hyperref}
\usepackage{nameref}

\title[]{Discovery of Two Consecutive Profile Change Events in the Millisecond Pulsar PSR J0437$-$4715}

\author[Rami~F. Mandow]{Rami~F. Mandow$^{1,2}$,\thanks{E-mail: rami.mandow@hdr.mq.edu.au}
Andrew Zic$^{2,3}$,
J.~R. Dawson$^{1,2}$,
Shuangqiang Wang$^{4,2}$,
Ma\l{}gorzata Cury\l{}o$^{5,3}$,
\newauthor 
Bhavnesh Bhat$^{6}$,
Michael J. Keith$^{6}$,
N. D. Ramesh Bhat$^{7}$,
Emma Carli$^{8,3}$,
Shi Dai$^{2}$,
Valentina Di Marco$^{9,3}$,
\newauthor 
Simon C.-C. Ho$^{13,8,3}$,
Agastya Kapur$^{2}$,
Wenhua Ling$^{2}$,
Marcus E. Lower$^{8}$,
Bradley W. Meyers$^{10,7}$,
\newauthor 
Saurav Mishra$^{8,3,2}$,
D.~J. Reardon$^{8,3}$,
Sparrow Roch$^{8,3,2}$,
Karolina Rozko,$^{11}$,
Christopher J. Russell$^{12}$,
\newauthor 
R.~M. Shannon$^{8,3}$
\\
\\
$^{1}$Department of Mathematics and Physical Sciences, Macquarie University, NSW 2109, Australia\\
$^{2}$Australia Telescope National Facility, CSIRO, Space and Astronomy, P.O. Box 76, Epping, NSW 1710, Australia\\
$^{3}$OzGrav: The ARC Center of Excellence for Gravitational Wave Discovery, Hawthorn VIC 3122, Australia\\
$^{4}$Xinjiang Astronomical Observatory, Chinese Academy of Sciences, Urumqi, Xinjiang 830011, People's Republic of China\\
$^{5}$School of Physics and Astronomy, Monash University, Clayton VIC 3800, Australia\\
$^{6}$Jodrell Bank Centre for Astrophysics, Department of Physics and Astronomy, The University of Manchester, Manchester M13 9PL, UK\\
$^{7}$International Centre for Radio Astronomy Research, Curtin University, Bentley, WA 6102, Australia\\
$^{8}$Centre for Astrophysics and Supercomputing, Swinburne University of Technology, Hawthorn VIC 3122, Australia\\
$^{9}$School of Physics, University of Melbourne, Parkville, VIC 3010, Australia\\
$^{10}$Australian SKA Regional Centre (AusSRC), Curtin University, Kent Street, Bentley, WA 6102, Australia\\
$^{11}$Janusz Gil Institute of Astronomy, University of Zielona Gora, ul. Prof. Z. Szafrana 2, 65-516 Zielona Gora, Poland\\
$^{12}$CSIRO Scientific Computing, PO Box 76, Epping, NSW, 1710, Australia\\
$^{13}$Research School of Astronomy and Astrophysics, The Australian National University, Canberra, ACT 2611, Australia
}

\date{Accepted XXX. Received YYY; in original form ZZZ}

\pubyear{\the\year{}}

\begin{document}
\label{firstpage}
\pagerange{\pageref{firstpage}--\pageref{lastpage}}
\maketitle

\begin{abstract}
The closest and brightest millisecond pulsar (MSP), PSR J0437$-$4715, is a high-priority pulsar for precision timing experiments, including the Parkes Pulsar Timing Array (PPTA) project, where it is used for both pulsar timing and for polarization calibration. Using approximately seven years of observations obtained with the Ultra-Wideband Low-Frequency receiver on Murriyang, CSIRO's Parkes radio telescope, we investigate the long-term spectro-polarimetric variability of PSR J0437$-$4715. We report the discovery of two consecutive discrete profile change events exhibiting the same phase-localised morphological variations, the first such behaviour reported in any MSP. Both events affect three phase-localised regions of the pulse profile that exhibit coupled and anti-correlated temporal evolution. We also identified, and corrected for, systematic profile variability induced by polarization calibration errors arising from long-term drift of the polarimetric properties of the receiver. The profile change exhibits frequency-dependent evolution in total intensity, together with variations in linear polarization, position angle and fractional linear polarization. These results confirm a magnetospheric origin for the events rather than propagation effects caused by the interstellar medium, and provide potential new constraints on MSP emission models.  
\end{abstract}




\section{Introduction}
\label{sec:int}
PSR J0437$-$4715 is a 5.75-millisecond-period pulsar in a binary system with a white dwarf companion \citep{1993Natur.361..613J}. It is the nearest and brightest millisecond pulsar (MSP) to the Solar System \citep{2024ApJ...971L..18R}. Due to its proximity, it has been broadly studied across the electromagnetic spectrum, including in radio \citep{1996MNRAS.280..331M,1998ApJ...498..365J,2011MNRAS.416..346K,2014MNRAS.441.3148O}, infrared \citep{2012ApJ...746....6D}, optical \citep{1993ApJ...411L..83B,1993Natur.364..603B}, ultraviolet \citep{2004ApJ...602..327K}, and high-energy wavelengths \citep{2013ApJ...762...96B,2024ApJ...971L..20C,2026ApJ..1000L..48M}. PSR J0437$-$4715 has also been used to probe the interstellar medium along the line of sight \citep{2020ApJ...904..104R}, to infer the neutron star equation of state of matter \citep{2024ApJ...971L..19R,2024ApJ...974..244T,2025ApJ...978L..14H}, and for dark matter studies \citep{2026A&A...706A.203S}.

Due to its southern declination and brightness, PSR J0437$-$4715 (hereon `J0437’) is also one of the highest-priority pulsars for the Parkes Pulsar Timing Array project (PPTA) \citep{2013PASA...30...17M}. In addition to its role in pulsar timing, J0437 is routinely used for polarization calibration purposes within the PPTA \citep{2004ApJS..152..129V,2013ApJS..204...13V}. Long-term timing observations of this MSP are also being carried out by other southern pulsar timing projects \citep{2021ApJ...908..158S,2021ApJ...911..137L,2022PASA...39...27S}, and it is also included in the Indian Pulsar Timing Array project (InPTA -- \citealt{2024PASA...41...36K}) and the North American Nanohertz Observatory for Gravitational Waves (NANOGrav) project \citep{2023ApJ...951L...9A}. 

Individual pulses from pulsars exhibit stochastic, pulse-to-pulse variability \citep[known as jitter --][]{2021MNRAS.502..407P,2024MNRAS.528.3658K,2024PASA...41...36K}. For many applications of pulsar timing and pulsar science, a stable integrated pulse profile is required. The integrated profile of a pulsar is formed by time-averaging many individual pulses aligned in rotational phase as the emission beam sweeps across the observer’s line of sight \citep{2025ApJ...991..135G}. This averaging suppresses the pulse-to-pulse variability in the integrated profile according to $t^{-1/2}$, where $t$ is the integration time \citep{2014MNRAS.443.1463S}.

Despite this averaging, variability of the integrated profile over time-scales of months to years is well established across the canonical pulsar population \citep{2016MNRAS.456.1374B,2022MNRAS.513.5861S,2023MNRAS.524.5904L,2024MNRAS.528.7458B,2025MNRAS.538.3104L}. In contrast, the integrated profiles of MSPs are remarkably stable over long time-scales \citep{2012MNRAS.420..361L}.

As the observed radio emission originates within the pulsar magnetosphere, the long-term stability of the integrated profiles is interpreted as a measure of magnetospheric stability \citep{1975ApJ...198..661H}. This profile stability, as well as the rotational stability of MSPs, is foundational to their use in sensitive precision timing experiments, including searching for nanohertz-regime gravitational waves \citep{2023ApJ...951L...6R,2023RAA....23g5024X,2023A&A...678A..48E,2023ApJ...951L...9A,2025MNRAS.536.1489M}, determining precise pulsar mass measurements \citep{2024ApJ...974..295D,2025PhRvD.111l3021G,2025ApJ...995...60M}, and tests of General Relativity \citep{2021MNRAS.504.2094K,2021PhRvX..11d1050K,2022A&A...667A.149H}. 

Despite this general stability, variations of the observed profile with respect to the long-term integrated pulse profile have been observed across the MSP population. For example, short-time-scale phenomena, known as mode-changing and nulling (occurring over time-scales ranging from individual rotations to days), have been reported \citep{2022MNRAS.510.5908M,2023MNRAS.523.4405N}. On longer time-scales, ranging from weeks to years, some MSPs have exhibited stochastic profile variability \citep{2018ApJ...868..122B,2021MNRAS.500.1178P,2025ApJ...984..139F}. Rare, discrete profile change events with long-term recovery time-scales have also been observed in the MSP population \citep{2016ApJ...828L...1S,2025PASA...42..142M}. These are generally different from the transient profile changes observed in the canonical pulsar population, which can exhibit continuous profile variability, long-term state switching, and gradual profile evolution \citep{2010Sci...329..408L,2011MNRAS.415..251K,2014ApJ...780L..31B,2016MNRAS.456.1374B,2022MNRAS.513.5861S,2025MNRAS.540.2486K}. Previous studies of J0437 have reported both short-term profile variability \citep{1997ApJ...475L..33A,2000ApJ...543..979V,2001MNRAS.326L..33V,2011MNRAS.418.1258O} and discrete profile changes \citep{2021MNRAS.502..478G}, indicating that this MSP exhibits multiple forms of profile instability. 

Pulsar radio emission is often highly polarized \citep{2023MNRAS.520.4961O,2024MNRAS.530.4839J}. Polarimetric information, including frequency-dependent linear and circular polarization, constrains the emission mechanisms and propagation effects within the pulsar magnetosphere \citep{2024MNRAS.532.3558K,2025A&A...695A.173X}. The observed linear polarization position angle (PA) can also be used to infer the magnetic field geometry when applying the ‘rotating vector model’ \citep{1969Natur.221..443R}. Changes in the polarization profile shape provide a means to probe the dynamics and stability of the pulsar magnetosphere.

In this paper, we present a wideband spectro-polarimetric analysis of profile variability in PSR J0437$-$4715, including the discovery of two consecutive discrete profile change events, their recovery and additional stochastic variations. We identify a primary event occurring on Modified Julian Date (MJD) 59670. We also analyse the recovery tail of an earlier event with onset MJD 58082 which occurred before the start of our dataset. To date, no spectro-polarimetric study of these events has been reported in the literature. Due to the brightness of J0437, these profile changes can be studied at high signal-to-noise ratio (S/N), allowing tight constraints on their temporal evolution, which can help characterize its impact on precision timing experiments. 

The structure of this paper is outlined as follows: Observation and Methods are presented in Section \ref{sec:obsmethods}. In our Results in Section \ref{sec:results}, we describe profile shape variability, polarimetric response to the profile change, Principle Component Analysis (PCA) of profile variability, and modelling of the profile change. In Section \ref{sec:discussion} we outline our discussions and interpretations.

\section{Observations and Methods}
\label{sec:obsmethods}
\subsection{Observations and Data Reduction}

PSR J0437-4715 has been observed at regular cadence with Murriyang, CSIRO’s Parkes Radio Telescope, as part of the PPTA observing program from 2004. Our analysis focuses on a seven-year dataset obtained using the Ultra-Wideband Low-Frequency receiver (UWL; \cite{2020PASA...37...12H}), which has been in operation for the PPTA since 2018. The UWL provides continuous wideband frequency coverage (704--4032\,MHz), enabling investigation of the frequency-dependent and temporal evolution of profile changes and, thereby, the magnetospheric response to these events and their recovery.

Observations used in this analysis were obtained as part of the PPTA programme (project code P456), which monitors a set of 37 MSPs at an approximate 3-week cadence. The data presented in this paper are part of the upcoming PPTA Fourth Data Release (DR4) (Wang et al. \textit{in prep.}). Here, we provide a summary of the data-processing procedure used for DR4 products as relevant to the analysis in this work. Details on the observing programme can be found in the PPTA Third Data Release publication \citep{2023PASA...40...49Z}. 

The PPTA observes J0437 for a duration of 64 minutes at 1\,MHz channel resolution, providing 3328 channels across the UWL band. On occasion, shorter integrations occur when observations are halted due to instrumentation or weather-related issues. The data are recorded and coherently de-dispersed during observations using a nominal dispersion measure (DM) value of 2.644983 pc cm$^{-3}$. The DM is subsequently updated using epoch-specific DM values derived from timing measurements with \textsc{TEMPO2} \citep{2006MNRAS.372.1549E} as part of the PPTA DR4 analysis, using the \textsc{PSRCHIVE}  tool \textsc{pam} \citep{2004PASA...21..302H,2011PASA...28....1V}. 
Additionally, we measure and correct for variations in interstellar and ionospheric rotation measure (RM). RM values are first estimated using the \textsc{PSRCHIVE} module \textsc{rmfit} using the brute force method, after which the data are corrected to an infinite reference frequency using the \textsc{pam} tool, with the \textsc{-\phantom{}-aux\_rm} option. We also applied the PPTA DR3 timing ephemeris to all observation files.

Before further analysis, we excised data out of the processed observations that were corrupted due to instrumental effects. These include digitiser synchronisation failures, packet loss, or aliasing at instrumental sub-band edges \citep[see][]{2020PASA...37...12H}. These issues affected segments of data localised in time and frequency, and will be described in more detail in Wang et al. (\textit{in prep.}).

We performed the majority of the analysis in this work on sub-banded data products, using the same frequency divisions as defined in the PPTA DR3. We list the details of these sub-bands in Table \ref{tab:uwlfreq}. 
\begin{table}
\centering
\caption{
UWL sub-band ranges (where $\nu_\mathrm{min}$ and $\nu_\mathrm{max}$ represent the minimum and maximum frequency), bandwidth (bw) and centre frequency ($\nu_{c})$}
\label{tab:uwlfreq}
    \begin{tabular}{ccccc}
    \hline
    \hline
    \textbf{Sub-band} & \textbf{$\nu_{\text{min}}$ (MHz)} & \textbf{$\nu_{\text{max}}$ (MHz)} & \textbf{BW} (MHz) & \textbf{$\nu_{\text{c}}$}\\
    \hline
    sbA & 704 & 768 & 64 & 736\\ 
    sbB & 768 & 864 & 96 & 816\\ 
    sbC & 864 & 960 & 96 & 912\\ 
    sbD & 960 & 1248 & 288 & 1104\\ 
    sbE & 1248 & 1504 & 256 & 1376\\ 
    sbF & 1504 & 2048 & 544 & 1776\\ 
    sbG & 2048 & 2592 & 544 & 2320\\ 
    sbH & 2592 & 4032 & 1440 & 3312\\
    \hline 
    \end{tabular}
\end{table}

To improve S/N, we frequency-average the sub-banded profiles into single-channel profiles, removing any that were completely flagged due to interference. On rare occasions, multiple observations occur within 24 hours of each other. For these cases, we combine the intra-day observations using the \textsc{PSRCHIVE} tool \textsc{psradd} into a single file per epoch. We then set a S/N threshold of 60, removing any observations below this value. The fractional portion of data removed is greatest in the lower S/N bands (sbA $\sim 40\%$; sbB $\sim 12\%$; sbC $\sim 24\%$; sbD $\sim 4\%$), whilst in the higher S/N bands this is minimal ($< 1\%$).

\subsection{Sub-banded Epoch-Customised Template Generation}
\label{subsec:template_create}

Our study focuses on profile residuals, defined as the difference between a single epoch profile and a reference template profile. We used the wide-band, polarized pulse portraits presented by \citet{2025arXiv251209220C} as our starting templates. These portraits contain 832 frequency channels (4\,MHz resolution across 704 -- 4032\,MHz), and were formed from UWL observations recorded from 2018 -- 2025 for use in the upcoming PPTA DR4. Further details on the construction of these templates can be found in \citet{2025arXiv251209220C}.

Frequency-dependent modulation of the intrinsic flux density, such as that caused by diffractive scintillation in the interstellar medium (ISM), can cause the flux density in individual frequency channels to vary, introducing channel-dependent amplitude variations that are not associated with changes in intrinsic profile shape \citep{2002ApJ...581..495B,2018ApJ...868..122B}.

In the presence of frequency-dependent profile evolution, variations in the relative flux density across frequency channels can alter the relative amplitudes of profile components once the channels are frequency-averaged. This can produce apparent profile shape variations that are not intrinsic to the pulsar. To mitigate this effect, we construct epoch-customised templates that replicate the relative flux densities of the scintillated profile across frequency channels. 

To achieve this, we take the full 3328-channel archive for each epoch and frequency-scrunch to 832-channel resolution using \textsc{PSRCHIVE}, to match the resolution of the portraits. We then measure the S/N in each channel using the \textsc{PSRCHIVE} tool \textsc{psrstat}, with the \texttt{snr=pdmp,snr} option. This estimates the S/N by smoothing the pulse profile with boxcar functions of various widths and selects the maximum integrated S/N\footnote{see \url{https://psrchive.sourceforge.net/manuals/psrstat/algorithms/snr/}}. These per-channel S/N measurements are converted to weights, with channels containing no signal assigned a weight of zero. The determined weights are then applied to the corresponding frequency channels of the portrait using \textsc{psredit}, rescaling the weight of each template channel to match the per-channel S/N in the individual epoch.

\begin{figure}
    \centering
    \includegraphics[width=\columnwidth]{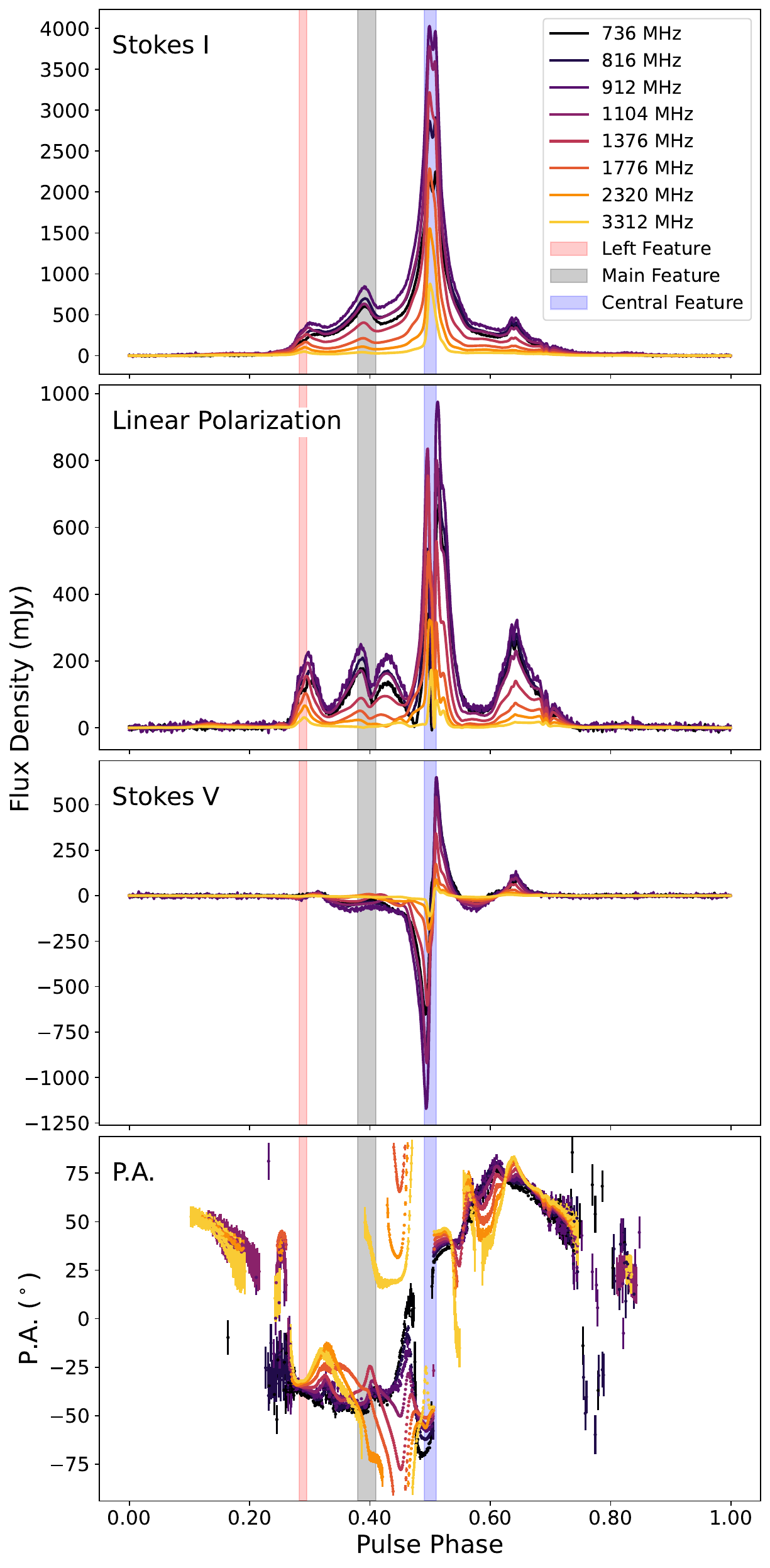}
    \caption{Non-normalized Stokes I, Linear Polarization, Stokes V and PA template profiles colour-mapped as a function of frequency, highlighting the frequency-dependent profile morphology observed for PSR J0437$-$4715. The three shaded regions indicate the phase locations of the profile features associated with the profile change events: Left Feature (pink), Main Feature (grey) and Central Feature (purple), which are introduced and discussed in Section \ref{ProVar}.}

    
    \label{fig:Temps}
\end{figure}

The final re-weighted epoch-customised portrait is then saved and frequency-scrunched into the eight sub-bands using the upper and lower boundaries as defined in Table \ref{tab:uwlfreq}. These sub-banded profiles form the customised templates for each epoch (hereon `epoch-customised templates'). We retain the full-Stokes parameters (I, Q, U, V), from which the linear polarization, position angle (PA), and associated uncertainties are derived in the subsequent analysis. We show the Stokes I, Linear Polarization, Stokes V, and PA templates for each of our eight portrait sub-bands in Figure \ref{fig:Temps}, and we use this polarization basis for profile residuals.

\subsection{Correction of Polarization Calibration Errors}
\label{subsec:MTM_correction}
\begin{figure}
    \centering
    \includegraphics[width=\columnwidth]{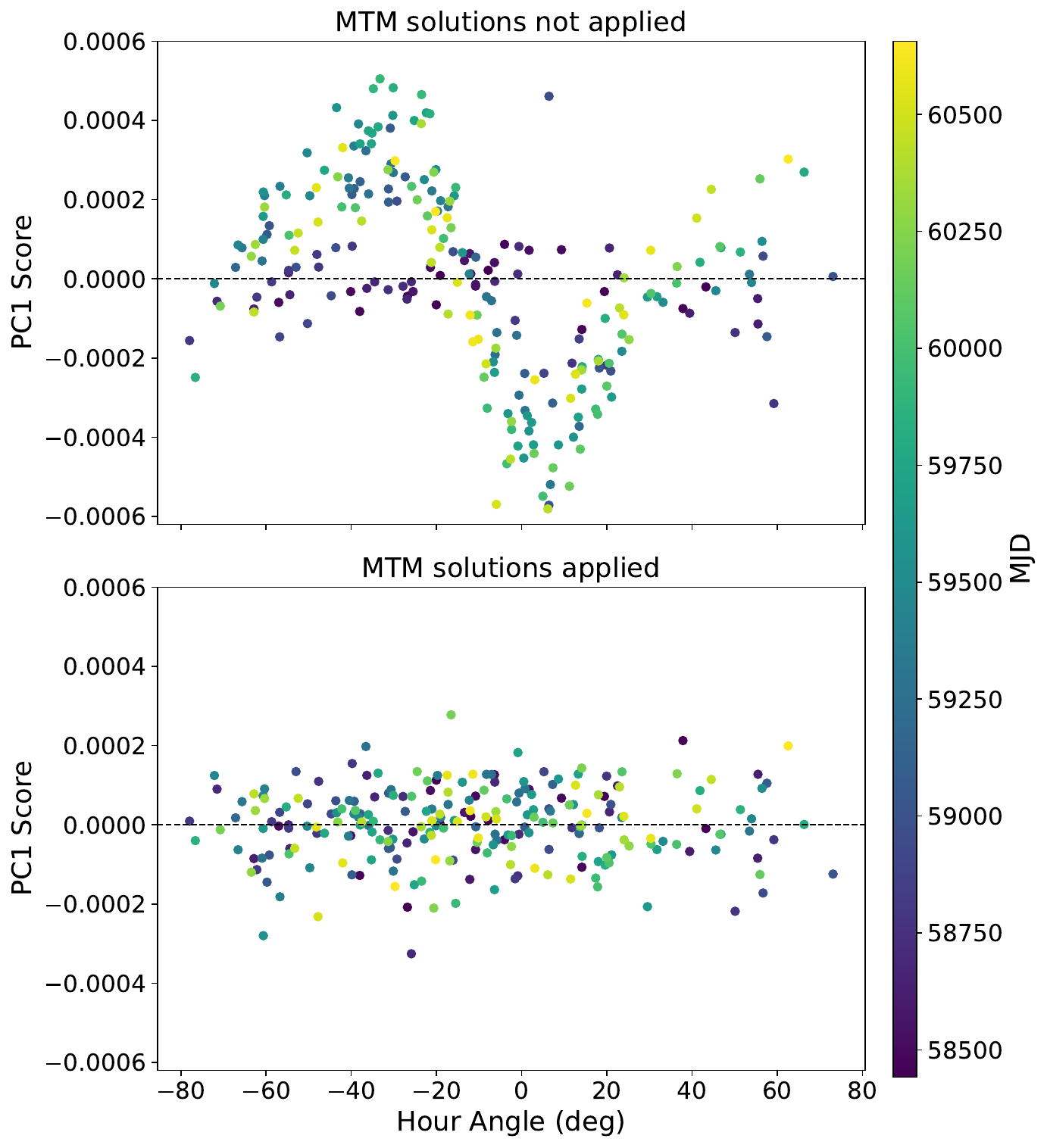}
    \caption {Principal Component Analysis (PCA) scores for the first principal component (PC1) on the sbG (2320\,MHz) Stokes V profile residuals, plotted as a function of hour angle before (top) and after (bottom) applying MTM corrections. Prior to the correction, the PC1 scores exhibit a clear dependence on telescope geometry, which is removed following the application of the MTM corrections.}

    
    \label{fig:MTM_Corr}
\end{figure}

During early testing of the time-series analysis pipeline, we identified unexpected temporal and phase-localised behaviour in the Stokes V profile residuals of the higher-frequency sub-bands (particularly sbG and sbH) in the outer regions of the pulse profile. To investigate whether this effect was astrophysical or due to systematics induced by calibration errors, we performed Principal Component Analysis (PCA) on the Stokes V residuals (methodology further described in \ref{PCAVar}), and plotted the scores as a function of observing hour angle and other observatory-specific parameters (see Fig. \ref{fig:MTM_Corr}). Prior to MJD 58900, the scores of the first principal component were distributed approximately around zero, as expected. However, later observations varied sinusoidally as a function of Hour Angle, indicating that the observed anomalous profile residuals indeed originated from polarization calibration errors. We investigated other MSPs in the PPTA sample and found this issue also affected PSR~J1744$-$1134, but not others, indicating the error is sensitive to specific details of the pulsar polarization state and/or S/N.

We determined that this observed systematic behaviour originated from the polarimetric calibration method used during the processing of the upcoming PPTA DR4, in which a full polarimetric calibration model (PCM), derived from the UWL receiver in 2018, was applied across the full span of the UWL observations. While the solution provided stable calibration during the first few years of the dataset, it became increasingly unstable at later epochs, suggesting that the receiver polarization characteristics (e.g., ellipticity and orthogonality of the feeds) drifted away from the state measured in 2018. Further details will be provided in the PPTA DR4 publication (Wang et al., \textit{in prep.}).

To correct these systematics, we use the Matrix Template Matching (MTM -- \citealt{2004ApJS..152..129V}) solutions generated for each epoch when the DR4 MTM ToAs were measured. To do so, we load the corresponding MTM solutions for each epoch and sub-band and extract the associated feed parameters, which describe the differential gain, differential phase, and the boost and rotation terms that characterize the feed ellipticity and orthogonality. Following the formalism from \cite{2000ApJ...532.1240B} we used these parameters to construct the corresponding boost and rotation matrices, which are then combined to form the full Mueller matrix (Eq. \ref{MuellerMatrix}) describing the transformation of the Stokes parameters:
\begin{equation}
\label{MuellerMatrix}
\begin{aligned}
\mathbf{M} =\;
&G\,
\mathbf{B}_{\hat{q}}(2\gamma)
\mathbf{R}_{\hat{q}}(2\phi_{I})
\mathbf{B}_{\hat{u}}(\delta_{\theta})
\mathbf{B}_{\hat{v}}(\delta_{\chi}) \\
&
\mathbf{R}_{\hat{u}}(\sigma_{\chi})
\mathbf{R}_{\hat{v}}(\sigma_{\theta})
\mathbf{R}_{\hat{v}}(2\zeta) 
\mathbf{R}_{\hat{v}}(2\phi_{\rm iono})
\mathbf{R}_{\hat{v}}(2\phi_{\rm ISM})\,\mathbf{S}
\end{aligned}
\end{equation}

\noindent Here, $G$ represents the absolute gain, however, we set this to 1 because we normalize our profiles later in our pipeline. $\gamma$ describes the differential gain between the orthogonal receptors, and $\phi_{I}$ represents the differential phase. The boost terms $B_{\hat{u}}(\delta_{\theta})$ and $B_{\hat{v}}(\delta_{\chi})$ describe the instrumental mixing between the Stokes parameters due to deviations in the receiver orthogonality ($\delta_\theta$) and ellipticity ($\delta_\chi$), while the rotation terms $R_{\hat{u}}(\sigma_{\chi})$ and $R_{\hat{v}}(\sigma_{\theta})$ account for rotation of the polarization state due to net deviations of the receiver ellipticity and orthogonality. The remaining terms, $R_{\hat{v}}(2\zeta)$, $R_{\hat{v}}(2\phi_{\rm iono})$, and $R_{\hat{v}}(2\phi_{\rm ISM})$, describe the parallactic angle rotation, ionospheric Faraday rotation, and interstellar Faraday rotation, respectively. As part of the corrections we apply, these rotations are absorbed into the fitted $R_{\hat{v}}(\sigma_{\theta})$ term and are not applied as separate rotation terms. 

We then invert the Mueller matrix and apply this inverted matrix to the observed Stokes vectors for every phase bin and frequency channel. This produces the final corrected Stokes vector using:

\begin{equation}
\mathbf{S}_{\rm corr}
=
\mathbf{M}^{-1}
\mathbf{S}_{\rm obs}
\end{equation}

where

\begin{equation}
\mathbf{S_{\rm obs}}
=
\begin{bmatrix}
I \\
Q \\
U \\
V
\end{bmatrix}
\end{equation}

This MTM correction process removes the systematic issues created by instrumental mixing between the Stokes parameters caused by the long-term application of the 2018-derived calibration solution and recovers the corrected polarimetric profiles. We exclude channels that contain invalid or non-finite calibration solutions from the analysis. While this MTM correction successfully removed the instrumental systematic in sbA--sbG, we found that it did not adequately correct sbH, where the calibration errors were most dominant at higher frequencies. To further mitigate these effects, we removed frequency channels in the top half of sbH. This reduced the telescope-geometry-dependent structure observed in the Stokes V residuals; however, some lower-level structure persisted in this band. This residual structure was not evident in the linear-polarization products after corrections, and across the band, we use the Stokes V PCA only as a diagnostic of the calibration systematics. We demonstrate the effectiveness of this correction in Figure \ref{fig:MTM_Corr}, where we compare the PC1 scores of the Stokes V residuals in sbG as a function of the hour angle before and after application of this correction. As can be noted, the MTM calibration corrections successfully remove the clear sinusoidal signature originating from polarization calibration errors, leaving PC scores that are randomly distributed around zero, with a comparable distribution to that prior to the correction. The corrected frequency channels are combined using the existing channel weights stored within each observation to produce the updated Stokes \textit{I}, \textit{Q}, \textit{U} and \textit{V} profiles. 

\subsection{Construction of Profile Residuals and Data Products}
\label{subsec:res_create}
After constructing the sub-banded epoch-customised templates, we generate profile residuals. Template and observation files are loaded using \textsc{PSRCHIVE} which contain all four Stokes parameters. For each template and observation, we derive polarimetric products, including the linear polarization profile and PA, following the method described in Section 2.3 of \cite{2025PASA...42..142M}. Here we utilise the image-based debiasing method from \cite{2017A&A...606A..41M}, modified for one-dimensional pulse profiles, to construct the linear polarization profile. This procedure mitigates the Ricean bias introduced when estimating the linear polarization intensity from Stokes Q and U profiles, while maintaining the underlying noise statistics.

A custom sigma-clipping routine is used to identify the on-pulse and off-pulse regions (hereon: `sigma clip’). In this routine, the Stokes I profile is iteratively clipped (up to 30 passes) by masking bins where the absolute amplitudes exceed 2$\sigma$, calculated from the remaining bins at each pass. This iteratively removes the pulse profile component and isolates the off-pulse baseline region. After convergence, the root mean square (RMS) of these remaining bins is used as the off-pulse noise estimate for Stokes I ($\sigma_{I}$). The excluded bins are taken as the on-pulse region, and a smoothing mask is applied to exclude noise spikes from the on-pulse mask, ensuring contiguous on-pulse regions which then define the on-pulse boundaries. The routine is repeated for Stokes Q and U (to also obtain $\sigma_{Q}$, $\sigma_{U}$). These values are used to compute the PA uncertainties:

\begin{equation}
\sigma_{\mathrm{PA}} = 
\frac{180}{\pi}
\sqrt{
\frac{
Q^{2}\sigma_{U}^{2}
+
U^{2}\sigma_{Q}^{2}
}{
4\left(Q^{2}+U^{2}\right)^{2}
}
}
\end{equation}

We exclude low S/N PA measurements by masking phase bins where the Stokes I signal falls below a 3.5$\sigma$ noise threshold. Stokes I provides a robust proxy for the presence of on-pulse emission across all Stokes profiles, restricting the PA analysis to the on-pulse region where polarization measurements can be reliably obtained. We then apply additional filtering by removing PA values with uncertainties above the 85th percentile of the PA error distribution. 

To account for variations in the absolute flux density in a single sub-band (e.g. as caused by scintillation), we normalize both the epoch-customised templates (described in Section \ref{subsec:template_create}) and the observation using the integrated flux density of the Stokes I profile over the on-pulse region, as defined by the sigma-clipping procedure. 

We then align the profiles to the Stokes I component of the templates using the Fourier phase-gradient algorithm of \cite{1992RSPTA.341..117T}, which determines the phase shift based on the Fourier phase gradient of the profile with respect to the template. The derived FFT phase-shift is applied to the full-Stokes normalized profiles, after which the linear polarization, PA, and PA uncertainties are derived from the shifted Stokes parameters. While producing PA residuals, we found small systematic offsets between the template PA and the epoch PA across pulse phase, which may be caused by differences in the RM correction methodology between the templates and the observations. As these offsets are not related to the PA variations investigated in this work, we remove them using an alignment process. For each epoch, we calculate the median PA difference between the template PA and observation PA across all on-pulse phase bins. We then apply this offset to the observation PA profile to align it with the template PA profile.

To construct the profile residuals, we subtract the Stokes I, Linear Polarization, and Stokes V epoch-customised template from each observation. PA residuals are constructed in an identical manner by subtracting the template PA from each observation. 

\subsection{Final Data Cleaning}
\label{subsec:cleaning}
We employ both an absolute off-pulse RMS noise cut and an S/N cut to remove poor-quality data that may bias downstream results. We achieve the first, absolute off-pulse RMS cleaning by measuring the distribution of off-pulse RMS noise, $\sigma_I$, determined using the sigma-clipping procedure described in section \ref{subsec:res_create} for each sub-band. For this distribution, we determine the width of the distribution using a robust median absolute deviation (MAD), and apply a threshold (defined as the median plus three times the robust standard deviation to remove outlying high-noise profiles). Observations are flagged when two criteria are simultaneously satisfied: (i) $\sigma_I$ exceeds this threshold, and (ii) the peak S/N falls below an empirically adopted value of S/N = 1880. We adopt this value based on the sbF data in J0437, which provides a consistently high S/N reference band for assessing observation quality. The S/N value is used as a secondary quality metric, ensuring that epochs are removed only when the elevated noise is accompanied by reduced signal strength. 

In a similar fashion, we derive the off-pulse RMS of profile residuals ($\sigma_{res}$) following normalization and alignment, and determine the excision threshold using the same MAD-based procedure. This excises profiles with suitable absolute off-pulse RMS but a low profile S/N that may bias downstream results.

Finally, we perform a final layer of outlier cleaning based on PCA scores vs. time (see Section \ref{PCAVar}). PCA scores are highly sensitive to dominant structures in data, such as those caused by outliers or bad data, and can be used to automatically detect remaining bad epochs to excise from our final sample. We achieve this by measuring the PCA scores vs time for the first ten PCA components. For each component, we fit a high-order polynomial and subtract the fit from the data, deriving PC score residuals. Then, similar to previously, we measure the distribution of these PC score residuals and identify data points more than 3$\sigma$ from the fitted polynomial, and excise these as outliers. We protect the primary event epoch (59670) and recovery interval from outlier removal when a PC exhibits the characteristic event-like signature. Further details of our identification of event-related eigenprofiles is provided in Section \ref{PCAVar}.

\section{Results}
\label{sec:results}
\subsection{Profile shape variability}
\label{ProVar}

\begin{figure*}
    \centering
    \includegraphics[width=\textwidth]{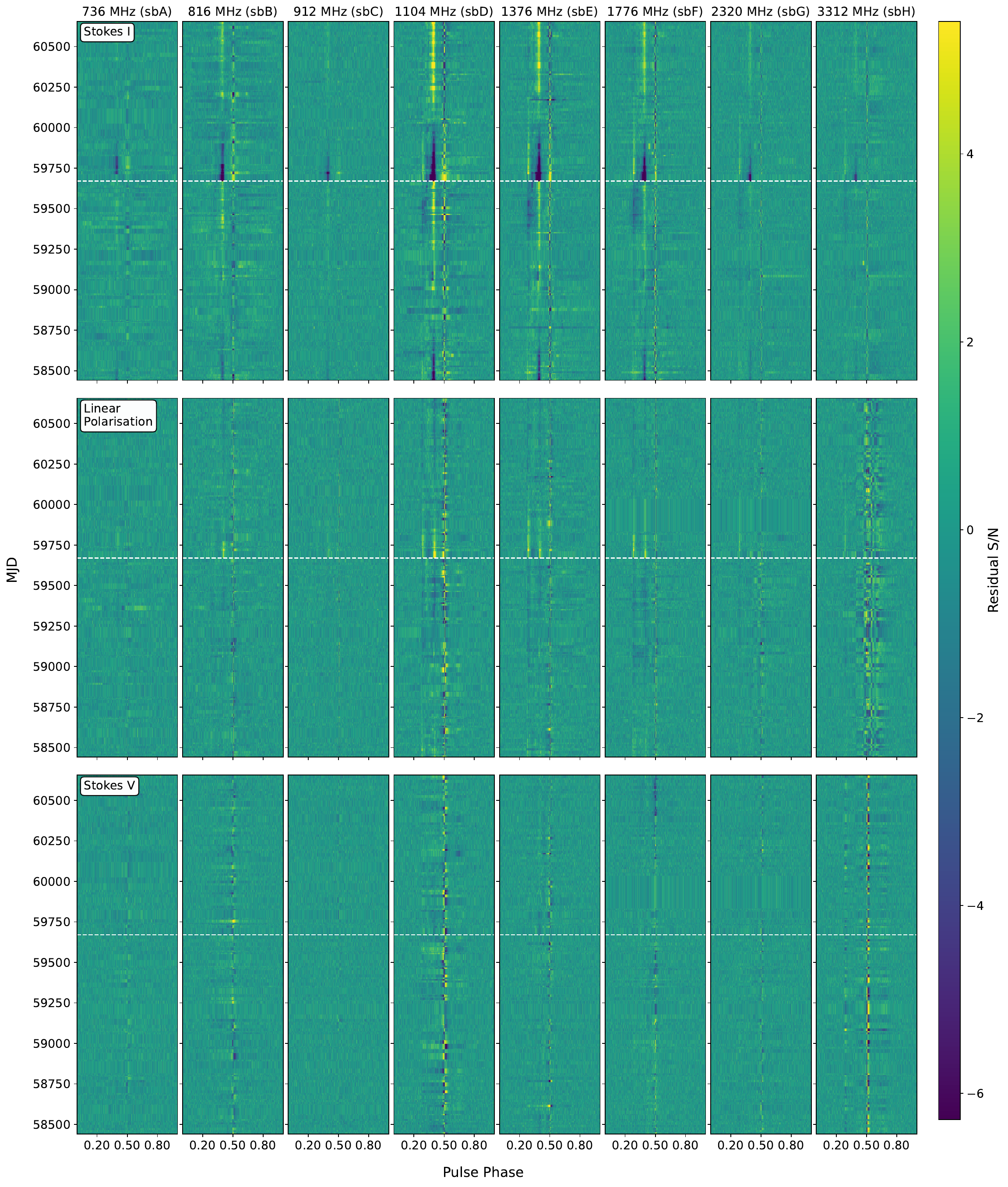}
    \caption{Residual S/N for all UWL sub-bands in Stokes I (top panels), Linear Polarization (middle panels), and Stokes V (bottom panels). For cosmetic purposes, rather than plotting the raw residuals, we display the residual S/N, obtained by normalizing each residual profile by the RMS of the off-pulse region (here, we specify the off-pulse regions as $\phi$ 0.00--0.14 and $\phi$ 0.85--1.00). For each sub-band, the mean residual profile is computed across all epochs and subtracted from the time series to remove any constant baseline offsets. The dashed white line marks the onset of the primary event at MJD 59670.}

    
    \label{fig:Res_Waterfalls}
\end{figure*}

In Figure \ref{fig:Res_Waterfalls}, we present the S/N of the profile residuals as a time series from October 2018 (MJD 58841) to January 2025 (MJD 60694). Residuals for Stokes I, Linear Polarization, and Stokes V are shown in separate rows, with each column corresponding to a sub-band. We exclude several epochs between MJD 59840 and 60025 in the Linear Polarization and Stokes V residuals for sbF and sbG, due to corruption in the raw polarimetric data. However, we retain these observations in Stokes I, where no corresponding issue is identified.

In Stokes I, we identify a primary profile change event (commencing at MJD 59670 -- marked by the dashed black line). Three phase-localised features exhibit frequency-dependent behaviour associated with this event. We define these as the ‘Left Feature’, centred at phase $\sim 0.28$ with a narrow phase extent (0.27–0.29), the ‘Main Feature’, centred at phase $\sim0.4$ with a broader phase extent (0.35--0.42), and the ‘Central Feature’, located near the profile centre (0.49--0.51). The phase locations and their phase extent of these three features are presented in Figure \ref{fig:Temps}. All three features show clear temporal evolution. At the event epoch (MJD 59670), the Main Feature residuals show a sharp transition from positive to negative values, with the negative residuals persisting until approximately MJD 60000. After passing through zero, the residuals become positive and continue to increase in magnitude through to the end of the dataset. The Left and Central Features follow similar temporal evolution, but with opposite sign. The Central Feature shows stochastic variability superposed on these long-term trends.

In addition to the primary event, we report the discovery of a second discrete profile change event that predates the UWL dataset. This is clearly evident in our Stokes I profile residuals, which show behaviour that mimics the primary event but in early epochs of our dataset, occurring at precisely the same phase locations as the primary event's Left, Main and Central Features. Similar evidence is also seen in the Linear Polarization residuals in the higher S/N subbands. To investigate when the onset of this prior event occurred, we analysed the pre-UWL observations from the PPTA archive and identified the date as MJD 58082. Therefore, the temporal evolution at the beginning of our dataset corresponds to the long-term recovery of this prior discrete profile change event. The Left, Main and Central Features all follow the same temporal morphology as the primary event, showing negative / positive residuals at early epochs, before transitioning to zero and becoming positive / negative prior to the sudden onset of the primary event. However, the Central Feature associated with the previous event exhibits less coherent evolution, as the contribution from stochastic variability becomes more dominant.

We observe this behaviour across the full UWL bandwidth for both the primary and preceding events, albeit with some differences that reflect both data quality the underlying frequency dependence. In the lowest-frequency sub-bands (sbA--sbC; 736--912\,MHz) in Figure \ref{fig:Res_Waterfalls}, the residual structures are present but less prominent due to lower S/N. Among these, sbB (816\,MHz) shows the clearest representation of both the Main and Central Features for both events, while the Left Feature is not clearly detected. The largest residual amplitudes are observed in the mid-frequency sub-bands (sbD--sbF; 1104--1776\,MHz), which also have the highest S/N. In these bands, the negative residuals of the Main Feature are detected over a broader phase range, and the Central Feature is clearly seen. At higher frequencies (sbG--sbH; 2320--3312\,MHz), the Main and Left Features decrease in both amplitude and phase extent. In these sub-bands, the Central Feature is not clearly detected. Residual structures associated with the earlier event are also weakly present in these bands.

\begin{figure}
    \centering
    \includegraphics[width=\columnwidth]{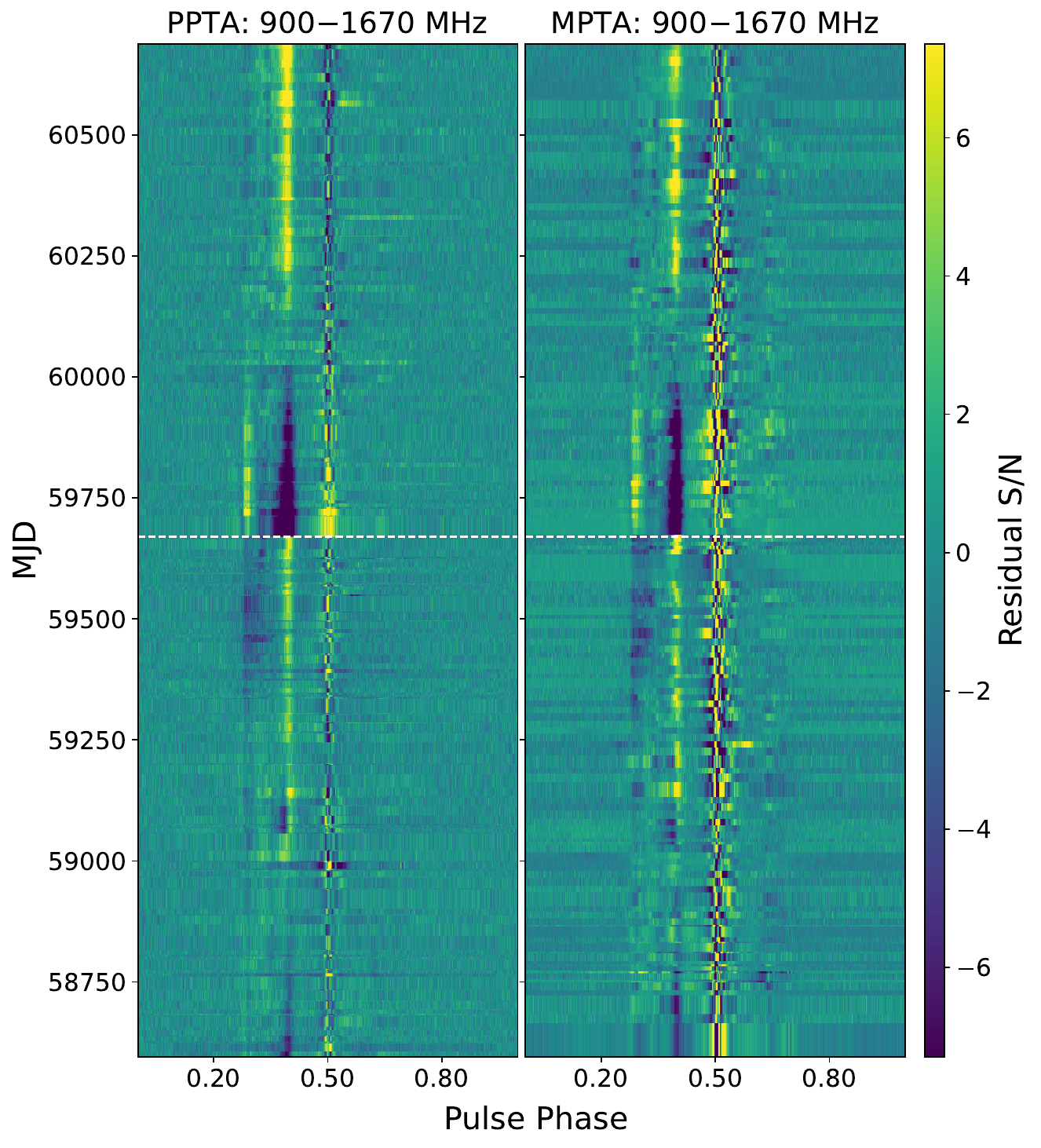}
    \caption{Stokes I profile residuals observed as part of the PPTA (left) and MPTA (right), plotted as a function of time. The dashed white line marks the onset of the primary event at MJD 59670. Features of the profile change exhibit closely matched temporal evolution in both datasets.}
    \label{fig:PPTA-MKT}
\end{figure}

We provide independent verification of the profile change events and variability on J0437 in Figure \ref{fig:PPTA-MKT}. This compares our results against profile residuals obtained with the MeerKAT Pulsar Timing Array (MPTA -- \citealt{2023MNRAS.519.3976M}) program, which uses the MeerKAT radio telescope in South Africa, operated by the Southern African Radio Astronomy Observatory (SARAO). These data were provided through ongoing collaboration with members of the MPTA and form part of an upcoming publication (Bhat et al., \textit{accepted}). The MPTA profile residuals are produced using an equivalent methodology to this work, and cover the MeerKAT L-Band frequency range (spanning 900--1670\,MHz). To enable direct comparison, we combined our UWL sub-bands spanning the same frequency range, mapped both data sets on to a common daily time grid, interpolating by carrying the most recent data forward on to the daily time grid between observing epochs, and generated residual S/N maps using the same plotting conventions. The comparison in Figure \ref{fig:PPTA-MKT} confirms both telescopes independently detect the primary event at MJD 59670, the recovery tail of the prior event, and the associated temporal evolution of the Main, Left and Central Features. This agreement between these independent datasets provides confidence that the detailed morphology and temporal evolution we observe in the PPTA dataset are robust, as they are independently reproduced by the MeerKAT observations.

Additionally, we observe a short-duration dip in the profile residuals in both datasets between approximately MJD 59000-59150. This small negative residual feature is located immediately left of the Main Feature at $\phi$ 0.38. Initially, we considered this a possible outlier in the PPTA data, but its presence at an almost identical pulse phase and epoch range in the MPTA observations also suggests this is a genuine astrophysical feature. We also note that the MPTA data exhibits stronger stochastic variability in the Central Feature, due to pulse jitter \citep{2021MNRAS.502..407P}, as the shorter per-epoch integration times of the MPTA observations ($\sim$256 seconds) result in less averaging of pulse-to-pulse variability, relative to the PPTA observations ($\sim$ 64 minutes).

\subsection{Polarimetric analysis of the profile change}
\label{sec:PolVar}
The three components identified in Stokes I (Main Feature, Left Feature and Central Feature) are also present in the linear polarization profile residuals. The Main Feature is detected in the primary event across sbA--sbF. Of these, sbA and sbC remain noise-dominated, whilst this feature is strongly present in sbB, sbD, sbE and sbF, exhibiting temporal evolution on the same time-scales as Stokes I. The Main Feature is not detected in sbG or sbH, reflecting a decrease in the linear polarization residual amplitude at higher frequencies.

The Left Feature is clearly detected in sub-bands sbD--sbH, with the strongest residual amplitudes observed in the high S/N bands sbD--sbF. Similar to the Main Feature, the Left Feature exhibits the same long-term temporal morphology associated with the primary event. Unlike Stokes I, however, the Main and Left Features exhibit residuals of the same sign in linear polarization, indicating that although these two components demonstrate an anti-correlated response to the profile change event in total intensity, their polarized emissions evolve coherently throughout the event.

Similarly to Stokes I, the Central Feature in linear polarization exhibits stochastic variability unrelated to either the primary or the previous profile change event. This feature is detected in all sub-bands, excluding sbA and sbC, where the lower S/N limits its visibility, and is most prominent in sbD. In sbH, the Central Feature broadens in phase, exhibiting additional structure, although no coherent temporal evolution is apparent.

The previous event is also detected in linear polarization. The Main and Left Features are visible in sbD--sbF, and the Central Feature is present in sbB and sbD--sbH. As observed in Stokes I, the temporal evolution of the previous event resembles that of the primary event.  

In Stokes V, the Main and Left Features are only weakly observed. Faint residuals associated with the Main Feature are present in sbE and sbF during the primary event, whilst the Left Feature is only marginally detected. The discrete profile change events are not observed in Stokes V outside of these bands. The dominant residual structure in Stokes V is the Central Feature, which is observed across all sub-bands except sbC (due to its lower S/N). The Central Feature of Stokes V does not exhibit any coherent temporal evolution, only the same stochastic variability seen in this component in Stokes I and linear polarization. 

The Central Feature in the profile residuals exhibits stochastic variability across all epochs in both linear and circular polarizations, distinct from the smooth temporal evolution associated with discrete profile change events. This stochastic behaviour in the same profile region was also noted in \cite{1983ApJ...274..333R,1990ApJ...352..247R,1997MNRAS.285..561G}, and \citealt{1998ApJ...501..823V}. In Stokes V, the Central Feature also shows additional phase structure in two sub-bands across all epochs. In sbD, broad residual structures are observed on either side of the feature, while in sbH, the residuals exhibit two broad artefacts before and after the central bins (approx. across $\phi$ 0.27--0.30, and $\phi$ 0.60-0.70). These phase regions coincide with the phase locations of the temporal polarimetric systematics identified prior to the correction of polarimetric calibration effects (see \ref{subsec:MTM_correction}). Although stochastic variability remains in these regions after corrections, we cannot confidently determine whether it is entirely astrophysical.

\begin{figure}
    \centering
    \includegraphics[width=\columnwidth]{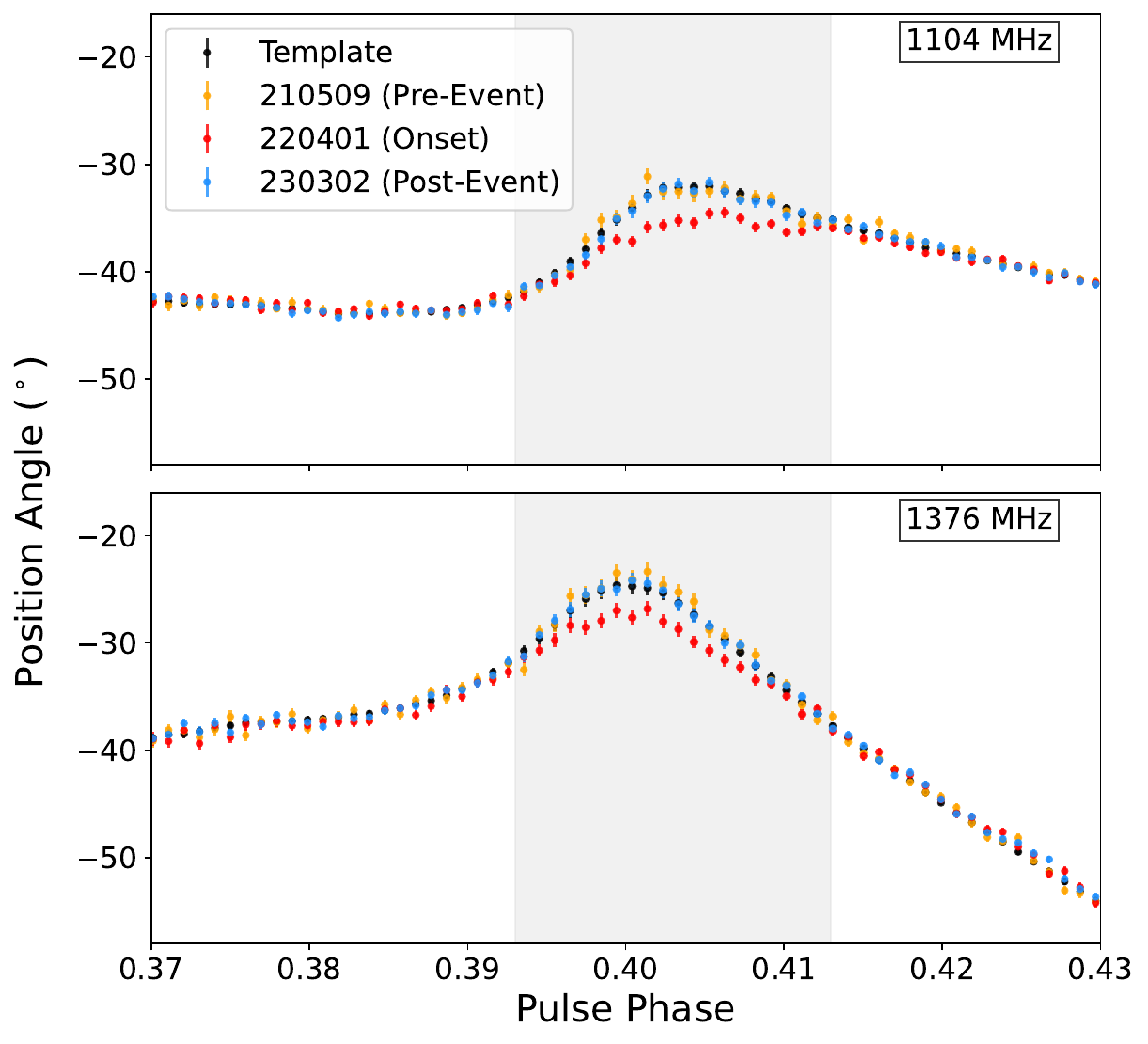}
    \caption{Temporal evolution of the polarization position angle (PA) across the Main Feature for sbD (1104\,MHz; top) and sbE (1376\,MHz; bottom), compared with the template PA. The template PA is represented by the black data points; yellow data points correspond to a high S/N epoch prior to the event occurring (MJD 59343); red data points correspond to the date of the primary event (MJD 59670), and blue data points correspond to a high S/N epoch occurring after the event (MJD 60005). The grey band marks the phase range over which the PA variations are observed within the broader Main Feature. The largest deviation occurs at the event onset (MJD 59670).}
    \label{fig:PA_shift}
\end{figure}

In Figure \ref{fig:PA_shift}, we show the PA as a function of pulse phase, in a phase range centred on the Main Feature, sbD (1104\,MHz) and sbE (1376\,MHz). The template PA is plotted together with three representative epochs selected for their high S/N: a pre-event observation (01 May 2021), the event onset (2022 April 01), and a post-event observation (2023 March 02). At the pre-event and post-event epochs, the PA values are consistent with the template. At the event epoch, the PA deviates from the template over a narrow phase range that corresponds to the Main Feature. We do not observe deviation outside this phase range. Although only the event epoch is shown, this PA deviation decreases in magnitude in the following epochs, with the PA returning to values consistent with the template by approximately MJD 59776.

\subsubsection{Fractional Polarization Analysis of the profile change}
\label{FracPol}
\begin{figure*}
    \centering
    \includegraphics[width=\textwidth,height=0.44\textheight,keepaspectratio]{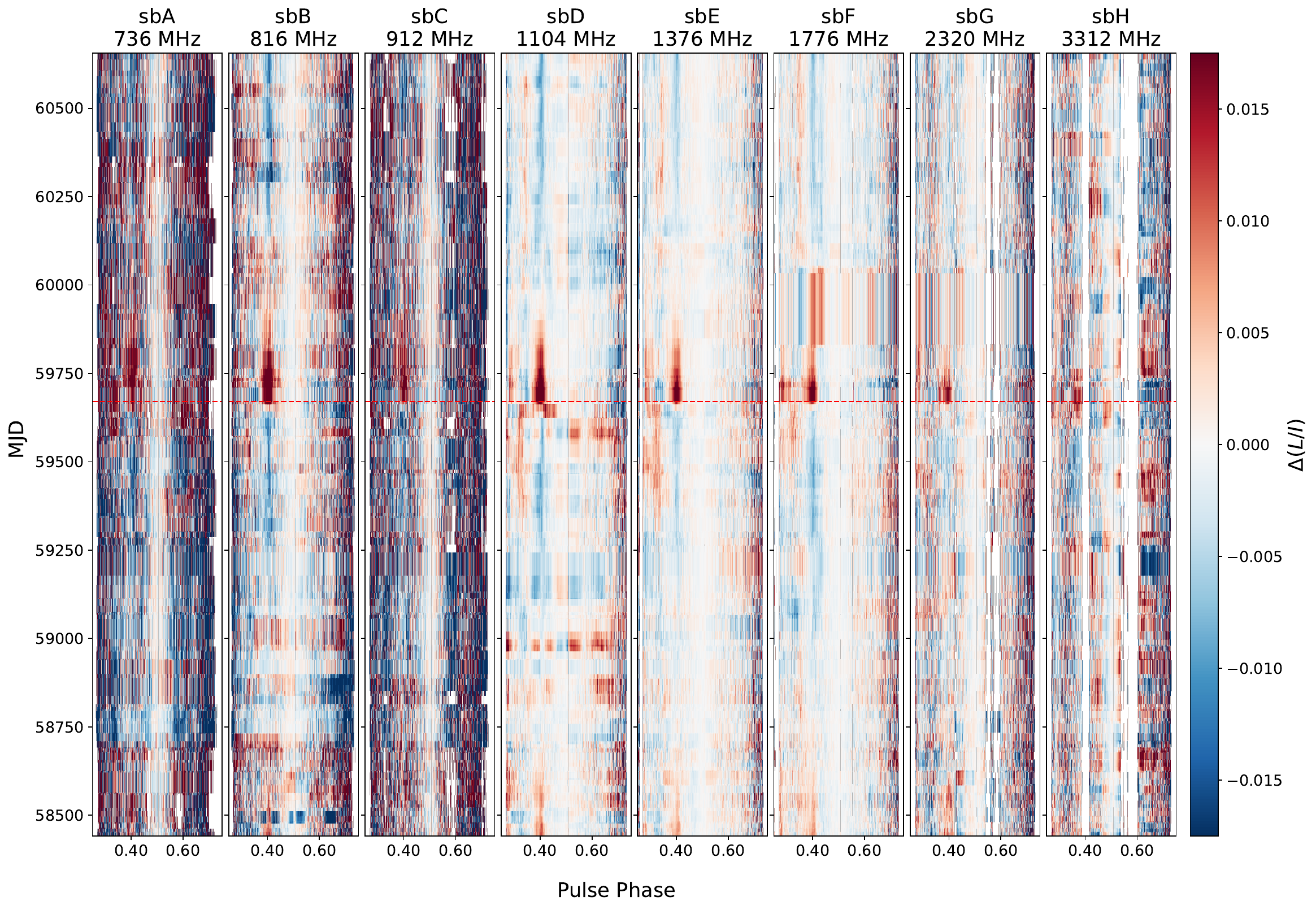}
    \caption{Fractional linear polarization ($L/I$) as a function of frequency and time, smoothed over 7 observing epochs (approximately 100 days). The red dashed line marks the primary event onset (MJD 59670). A sudden change occurs at the Main Feature across the full bandwidth where the ($L/I$) increases, followed by a gradual recovery before sign reversal. Similar behaviour is observed at the same phase location during the recovery of the prior event.}
    \label{fig:LI_FracPol}
\end{figure*}

We further investigate the polarimetric behaviour of the profile change events by calculating the fractional linear polarization ($L/I$), as a function of pulse phase and time. For each phase bin, we subtract the mean $L/I$ from the data, producing $\Delta(L/I)$, representing deviations from the long-term average fractional polarization. We note that the epochs removed from the linear polarization and Stokes V waterfall grid in sbF and sbG in Figure \ref{fig:Res_Waterfalls} (MJD 59840--60025) are also excluded from the fractional polarization analysis, for the same reasons -- these data were affected by corruption in the raw polarimetric products. We present the $\Delta(L/I)$ time series in Figure \ref{fig:LI_FracPol}. Positive values correspond to epochs where the fractional linear polarization is enhanced relative to the mean, and negative values indicate reduced fractional linear polarization. The highest S/N sub-bands are (sbD--sbF), and we focus primarily on those here. 

At the phase location of the Main Feature ($\phi \sim 0.4$), the fractional linear polarization exhibits a sharp increase at the onset of the primary event. This gradually decreases with time, returning to the mean state (approx. MJD 60000) before evolving into a weakly negative fractional polarization state at the end of the dataset. The same phase location also exhibits a similar response during the previous event, though only weakly detected. The amplitude of this feature shows an apparent frequency dependence, decreasing with increasing frequency. This broadly follows the behaviour observed in the Stokes I residuals.

We also examined the fractional circular polarization \textit{V/I} and the absolute fractional circular polarization |\textit{V}|\textit{/I} using the same procedure. In contrast to $\Delta$ \textit{L/I}, we find no evidence that the discrete profile change events modified the fractional circular polarization of the emission.

\subsection{Principal Component Analysis on Profile Variability}
\label{PCAVar}
We performed PCA on the Stokes I residuals for each sub-band, producing eigenprofiles and corresponding PC score time series for each epoch. This allows us to decompose the different modes of profile variability and isolate the component associated with the discrete profile change event, thereby enabling the study of the temporal evolution of the profile change independently of other sources of variability encoded in the profile residuals. The PCA was implemented using the \textsc{sklearn} library (\citealt{scikit-learn}), following the method described in \citealt{2025PASA...42..142M} (Section 3.1.1).

As the primary focus of this work is the discrete profile change event, we analyse the first 10 principal components and visually inspect their eigenprofiles and PC score time series to determine their association with specific profile features (Main, Left, Central). Event-like PCs are identified by searching for the characteristic temporal morphology of the profile change, which consists of a sudden deviation at the event epoch, followed by a gradual recovery towards the pre-event baseline. We note that in sbA, the profile change signal is distributed between two PCs with similar eigenprofiles and temporal behaviour. This is expected as this is a low S/N band. These components are therefore combined to improve the S/N of the profile change signal and decouple it from the background noise.

\begin{figure*}
    \centering
    \includegraphics[width=\textwidth,height=0.44\textheight,keepaspectratio]{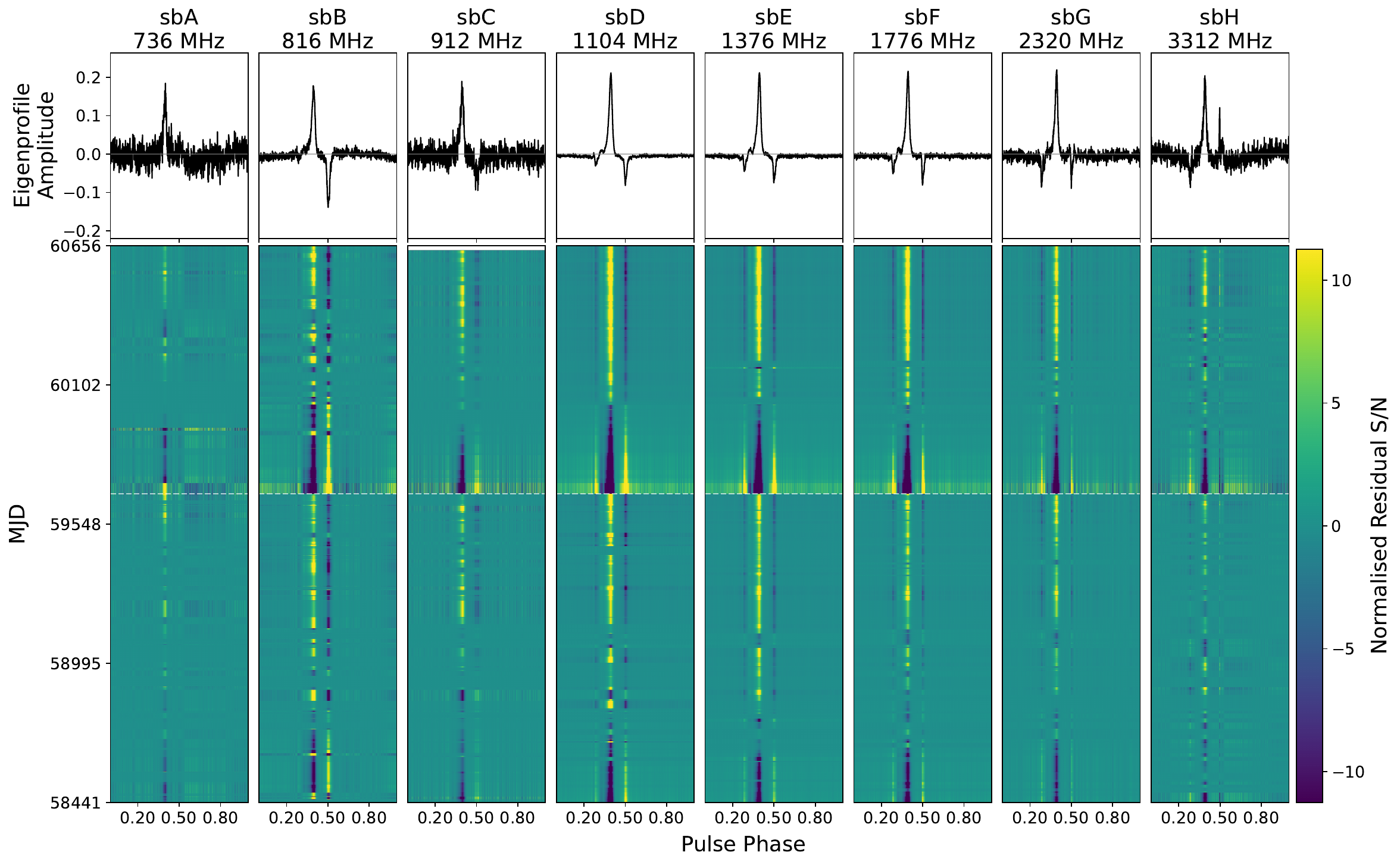}
    \caption{PCA decomposition of the Stokes I profile residuals associated with the profile change event component as a function of frequency. Top: PC1 eigenprofiles showing the three phase-localised features of the profile change. Bottom: normalized residuals reconstructed from PC1. The prior and primary events exhibit similar temporal evolution at the same three phase locations. The white dashed line represents the onset of the primary event at MJD 59670.}
    \label{fig:Eig_WFs}
\end{figure*}

In Figure \ref{fig:Eig_WFs} we present the eigenprofiles of the PCs associated with the profile change event for each sub-band (top row) with corresponding waterfall residual maps constructed from the PCs. These PC waterfall residual maps are generated by multiplying each eigenprofile by its associated PC score at each epoch, producing a time series representation of the component as a function of pulse phase and time in the same format as the profile residual maps in Figure \ref{fig:Res_Waterfalls}. The dominant component in the eigenprofiles is the Main Feature, centred at phase $\sim 0.4$, which appears as a positive peak across all sub-bands. A weaker trough in the eigenprofiles at phase $\sim 0.3$ corresponds to the Left Feature. A third component is also present around phase $\sim 0.5$, corresponding to the Central Feature. This appears as a negative feature in the eigenprofiles, before changing sign in sbH. 

By isolating the PC associated with the profile change, we have effectively removed the remaining variability modes, providing the clearest representation of the temporal evolution of the discrete profile change events across all sub-bands, including both the primary and previous events. The Main, Left and Central Features all exhibit the same temporal morphology, consisting of a gradual recovery from the previous event, a sudden transition at the onset of the primary event, and long-term recovery thereafter. 

The Main Feature exhibits PC residuals of opposite sign to both the Left and Central Features, except in sbH. This anti-correlated behaviour is observed in both the primary and preceding events and is clearest in sbD--sbF, where the S/N is highest. 

We also observe a clear frequency dependence in the morphology of the profile change in this figure. The Left Feature is weak or absent in lower-frequency sub-bands but becomes increasingly prominent towards higher frequencies. In contrast, the Central Feature is strongest at lower frequencies and decreases in amplitude as frequency increases. In sbH, the Central Feature appears to change sign and no longer exhibits the anti-correlated behaviour observed at lower frequencies; however, given the reduced S/N at high frequencies, this apparent reversal may be an artefact of the PCA decomposition rather than a genuine variation of the profile morphology. The Main Feature remains clearly visible across the full observing band and shows little frequency dependence in the amplitude of the eigenprofiles or PC scores. 

\begin{figure*}
    \centering
    \includegraphics[width=\textwidth,height=0.38\textheight,keepaspectratio]{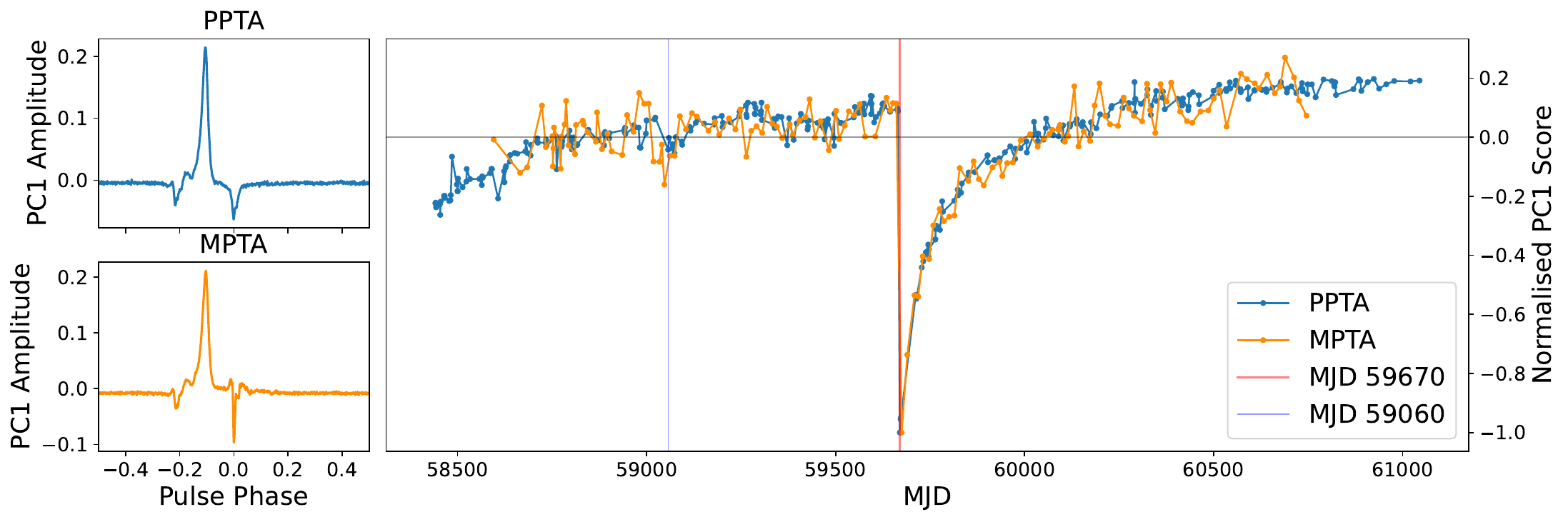}
    \caption{PCA decomposition of the Stokes I profile residuals associated with the profile change event component observed by the PPTA and MPTA. Left: PC1 eigenprofiles, exhibiting near-identical phase-localised features. The additional structure in the Central Feature of the MeerKAT eigenprofile is likely associated with pulse jitter due to shorter integration times of the MPTA observations ($\sim$256 seconds). Right: normalized PC1 scores showing consistent temporal evolution. The additional white-noise-like structure in the MeerKAT data is likely due to the influence of pulse jitter. The red line marks the primary onset (MJD 59670); the faint blue line (MJD 59060) marks a short-duration negative residual feature marginally detected by both telescopes, located at $\phi \sim 0.38$.} 
    \label{fig:PCA_PKS_MKT}
\end{figure*}

Figure \ref{fig:PCA_PKS_MKT} compares the first PC eigenprofiles and the corresponding normalized PC score time series from our PPTA data and the MPTA data discussed in Bhat et al. 2026 (\textit{accepted}). Our residuals are cut to match the MPTA L-band range (960--1670\,MHz), and we normalize by the minimum PC score in each dataset. The eigenprofiles exhibit nearly identical morphology, confirming again that both telescopes detect the same pattern of profile variation. The PC scores closely track one another throughout the observing span, reproducing both the primary profile change event, and the recovery tail of the prior event. We also observe a common deviation in the PC scores near MJD $\sim$59060. We suggest this small deviation corresponds with the marginal, short-duration negative profile residuals identified in Section \ref{ProVar} and Figure \ref{fig:PPTA-MKT}. The MeerKAT PC scores exhibit larger epoch-to-epoch scatter than those from the PPTA, consistent with the shorter integration times providing less averaging over intrinsic pulse variability (jitter). Regardless, the close agreement between the independent datasets provides further confidence in our results, particularly in the temporal morphology and recovery of both these profile change events.

\subsection{Modelling the Profile Change}
\label{sec:PCA_MCMC}

To quantify the temporal and frequency evolution of the profile change events, we simultaneously fit the PC scores from each sub-band using a two-component model that describes both the preceding and primary events. The preceding event is modelled as an exponential decay with its onset fixed at MJD 58082, while the primary event is modelled as a power-law recovery, with an additional linear term to capture long-term evolution. We fix the onset of the primary event at MJD 59670. We selected these functional forms through model selection tests that first compared combinations of exponential and power-law recoveries for the two events, followed by tests of additional model components and the complexity of their frequency dependence. Model selection was based on the relative Akaike information criterion (AIC) and Bayesian information criterion (BIC), with reduced $\chi^2$ values and visual inspection of the fitted models used as additional diagnostics. The AIC and BIC values are calculated as:

\begin{equation}
{\rm AIC} = 2k - 2\ln(\hat{\mathcal{L}}),
\end{equation}

\begin{equation}
{\rm BIC} = k\ln(n) - 2\ln(\hat{\mathcal{L}}),
\end{equation}

\noindent where $k$ is the number of free parameters in the model, $n$ is the number of data points, and $\hat{\mathcal{L}}$ is the maximum likelihood of the fitted model. For each model comparison, we calculate the $\Delta$ AIC and BIC values as: 

\begin{equation}
\Delta{\rm AIC} = {\rm AIC}_{\rm test} - {\rm AIC}_{\rm ref},
\end{equation}

\begin{equation}
\Delta{\rm BIC} = {\rm BIC}_{\rm test} - {\rm BIC}_{\rm ref},
\end{equation}

\noindent where ${\rm AIC}_{\rm test}$ denotes the model being tested and ${\rm AIC}_{\rm ref}$ is the reference model. For the preceding event, the exponential recovery model (M01) is used as the reference model, while for the primary event the exponential recovery model (M03) is used as the reference model. The resulting $\Delta{\rm AIC}$ and $\Delta{\rm BIC}$ values favour the exponential model for the preceding event and the power-law model (M04) for the primary event. For the subsequent model-selection tests, M04 is adopted as the reference model, with the $\Delta{\rm AIC}$ and $\Delta{\rm BIC}$ values favouring M04 over each of the alternative models tested. The results of this model selection can be seen in the Appendix, Table \ref{tab:model_selection_recovery} and Table \ref{tab:model_selection_additional}. 

In both cases, we allow the amplitudes and offsets to vary independently for each sub-band, while the remaining parameters are fitted simultaneously across all sub-bands.  

The exponential decay model for the preceding event is given by:

\begin{equation}
y_{\mathrm{pre}}(t, \nu) = A_{\mathrm{pre}}(\nu)\,\exp\left(-\frac{t - t_{0,\mathrm{pre}}}{\tau_{\mathrm{pre}}(\nu)}\right) + C_{\mathrm{pre}}(\nu),
\end{equation}

\noindent while the power-law recovery model for the primary event is given by:

\begin{equation}
\begin{split}
y_{\mathrm{post}}(t,\nu)
={}&
A_{\mathrm{post}}(\nu)
\left(
1+\frac{t-t_{0,\mathrm{post}}}
{\tau_{\mathrm{post}}(\nu)}
\right)^{-\alpha} \\
&+ C_{\mathrm{post}}(\nu)
+ m_{\mathrm{post}}
\left(t-t_{c,\mathrm{post}}\right).
\end{split}
\end{equation}

\noindent Here, $A_{\rm pre/post}(\nu)$ and $C_{\rm pre/post}(\nu)$ are the frequency-dependent amplitudes and constant offsets, modelled as free parameters for each sub-band. The parameters $t_{0,\mathrm{pre}}$ and $t_{0,\mathrm{post}}$ are the fixed event onset epochs, and $\alpha$ is the power-law index governing the recovery behaviour of the primary event. The term $m_{\mathrm{post}}(t-t_{c,\mathrm{post}})$ represents an additional linear evolution in the post-event baseline $C_{\rm post}(\nu)$, where $m_{\mathrm{post}}$ is a single frequency-independent fitted slope across all sub-bands, and $t_{c,\mathrm{post}}$ is the median epoch of the post-event observations. During model selection testing, we found the inclusion of the $m_{\mathrm{post}}$ is strongly favoured, with its non-inclusion increasing the AIC by 162.03 and BIC by 157.49 (see Table \ref{tab:model_selection_additional}). We centre the linear trend on the median epoch to reduce the covariance between the $m_{\mathrm{post}}$ and fitted $C_{\mathrm{post}}(\nu)$ values.

The frequency dependence of the time-scales is defined as
\begin{equation}
\tau_{\mathrm{pre/post}}(\nu) = \sum_{k=0}^{n} b_{k}(\nu / \nu_{\mathrm{ref}})^k~,
\end{equation}
where $\nu$ is the sub-band centre frequency, $b_{k}$ are the temporal evolution frequency-dependent coefficients, and $\nu_{\mathrm{ref}}$ is the reference frequency (chosen to be 1376\,MHz, corresponding to sbE). We determined the polynomial orders through model selection tests described above, which favoured a linear ($n=1$) frequency dependence for the prior event, and a cubic ($n=3$) frequency dependence for the primary event (see Table \ref{tab:model_selection_additional})

We perform parameter estimation using the affine-invariant Markov Chain Monte Carlo (MCMC) ensemble sampler implemented in \textsc{emcee} with priors restricting the $\tau$ and $\alpha$ parameters to positive values. We initialised the posterior sampling chains using maximum-likelihood estimates of the model parameters derived using \textsc{Scipy}'s \texttt{curve\_fit} routine with the Trust Region Reflective algorithm. We initially estimate the uncertainty on the PC scores from the off-pulse RMS of the corresponding profile residuals ($\sigma_{res}$). We then rescale $\sigma_{res}$ independently for each band using an estimate of the scatter in the normalized model residuals, with the scale factor calculated as $1.4826 \cdot {\rm median}(|r/\sigma_{res}|)$ where $r$ represents the model residuals. This median-based estimator provides a robust measure of the scatter in the data in the presence of outliers. The scale factor 1.4826 converts the median absolute residual scale to a comparable scale of the standard deviation of a normal distribution. The rescaled uncertainties are then used in the Gaussian log-likelihood for the MCMC parameter estimation, while the reduced $\chi^2$ score is calculated from the final model to evaluate the quality of the fit. Finally, we derive median parameter values and 68\% credible intervals from the MCMC posterior distributions.

\begin{figure*}
    \centering   \includegraphics[width=\textwidth,height=1.0\textheight,keepaspectratio]{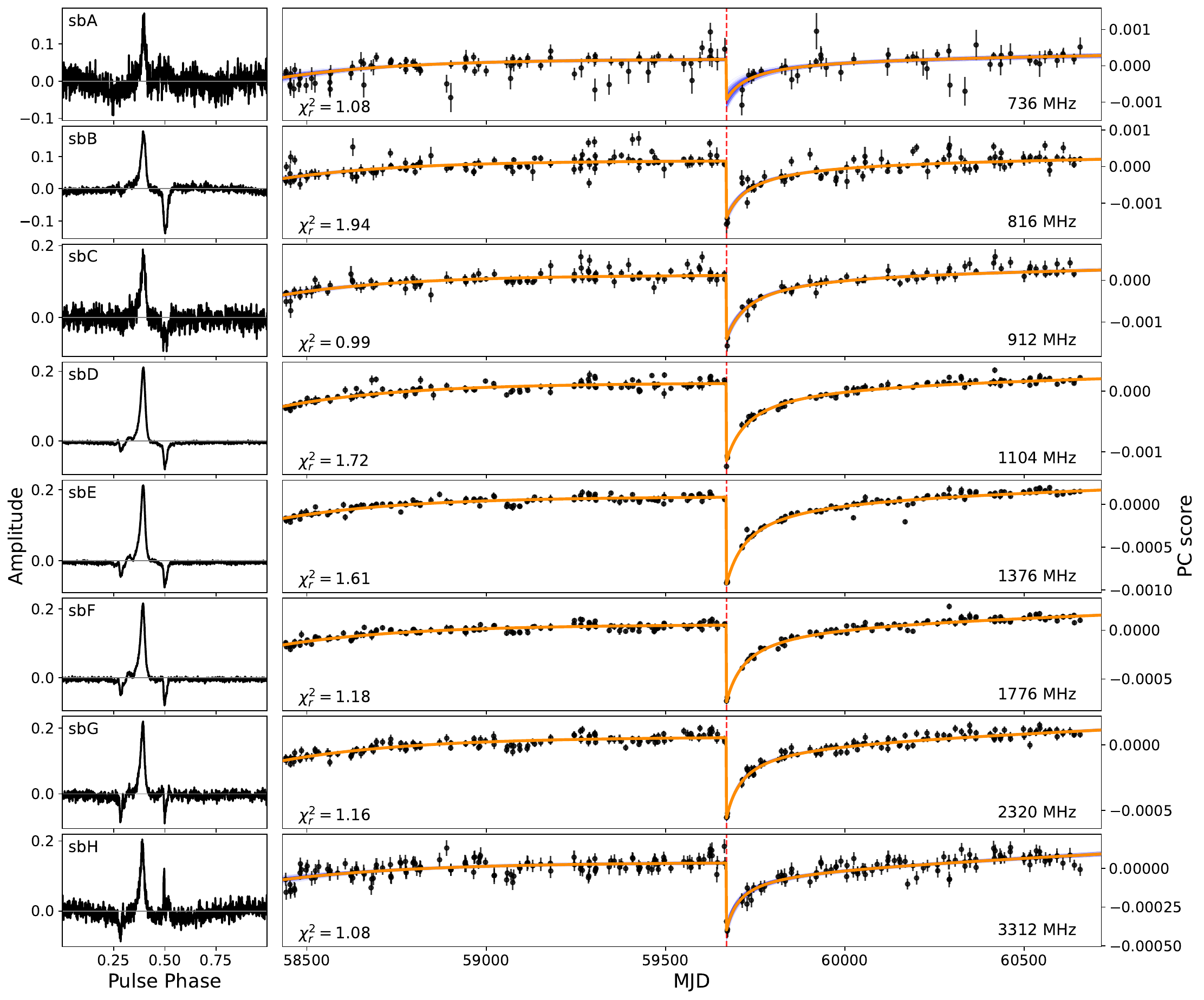}
    \caption{PCA decomposition and modelling of the Stokes I profile residuals associated with the profile change component. Left: eigenprofiles associated with the profile change. Right: corresponding PC scores for the two consecutive events, with the MCMC model fits (orange) representing the median posterior parameter values. The blue curves show 200 randomly selected draws from the MCMC posterior distributions, illustrating the uncertainty in the fitted models. Reduced $\chi^2$ values indicate the goodness of fit in each sub-band. The red dashed line corresponds to the onset of the primary event (MJD 59670)}
    \label{fig:MCMC_Grid}
\end{figure*}

Figure \ref{fig:MCMC_Grid} presents the PC scores as a function of time, with the median posterior parameter values used to evaluate the model, which is overlaid on the data. The models provide a reasonable fit to the data with reduced $\chi^2$ scores ranging from 0.99 to 1.94.

\begin{figure}
    \centering
    \includegraphics[width=\columnwidth]{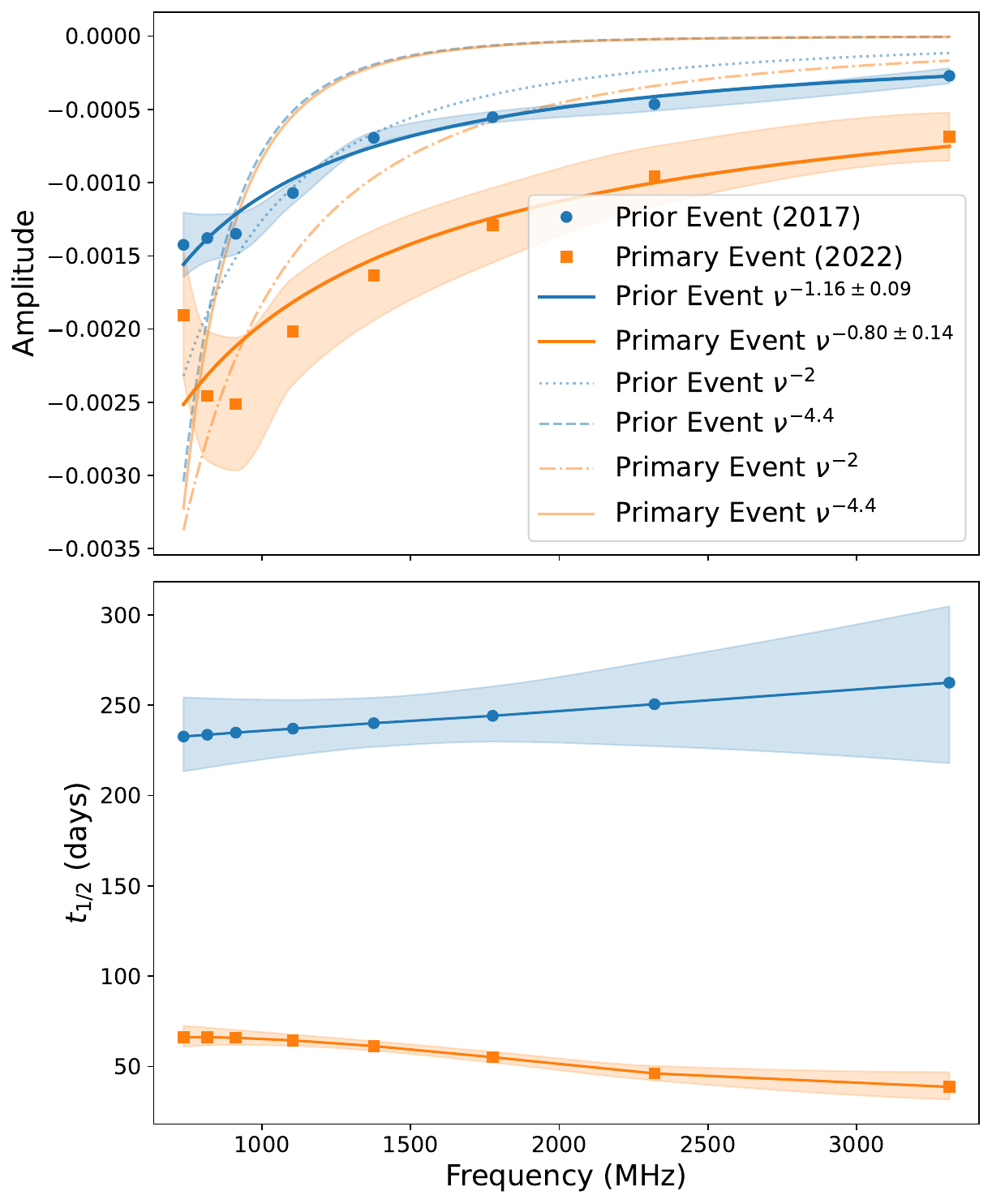}
    \caption{Frequency dependence of the amplitude and recovery half-life derived from MCMC modelling of the Stokes I PC1 score evolution for the prior and primary profile change events. Top: fitted amplitudes in each sub-band (bold blue and orange fits), with shaded regions indicating uncertainties for the prior and primary events, exhibiting similar frequency-dependent behaviour. Power-law fits of the amplitudes are shown in the legend, together with reference $\nu^{-2}$ and $\nu^{-4.4}$ scaling for comparison with dispersive and scattering-like chromatic behaviour. The blue dotted and blue dashed curves respectively represent the prior and primary event $\nu^{-2}$ fits, and the orange dot-dashed and orange solid represent the prior and primary event $\nu^{-4.4}$ fits, respectively. Bottom: recovery half-lives ($t_{1/2}$), derived from the exponential model for the prior event and power-law model for the primary event, providing a common measure of recovery time-scale. Neither event shows strong frequency dependence in recovery time-scales, although the primary event recovers substantially faster across the band.}
    \label{fig:PCA_MCMC_Params}
\end{figure}

In the top panel of Figure \ref{fig:PCA_MCMC_Params}, we present the frequency dependence of the fitted amplitudes. Between sbC--sbH, both events exhibit a systematic decrease in absolute amplitude with increasing frequency. However, in the primary event, the amplitude shows a marginal increase between sbA--sbB, although the values remain consistent within their uncertainties, whereas in the prior event the amplitude remains constant in these sub-bands to within the uncertainties. To quantify this frequency dependence, we fitted the median absolute amplitude parameters recovered from the MCMC modelling for each sub-band, together with their 68\% credible intervals, with a power-law relation, $|A|\propto\nu^{-x}$, obtaining $x=1.16\pm0.09$ for the prior event and $x=0.80\pm0.14$ for the primary event, demonstrating that both events exhibit similar amplitude-frequency scaling. For comparison, we also show the $\nu^{-2}$ and $\nu^{-4.4}$ scaling associated with dispersion and scattering, respectively. For these fixed relations, we used 1104\,MHz (sbD) as the reference frequency and fitted a normalization value ($K$) to the amplitudes of each event, allowing their frequency dependence to be compared directly with the observed amplitude evolution. We discuss the interpretation of these results further in Section \ref{sec:PhysOrg}.

In the bottom panel of Figure \ref{fig:PCA_MCMC_Params}, we present the comparable recovery half-life, $t_{1/2}$, defined as the time-scale at which the modelled amplitude of the profile residuals decays to half the initial value. These are expressed for both models as:

\begin{equation}
t_{\mathrm{1/2,pre}}(\nu) = \tau_{pre} (\nu)\cdot \ln(2),
\end{equation}

\begin{equation}
t_{\mathrm{1/2,post}}(\nu) = \tau_{post} (\nu)\cdot 2^{(1 / \alpha) - 1},
\end{equation}

We evaluate $t_{1/2}$ as a function of observing frequency for both events. Neither event shows strong evidence for frequency dependence across the band. However, the two events exhibit different recovery time-scales, with the prior event having a longer half-life than the primary event across all frequencies. The full set of fitted MCMC parameter values and their uncertainties are provided in the Appendix (Table \ref{tab:MCMC_global_params} and Table \ref{tab:MCMC_subband_params}).

\section{Discussion}
\label{sec:discussion}
Profile variations across the MSP population have historically been considered rare, with relatively few examples of discrete profile change events reported in the literature \citep{2016ApJ...828L...1S,2021MNRAS.502..478G,2024ApJ...961...48W,2024ApJ...964..179J}. In contrast, transient profile changes are well established in the canonical pulsar population, which exhibits stochastic variability, continuous profile variability, long-term state switching, and gradual profile evolution \citep{2010Sci...329..408L,2011MNRAS.415..251K,2014ApJ...780L..31B,2016MNRAS.456.1374B,2022MNRAS.513.5861S,2025MNRAS.540.2486K}. The following discussion focuses on MSPs, for which the behaviours observed in the canonical pulsar population are generally distinct from the discrete profile changes reported to date. Instead, discrete profile change events across MSPs exhibit similar temporal morphology, characterized by a sudden onset followed by a gradual recovery, and in some cases a potential permanent reconfiguration of the profile has been proposed. The physical origin of these profile change events remains uncertain, with previous cases exhibiting diverse behaviour. For example, the event in PSR J1643$-$1224 showed no significant polarimetric changes \citep{2016ApJ...828L...1S}, whereas the 2021 event in PSR J1713$+$0747 exhibited strong polarimetric variations \citep{2025PASA...42..142M}. Despite the absence of temporal polarimetric variations in PSR J1643$-$1224, \cite{2016ApJ...828L...1S} concluded it was magnetospheric in origin, ruling out an ISM origin because the observed profile variations were strongest at higher frequencies, as opposed to the expected frequency-dependent effects of profile changes induced by ISM propagation. The authors further supported this interpretation by reporting the appearance of a new profile component that persisted for several months before decaying, after which the profile remained in a reconfigured state. In contrast, the polarimetric response to the profile change event in PSR J1713$+$0747 provided clear evidence of the magnetospheric origin. A previous profile change event (MJD 57070) for J0437 was also reported in \citet{2021MNRAS.502..478G}. While this event pre-dates the commencement of our dataset, we investigated and recovered this event as part of this analysis, finding that the profile variations occur at a different phase location (central region), exhibit different morphology and temporal evolution, and do not show the recurrent behaviour observed for the two events reported here.

\subsection{Physical Origin of the Profile Change}
\label{sec:PhysOrg}
As we found for the similar event observed in PSR J1713$+$0747 \citep{2025PASA...42..142M}, the results we present in this work collectively favour a magnetospheric origin for the two consecutive discrete profile change events in J0437. Several lines of evidence support this interpretation, including the observed polarimetric response and the frequency dependence of the profile evolution. 

Firstly, our polarimetric analysis demonstrates that the discrete profile change events affect not only the total intensity profile, but also the linear polarization profile, PAs, and fractional linear polarization of the emission. At the onset of the primary event, the PA deviates only in the narrow phase window coincident with the Main Feature, while the global PA structure remains unchanged. This indicates that the large-scale magnetic field geometry is largely unaffected, and that the observed polarimetric changes instead arise from localised variations in the emission, providing evidence that the origin of these events is associated with the MSP magnetosphere, rather than changes in Faraday rotation caused by discrete structures in the ISM. We also observe significant frequency-dependent evolution in the fractional linear polarization ($\Delta$ \textit{L/I}) at the same phase location, with the strongest changes occurring at the onset of the primary event.  

In the profile change events observed in PSR J1643$-$1224 \citep{2016ApJ...828L...1S} and PSR J1713$+$0747 \citep{2024ApJ...964..179J}, as well as in our analysis of J0437$-$4715, the observed frequency dependence and complex phase-localised morphology of the profile change are inconsistent with the expected chromatic behaviour of ISM propagation effects. For example, dispersive delays scale as $\nu^{-2}$, and applying the wrong DM can lead to smearing, causing frequency-averaged profiles to become distorted. However, the non-monotonic frequency dependence of the profile residuals in the Primary Event, together with correlated temporal evolution of phase-localised profile components, are inconsistent with an ISM dispersive origin. Similarly, scatter-induced profile broadening \citep{2013MNRAS.434...69L,2017ApJ...846..104K} scales approximately as $\nu^{-4.4}$, producing asymmetric trailing profile structures, which we do not observe. This conclusion is further supported by the frequency dependence of the fitted amplitudes for each event. The spectral index values we observe are substantially shallower than the dispersion and scattering scalings expected from ISM propagation (see Figure \ref{fig:PCA_MCMC_Params}). Additionally, our results on J0437 disfavour plasma lensing as these effects can produce delayed replicas of existing pulsar components and changes in pulse intensity, rather than the fixed, phase-localised behaviour we observe in J0437  \citep{2011MNRAS.410..499G}. We also rule out scintillation effects \citep{2018ApJ...868..122B}, since these were explicitly mitigated as described in Section \ref{subsec:template_create}. We find no evidence for sudden changes to the profile as a result of precession \citep{1989ApJ...347.1030W,2000Natur.406..484S}, which would produce secular, cyclical changes of the observed profile shape as the viewing geometry changes. This is inconsistent with the sudden, phase-localised variations in the profile we observe. We also suggest that the profile changes are unlikely to be associated with glitch-related events \citep{2011MNRAS.411.1917W}, as no glitches have been reported to date for J0437 from either the PPTA or the MPTA, although this could be the focus of a future, detailed study. We rule out pulse-to-pulse jitter \citep{2024MNRAS.528.3658K} as the origin of the observed behaviour, given the long-duration and coherent temporal evolution observed across both events in J0437. Several works have suggested that profile changes may be caused by the evaporation of asteroid material interacting with the magnetosphere \citep{2013ApJ...766....5S, 2014ApJ...780L..31B,2023MNRAS.524.5904L,2025MNRAS.538.3104L}. Such interactions could also produce a transient change in the MSP spin-down rate through associated changes in the magnetospheric charged-particle density. Potential changes in spin-down rate associated with profile variations have recently been explored for three MSPs in Bhat et al. 2026 (\textit{accepted}); however, evidence for these changes remains largely unresolved. We leave it to future work to determine whether there were any transient spin-down variations for the J0437 profile events reported here, and to assess the plausibility of two asteroid encounters with an MSP \citep{2013ApJ...766....5S,2016ApJ...828L...1S} in a relatively short time period.

\subsection{Implications for MSP Emission Models}
\label{sec:Interp} 
The two consecutive discrete profile change events we report here both exhibit similar temporal evolution, characterized by a sudden onset followed by a gradual recovery and subsequent overshoot of the template. The recurrence of this behaviour is an important constraint on the physical origin: whatever the underlying cause that produces these events has acted on the same regions of the pulse profile in a similar manner on at least two occasions. Both events are confined to three phase-localised regions associated with the Main Feature ($\phi \sim 0.4$), the Left Feature ($\phi \sim 0.3$), and the Central Feature ($\phi \sim 0.5$). Our PCA decomposition results -- robustly verified via independent analysis of MeerKAT data -- also demonstrate that these three features share a common origin, associated with the same principal component.

Several ideas have been proposed to associate pulse profile components with different emission geometries, based on numerous models and observations, such as the core/cone model \citep{1993ApJ...405..285R,1997MNRAS.285..561G,2001ApJ...555...31G,2003ApJ...584..418G,2017ApJ...845...23R}; the patchy beam model \citep{1988MNRAS.234..477L,1999ApJ...526..957K}; mapping of emission to a range of different magnetospheric altitudes \citep{1978ApJ...222.1006C,1998MNRAS.299..855K,2007MNRAS.380.1678K}; or a narrow region of altitude in particular for MSPs \citep{1998ApJ...501..286X, 1999ApJ...526..957K}, which has recently been supported by observational evidence presented by \citet{2026MNRAS.549ag835C}; the fan beam model \citep{2017MNRAS.471L.131D,2025A&A...704A.214J}; and multiple emission regions beyond the polar cap \citep{2015MNRAS.449.3223D,2026MNRAS.547f2258K}. The profile change events we observe in J0437 provide an opportunity to investigate how the varying components might relate to several of these proposed pulsar emission geometries. For example, the anti-correlated behaviour of the Left and Central Features relative to the Main Feature could represent emission variations across multiple regions of the polar cap. Recently, coupled, anti-correlated emission variations between two spatially distinct profile components (main pulse and interpulse) have been reported for two canonical pulsars in \citet{2026A&A...706A.160S}, with the evidence interpreted as either communication between different regions of the magnetosphere or a global magnetospheric process affecting both regions. The coupled, anti-correlated behaviour we observe in J0437 may similarly indicate that the corresponding emission regions are responding to a common magnetospheric process. An alternative explanation could instead consider changes in the relative emission strength of the different conal regions intersected by our line of sight. This latter interpretation requires the event to modify the beam non-uniformly to explain why the Main Feature decreases in power while the Left and Central Features increase. 

Changes in the density or distribution of plasma can modify radio-wave propagation through the magnetosphere \citep{1984ApJS...55..247S, 2015A&A...576A..62N}. The distribution of magnetospheric plasma is intrinsically linked to magnetic field lines and regions of pair production that are associated with the generation of radio emission \citep{2008ApJ...683L..41B}. Recent global particle-in-cell simulations from \cite{2023ApJ...958L...9B} demonstrated that pair production can occur through episodic discharges associated with spatially distinct magnetospheric regions, with the discharges exhibiting coupled behaviour. These distinct discharge regions may contribute to different components of the observed radio profile. Such changes in plasma production across different emission regions could provide a mechanism for producing the correlated profile variations we observe in J0437. However, the occurrence of two consecutive events with temporal coupling and anti-correlated behaviour we observe affecting the same phase-localised regions disfavours a purely stochastic plasma redistribution origin. Instead, the repeated behaviour may indicate the plasma redistribution is driven by an underlying repeatable magnetospheric process associated with quasi-stable field configurations, similar to discussions presented in \cite{2010MNRAS.408L..41T,2010MNRAS.408.2092T}, and \citet{2017MNRAS.469.2049Y}. 

The profile change events could have also arisen from a change in the relative contributions of orthogonal polarization modes \citep{2015MNRAS.448..771W}, producing variations observed in both intensity and polarization. We note that the phase location of the Main Feature in both events ($\sim\phi$ 0.4) coincides with the transition boundary of an OPM jump observed at lower frequencies (325\,MHz \citep{2016ApJ...833L..10D} and 438\,MHz \citep{1997ApJ...486.1019N}) occuring at approximately $\phi \sim 0.39-0.40$). The discrete profile change event in PSR J1713$+$0747 was similarly observed at an OPM transition boundary \citep{2025PASA...42..142M}. Whether this similarity reflects a common physical origin remains uncertain, but it may indicate OPM transition regions are particularly susceptible to discrete profile change events.

The consistent phase alignment of the preceding and primary events provides important constraints on the physical origin and suggests that they may represent a recurrent process that can be associated with a specific magnetospheric configuration that can return to, or undergo, a similar state of re-organisation. However, the temporal extent of our dataset is too short to provide evidence of additional events beyond the two reported, and no similar events have previously been identified in the PPTA observations of J0437. Regardless, this is the first reported work of consecutive discrete profile change events with phase-localised behaviour in any MSP across the literature.

\subsection{Implications for Polarization Calibration}
PSR J0437$-$4715 is a key MSP within the PPTA and due to its brightness, it has been routinely used as a polarization calibrator. As outlined in \cite{2004ApJS..152..129V,2013ApJS..204...13V}, rise-to-set observations of J0437 are used to characterize and correct for instrument polarization effects, including differential gain and phase and cross-coupling between the orthogonal receptors, which assumes that the (polarized) pulse profile remains stable over time.  Our results show this assumption is not always valid for J0437, with phase-localised variations in linear polarization and the position angle associated with discrete profile change events, as well as stochastic variability in the linear polarization, and in Stokes V, particularly around the peak of the profile. Although regular observations have been advocated for this calibration procedure, in practice these rise-to-set observations have been obtained at less regular intervals. Observing programs outside the PPTA that similarly use J0437 to derive polarimetric calibration solutions should also consider accounting for the intrinsic variations in its polarization profile when constructing these solutions. These findings motivate further investigation to determine the impact of these variations on polarization calibrators and whether to develop alternative strategies to mitigate this. A more detailed assessment of the implications of this behaviour for timing and polarization calibration is beyond the scope of this work, but should be considered in the future.

\subsection{Implications for Pulsar Timing Arrays and Future Work}
PTAs operate by combining timing data from a suite of MSPs, using cross-correlation techniques \citep{1990ApJ...361..300F} to detect the GWB. As a high-priority MSP for southern-hemisphere PTAs with a long-term dataset and favourable sky location, profile variations in J0437 warrant careful consideration for their impact on PTA science objectives. These experiments rely on precise modelling of all noise contributions \citep{2021MNRAS.502..478G,2023ApJ...951L...7R}, and both stochastic variability and discrete profile change events, such as those presented in this work, could introduce additional unmodelled noise components, motivating their characterization and inclusion in pulsar timing noise models.

Previously, methods for mitigating the effects of pulse-profile variations on pulsar timing have been developed using profile-domain analysis \citep{2017MNRAS.466.3706L}. More recently, new methods of mitigating the effects of anomalous shape changes on timing residuals specifically for PSR J1713$+$0747 have been proposed by \cite{2026arXiv260712038N} and \citet{2026arXiv260804108C}. Additionally, \cite{2026ApJ..1005...58J} have used PCA to identify similar profile change events in this, and other MSPs.

The increasing timespan of PTA datasets, now spanning decades, combined with improvements in instrumental sensitivity, makes this an opportune time to develop and implement models that characterize profile change events. This will be particularly important in preparation for the next generation of radio observatories, including the SKA \citep{2025OJAp....854638J,2025OJAp....854244O,2025OJAp....854243S}, ngVLA \citep{2018ASPC..517..751C}, and existing/future FAST science objectives \citep{2019RAA....19...20H}. These observatories will be capable of detecting lower-amplitude profile variations with significantly higher telescope sensitivity, although the achievable sensitivity to such variations will also depend on the intrinsic sources of profile variability, such as pulse jitter.

An intriguing aspect of the J0437 events is the relatively small characteristic PCA amplitudes ($\sim 10^{-4}$ -- $10^{-3}$), approximately one-to-two orders of magnitude difference in scale, when compared to the PCA amplitudes ($\sim 10^{-2}$) we reported for PSR J1713$+$0747 \citep{2025PASA...42..142M}. This raises the question of whether similar smaller-amplitude profile variations may be present in other MSPs but remain undetected due to lower S/N. To investigate this across the full PPTA sample, we are currently conducting a systematic profile stability analysis, with results to be presented in a forthcoming study (Mandow et al. \textit{in prep}.), while a similar analysis of this nature for MeerKAT will be presented in Bhat et al. (\textit{accepted}). 

The proximity and brightness of J0437 has provided us with this unique opportunity to probe such behaviour in detail, offering a potential view into magnetospheric variability that may be common across the MSP population. Alternatively, these events may be rare and pulsar-dependent. Future observations of other bright MSPs exhibiting discrete profile change events will be important in distinguishing between these possibilities.

\section{CONCLUSION}
\label{sec:conclusion}
We present a detailed analysis of profile variability in PSR J0437$-$4715, the closest and brightest millisecond pulsar, using approximately seven years of observations obtained with the UWL receiver (704--4032\,MHz) on Murriyang. We report the discovery of two consecutive discrete profile change events in PSR J0437$-$4715: a primary event with onset at MJD 59670, and a preceding event beginning at MJD 58082. This is the first reported instance of two consecutive profile change events exhibiting the same phase-localised morphological variations on any millisecond pulsar. Both events affect the same phase-localised regions of the profile, associated with the Main Feature (pulse phase $\sim 0.4$), the Left Feature (pulse phase $\sim 0.3$), and the Central Feature (pulse phase $\sim 0.5$), which exhibit anti-correlated and coupled temporal behaviour suggesting that the same underlying process affects multiple components of the emission beam. The two consecutive events suggest a potentially recurrent discrete profile change event in this MSP.

The profile changes exhibit a clear frequency-dependent response in Stokes I, with additional weaker signatures present in the linear polarization. We do not detect variations as a result of these profile change events in Stokes V. The position angle and fractional linear polarization also show correlated phase-localised deviations coincident with the Main Feature, with the same temporal evolution of the Stokes I variations, confirming the magnetospheric origin of these events.

Using PCA, we isolate the dominant profile variability associated with the profile change event and demonstrate that the three features are part of the same principal component eigenprofile. We develop a joint model to describe both discrete profile change events, finding that the preceding event is well characterized by an exponential decay, while the primary event follows a power-law recovery, exhibiting frequency-dependent amplitudes, but no strong evidence for frequency dependence in the recovery time-scales. We observe the late-time evolution of both events to show similar departures from the pre-event template, potentially suggesting that this behaviour might be a feature of the long-term recovery of these events, or potential reconfigurations of the profile.

These results represent the most detailed characterization of profile variations in PSR J0437$-$4715 to date. Owing to its proximity and high S/N, these events on J0437$-$4715 provide a unique opportunity to probe magnetospheric variability in an MSP. Continued investigation of these events will be important for understanding pulsar emission physics and for quantifying their impact on high-precision pulsar timing experiments, particularly within PTA experiments.

\section*{Acknowledgements}
Murriyang, CSIRO’s Parkes radio telescope, is part of the Australia Telescope National Facility (\url{https://ror.org/05qajvd42}) which is funded by the Australian Government for operation as a National Facility managed by CSIRO. We acknowledge the Wiradjuri people as the Traditional Owners of the Observatory site. We acknowledge the Wallumattagal people as the Traditional Owners of the land where this work was carried out. This paper includes archived data obtained through the CSIRO Data Access Portal (http://data.csiro.au). Work at NRL is supported by NASA. MEL is supported by an Australian Research Council (ARC) Discovery Early Career Research Award DE250100508. Parts of this research were conducted by the Australian Research Council Centre of Excellence for Gravitational Wave Discovery (OzGrav), project number CE230100016. The MeerKAT telescope is operated by the South African Radio Astronomy Observatory (SARAO), which is a facility of the National Research Foundation (NRF), an agency of the Department of Science and Innovation (DSI)

\section*{Data Availability}
Profiles, profile residuals, and analysis scripts used in this paper will be available at https://github.com/CosmicRami/

Software: PSRCHIVE \citep{2004PASA...21..302H,2011PASA...28....1V}, TEMPO2 \citep{2006MNRAS.372.1549E}\\



\bibliographystyle{mnras}
\bibliography{example} 

@ARTICLE{1993Natur.361..613J,
       author = {{Johnston}, Simon and {Lorimer}, D.~R. and {Harrison}, P.~A. and {Bailes}, M. and {Lynet}, A.~G. and {Bell}, J.~F. and {Kaspi}, V.~M. and {Manchester}, R.~N. and {D'Amico}, N. and {Nleastrol}, L. and {Shengzhen}, Jin},
        title = "{Discovery of a very bright, nearby binary millisecond pulsar}",
      journal = {\nat},
         year = 1993,
        month = feb,
       volume = {361},
       number = {6413},
        pages = {613-615},
          doi = {10.1038/361613a0},
       adsurl = {https://ui.adsabs.harvard.edu/abs/1993Natur.361..613J}
}

@ARTICLE{2024ApJ...971L..18R,
       author = {{Reardon}, Daniel J. and {Bailes}, Matthew and {Shannon}, Ryan M. and {Flynn}, Chris and {Askew}, Jacob and {Bhat}, N.~D. Ramesh and {Chen}, Zu-Cheng and {Cury{\l}o}, Ma{\l}gorzata and {Feng}, Yi and {Hobbs}, George B. and {Kapur}, Agastya and {Kerr}, Matthew and {Liu}, Xiaojin and {Manchester}, Richard N. and {Mandow}, Rami and {Mishra}, Saurav and {Russell}, Christopher J. and {Shamohammadi}, Mohsen and {Zhang}, Lei and {Zic}, Andrew},
        title = "{The Neutron Star Mass, Distance, and Inclination from Precision Timing of the Brilliant Millisecond Pulsar J0437-4715}",
      journal = {\apjl},
         year = 2024,
        month = aug,
       volume = {971},
       number = {1},
          eid = {L18},
        pages = {L18},
          doi = {10.3847/2041-8213/ad614a},
archivePrefix = {arXiv},
       eprint = {2407.07132},
 primaryClass = {astro-ph.HE},
       adsurl = {https://ui.adsabs.harvard.edu/abs/2024ApJ...971L..18R}
}

@ARTICLE{1996MNRAS.280..331M,
       author = {{McConnell}, D. and {Ables}, J.~G. and {Bailes}, M. and {Erickson}, W.~C.},
        title = "{Observations of the millisecond pulsar J0437-4715 at 76 MHz.}",
      journal = {\mnras},
         year = 1996,
        month = may,
       volume = {280},
       number = {1},
        pages = {331-334},
          doi = {10.1093/mnras/280.2.331},
       adsurl = {https://ui.adsabs.harvard.edu/abs/1996MNRAS.280..331M}
}

@ARTICLE{1998ApJ...498..365J,
       author = {{Jenet}, F.~A. and {Anderson}, S.~B. and {Kaspi}, V.~M. and {Prince}, T.~A. and {Unwin}, S.~C.},
        title = "{Radio Pulse Properties of the Millisecond Pulsar PSR J0437-4715. I. Observations at 20 Centimeters}",
      journal = {\apj},
         year = 1998,
        month = may,
       volume = {498},
       number = {1},
        pages = {365-372},
          doi = {10.1086/305529},
archivePrefix = {arXiv},
       eprint = {astro-ph/9803028},
 primaryClass = {astro-ph},
       adsurl = {https://ui.adsabs.harvard.edu/abs/1998ApJ...498..365J}
}

@ARTICLE{2011MNRAS.416..346K,
       author = {{Keith}, M.~J. and {Johnston}, S. and {Levin}, L. and {Bailes}, M.},
        title = "{17- and 24-GHz observations of southern pulsars}",
      journal = {\mnras},
         year = 2011,
        month = sep,
       volume = {416},
       number = {1},
        pages = {346-354},
          doi = {10.1111/j.1365-2966.2011.19041.x},
archivePrefix = {arXiv},
       eprint = {1105.3961},
 primaryClass = {astro-ph.SR},
       adsurl = {https://ui.adsabs.harvard.edu/abs/2011MNRAS.416..346K}
}

@ARTICLE{2014MNRAS.441.3148O,
       author = {{Os{\l}owski}, S. and {van Straten}, W. and {Bailes}, M. and {Jameson}, A. and {Hobbs}, G.},
        title = "{Timing, polarimetry and physics of the bright, nearby millisecond pulsar PSR J0437-4715 - a single-pulse perspective}",
      journal = {\mnras},
         year = 2014,
        month = jul,
       volume = {441},
       number = {4},
        pages = {3148-3160},
          doi = {10.1093/mnras/stu804},
archivePrefix = {arXiv},
       eprint = {1405.2638},
 primaryClass = {astro-ph.SR},
       adsurl = {https://ui.adsabs.harvard.edu/abs/2014MNRAS.441.3148O}
}

@ARTICLE{2012ApJ...746....6D,
       author = {{Durant}, Martin and {Kargaltsev}, Oleg and {Pavlov}, George G. and {Kowalski}, Piotr M. and {Posselt}, Bettina and {van Kerkwijk}, Marten H. and {Kaplan}, David L.},
        title = "{The Spectrum of the Recycled PSR J0437-4715 and Its White Dwarf Companion}",
      journal = {\apj},
         year = 2012,
        month = feb,
       volume = {746},
       number = {1},
          eid = {6},
        pages = {6},
          doi = {10.1088/0004-637X/746/1/6},
archivePrefix = {arXiv},
       eprint = {1111.2346},
 primaryClass = {astro-ph.HE},
       adsurl = {https://ui.adsabs.harvard.edu/abs/2012ApJ...746....6D}
}

@ARTICLE{1993ApJ...411L..83B,
       author = {{Bailyn}, Charles D.},
        title = "{The Optical Counterpart of the Bright Nearby Millisecond Pulsar PSR J0437-4715}",
      journal = {\apjl},
         year = 1993,
        month = jul,
       volume = {411},
        pages = {L83},
          doi = {10.1086/186918},
       adsurl = {https://ui.adsabs.harvard.edu/abs/1993ApJ...411L..83B}
}

@ARTICLE{1993Natur.364..603B,
       author = {{Bell}, J.~F. and {Bailes}, M. and {Bessell}, M.~S.},
        title = "{Optical detection of the companion of the millisecond pulsar J0437-4715}",
      journal = {\nat},
         year = 1993,
        month = aug,
       volume = {364},
       number = {6438},
        pages = {603-605},
          doi = {10.1038/364603a0},
       adsurl = {https://ui.adsabs.harvard.edu/abs/1993Natur.364..603B}
}

@ARTICLE{2004ApJ...602..327K,
       author = {{Kargaltsev}, Oleg and {Pavlov}, George G. and {Romani}, Roger W.},
        title = "{Ultraviolet Emission from the Millisecond Pulsar J0437-4715}",
      journal = {\apj},
         year = 2004,
        month = feb,
       volume = {602},
       number = {1},
        pages = {327-335},
          doi = {10.1086/380993},
archivePrefix = {arXiv},
       eprint = {astro-ph/0310854},
 primaryClass = {astro-ph},
       adsurl = {https://ui.adsabs.harvard.edu/abs/2004ApJ...602..327K}
}

@ARTICLE{2013ApJ...762...96B,
       author = {{Bogdanov}, Slavko},
        title = "{The Nearest Millisecond Pulsar Revisited with XMM-Newton: Improved Mass-radius Constraints for PSR J0437-4715}",
      journal = {\apj},
         year = 2013,
        month = jan,
       volume = {762},
       number = {2},
          eid = {96},
        pages = {96},
          doi = {10.1088/0004-637X/762/2/96},
archivePrefix = {arXiv},
       eprint = {1211.6113},
 primaryClass = {astro-ph.HE},
       adsurl = {https://ui.adsabs.harvard.edu/abs/2013ApJ...762...96B}
}

@ARTICLE{2024ApJ...971L..20C,
       author = {{Choudhury}, Devarshi and {Salmi}, Tuomo and {Vinciguerra}, Serena and {Riley}, Thomas E. and {Kini}, Yves and {Watts}, Anna L. and {Dorsman}, Bas and {Bogdanov}, Slavko and {Guillot}, Sebastien and {Ray}, Paul S. and {Reardon}, Daniel J. and {Remillard}, Ronald A. and {Bilous}, Anna V. and {Huppenkothen}, Daniela and {Lattimer}, James M. and {Rutherford}, Nathan and {Arzoumanian}, Zaven and {Gendreau}, Keith C. and {Morsink}, Sharon M. and {Ho}, Wynn C.~G.},
        title = "{A NICER View of the Nearest and Brightest Millisecond Pulsar: PSR J0437─4715}",
      journal = {\apjl},
         year = 2024,
        month = aug,
       volume = {971},
       number = {1},
          eid = {L20},
        pages = {L20},
          doi = {10.3847/2041-8213/ad5a6f},
archivePrefix = {arXiv},
       eprint = {2407.06789},
 primaryClass = {astro-ph.HE},
       adsurl = {https://ui.adsabs.harvard.edu/abs/2024ApJ...971L..20C}
}

@ARTICLE{2026ApJ..1000L..48M,
       author = {{Miller}, M.~C. and {Dittmann}, A.~J. and {Holt}, I.~M. and {Lamb}, F.~K. and {Chirenti}, C. and {Arzoumanian}, Z. and {Berteaud}, J. and {Bogdanov}, S. and {Gendreau}, K.~C. and {Ho}, W.~C.~G. and {Morsink}, S.~M. and {Ray}, P.~S. and {Remillard}, R.~A. and {Wadiasingh}, Z. and {Wolff}, M.~T.},
        title = "{The Radius of PSR J0437─4715 from NICER Data}",
      journal = {\apjl},
         year = 2026,
        month = apr,
       volume = {1000},
       number = {2},
          eid = {L48},
        pages = {L48},
          doi = {10.3847/2041-8213/ae5057},
archivePrefix = {arXiv},
       eprint = {2512.08790},
 primaryClass = {astro-ph.HE},
       adsurl = {https://ui.adsabs.harvard.edu/abs/2026ApJ..1000L..48M}
}

@ARTICLE{2020ApJ...904..104R,
       author = {{Reardon}, Daniel J. and {Coles}, William A. and {Bailes}, Matthew and {Bhat}, N.~D. Ramesh and {Dai}, Shi and {Hobbs}, George B. and {Kerr}, Matthew and {Manchester}, Richard N. and {Os{\l}owski}, Stefan and {Parthasarathy}, Aditya and {Russell}, Christopher J. and {Shannon}, Ryan M. and {Spiewak}, Ren{\'e}e and {Toomey}, Lawrence and {Tuntsov}, Artem V. and {van Straten}, Willem and {Walker}, Mark A. and {Wang}, Jingbo and {Zhang}, Lei and {Zhu}, Xing-Jiang},
        title = "{Precision Orbital Dynamics from Interstellar Scintillation Arcs for PSR J0437-4715}",
      journal = {\apj},
         year = 2020,
        month = dec,
       volume = {904},
       number = {2},
          eid = {104},
        pages = {104},
          doi = {10.3847/1538-4357/abbd40},
archivePrefix = {arXiv},
       eprint = {2009.12757},
 primaryClass = {astro-ph.HE},
       adsurl = {https://ui.adsabs.harvard.edu/abs/2020ApJ...904..104R}
}

@ARTICLE{2024ApJ...971L..19R,
       author = {{Rutherford}, Nathan and {Mendes}, Melissa and {Svensson}, Isak and {Schwenk}, Achim and {Watts}, Anna L. and {Hebeler}, Kai and {Keller}, Jonas and {Prescod-Weinstein}, Chanda and {Choudhury}, Devarshi and {Raaijmakers}, Geert and {Salmi}, Tuomo and {Timmerman}, Patrick and {Vinciguerra}, Serena and {Guillot}, Sebastien and {Lattimer}, James M.},
        title = "{Constraining the Dense Matter Equation of State with New NICER Mass─Radius Measurements and New Chiral Effective Field Theory Inputs}",
      journal = {\apjl},
         year = 2024,
        month = aug,
       volume = {971},
       number = {1},
          eid = {L19},
        pages = {L19},
          doi = {10.3847/2041-8213/ad5f02},
archivePrefix = {arXiv},
       eprint = {2407.06790},
 primaryClass = {astro-ph.HE},
       adsurl = {https://ui.adsabs.harvard.edu/abs/2024ApJ...971L..19R}
}

@ARTICLE{2024ApJ...974..244T,
       author = {{Tang}, Shao-Peng and {Huang}, Yong-Jia and {Han}, Ming-Zhe and {Fan}, Yi-Zhong},
        title = "{Upper Limit of Sound Speed in Nuclear Matter: A Harmonious Interplay of Transport Calculation and Perturbative Quantum Chromodynamic Constraint}",
      journal = {\apj},
         year = 2024,
        month = oct,
       volume = {974},
       number = {2},
          eid = {244},
        pages = {244},
          doi = {10.3847/1538-4357/ad7503},
archivePrefix = {arXiv},
       eprint = {2404.09563},
 primaryClass = {astro-ph.HE},
       adsurl = {https://ui.adsabs.harvard.edu/abs/2024ApJ...974..244T}
}

@ARTICLE{2025ApJ...978L..14H,
       author = {{Huang}, Chun},
        title = "{Equation of State Independent Determination on the Radius of a 1.4 M$_{{\ensuremath{\odot}}}$ Neutron Star Using Mass─Radius Measurements}",
      journal = {\apjl},
         year = 2025,
        month = jan,
       volume = {978},
       number = {1},
          eid = {L14},
        pages = {L14},
          doi = {10.3847/2041-8213/ad9f3c},
archivePrefix = {arXiv},
       eprint = {2412.10242},
 primaryClass = {astro-ph.HE},
       adsurl = {https://ui.adsabs.harvard.edu/abs/2025ApJ...978L..14H}
}

@ARTICLE{2026A&A...706A.203S,
       author = {{Shahrbaf}, Mahboubeh and {Rafiei Karkevandi}, Davood and {Ayriyan}, Alexander and {Typel}, Stefan},
        title = "{Observational probes of the neutron star equation of state with hyperons, bosonic dark matter, and quark matter}",
      journal = {\aap},
         year = 2026,
        month = feb,
       volume = {706},
          eid = {A203},
        pages = {A203},
          doi = {10.1051/0004-6361/202556315},
archivePrefix = {arXiv},
       eprint = {2402.18686},
 primaryClass = {nucl-th},
       adsurl = {https://ui.adsabs.harvard.edu/abs/2026A&A...706A.203S}
}

@ARTICLE{2013PASA...30...17M,
       author = {{Manchester}, R.~N. and {Hobbs}, G. and {Bailes}, M. and {Coles}, W.~A. and {van Straten}, W. and {Keith}, M.~J. and {Shannon}, R.~M. and {Bhat}, N.~D.~R. and {Brown}, A. and {Burke-Spolaor}, S.~G. and {Champion}, D.~J. and {Chaudhary}, A. and {Edwards}, R.~T. and {Hampson}, G. and {Hotan}, A.~W. and {Jameson}, A. and {Jenet}, F.~A. and {Kesteven}, M.~J. and {Khoo}, J. and {Kocz}, J. and {Maciesiak}, K. and {Oslowski}, S. and {Ravi}, V. and {Reynolds}, J.~R. and {Sarkissian}, J.~M. and {Verbiest}, J.~P.~W. and {Wen}, Z.~L. and {Wilson}, W.~E. and {Yardley}, D. and {Yan}, W.~M. and {You}, X.~P.},
        title = "{The Parkes Pulsar Timing Array Project}",
      journal = {\pasa},
         year = 2013,
        month = jan,
       volume = {30},
          eid = {e017},
        pages = {e017},
          doi = {10.1017/pasa.2012.017},
archivePrefix = {arXiv},
       eprint = {1210.6130},
 primaryClass = {astro-ph.IM},
       adsurl = {https://ui.adsabs.harvard.edu/abs/2013PASA...30...17M}
}

@ARTICLE{2004ApJS..152..129V,
       author = {{van Straten}, W.},
        title = "{Radio Astronomical Polarimetry and Point-Source Calibration}",
      journal = {\apjs},
         year = 2004,
        month = may,
       volume = {152},
       number = {1},
        pages = {129-135},
          doi = {10.1086/383187},
archivePrefix = {arXiv},
       eprint = {astro-ph/0401536},
 primaryClass = {astro-ph},
       adsurl = {https://ui.adsabs.harvard.edu/abs/2004ApJS..152..129V}
}

@ARTICLE{2013ApJS..204...13V,
       author = {{van Straten}, W.},
        title = "{High-fidelity Radio Astronomical Polarimetry Using a Millisecond Pulsar as a Polarized Reference Source}",
      journal = {\apjs},
         year = 2013,
        month = jan,
       volume = {204},
       number = {1},
          eid = {13},
        pages = {13},
          doi = {10.1088/0067-0049/204/1/13},
archivePrefix = {arXiv},
       eprint = {1212.3446},
 primaryClass = {astro-ph.IM},
       adsurl = {https://ui.adsabs.harvard.edu/abs/2013ApJS..204...13V}
}

@ARTICLE{2022PASA...39...27S,
       author = {{Spiewak}, R. and {Bailes}, M. and {Miles}, M.~T. and {Parthasarathy}, A. and {Reardon}, D.~J. and {Shamohammadi}, M. and {Shannon}, R.~M. and {Bhat}, N.~D.~R. and {Buchner}, S. and {Cameron}, A.~D. and {Camilo}, F. and {Geyer}, M. and {Johnston}, S. and {Karastergiou}, A. and {Keith}, M. and {Kramer}, M. and {Serylak}, M. and {van Straten}, W. and {Theureau}, G. and {Venkatraman Krishnan}, V.},
        title = "{The MeerTime Pulsar Timing Array: A census of emission properties and timing potential}",
      journal = {\pasa},
         year = 2022,
        month = jul,
       volume = {39},
          eid = {e027},
        pages = {e027},
          doi = {10.1017/pasa.2022.19},
archivePrefix = {arXiv},
       eprint = {2204.04115},
 primaryClass = {astro-ph.HE},
       adsurl = {https://ui.adsabs.harvard.edu/abs/2022PASA...39...27S}
}

@ARTICLE{2021ApJ...908..158S,
       author = {{Sosa Fiscella}, V. and {del Palacio}, S. and {Combi}, L. and {Lousto}, C.~O. and {Combi}, J.~A. and {Gancio}, G. and {Garc{\'\i}a}, F. and {Guti{\'e}rrez}, E. and {Hauscarriaga}, F. and {Kornecki}, P. and {L{\'o}pez Armengol}, F.~G. and {Mancuso}, G.~C. and {M{\"u}ller}, A.~L. and {Simaz Bunzel}, A.},
        title = "{PSR J0437-4715: The Argentine Institute of Radioastronomy 2019-2020 Observational Campaign}",
      journal = {\apj},
         year = 2021,
        month = feb,
       volume = {908},
       number = {2},
          eid = {158},
        pages = {158},
          doi = {10.3847/1538-4357/abceb3},
archivePrefix = {arXiv},
       eprint = {2010.00010},
 primaryClass = {astro-ph.GA},
       adsurl = {https://ui.adsabs.harvard.edu/abs/2021ApJ...908..158S}
}

@ARTICLE{2021ApJ...911..137L,
       author = {{Lam}, M.~T. and {Hazboun}, J.~S.},
        title = "{Precision Timing of PSR J0437-4715 with the IAR Observatory and Implications for Low-frequency Gravitational Wave Source Sensitivity}",
      journal = {\apj},
         year = 2021,
        month = apr,
       volume = {911},
       number = {2},
          eid = {137},
        pages = {137},
          doi = {10.3847/1538-4357/abeb64},
archivePrefix = {arXiv},
       eprint = {2007.00260},
 primaryClass = {astro-ph.HE},
       adsurl = {https://ui.adsabs.harvard.edu/abs/2021ApJ...911..137L}
}

@ARTICLE{2024PASA...41...36K,
       author = {{Kikunaga}, Tomonosuke and {Hisano}, Shinnosuke and {Batra}, Neelam Dhanda and {Desai}, Shantanu and {Joshi}, Bhal Chandra and {Bagchi}, Manjari and {Prabu}, T. and {Takahashi}, Keitaro and {Arumugam}, Swetha and {Bathula}, Adarsh and {Dandapat}, Subhajit and {Deb}, Debabrata and {Dwivedi}, Churchil and {Gupta}, Yashwant and {Jacob}, Shebin Jose and {Kareem}, Fazal and {Nobleson}, K. and {Mamidipaka}, Pragna and {Paladi}, Avinash Kumar and {Pandian}, B. Arul and {Rana}, Prerna and {Singha}, Jaikhomba and {Srivastava}, Aman and {Surnis}, Mayuresh and {Tarafdar}, Pratik},
        title = "{Low-frequency pulse-jitter measurement with the uGMRT I: PSR J0437-4715}",
      journal = {\pasa},
         year = 2024,
        month = may,
       volume = {41},
          eid = {e036},
        pages = {e036},
          doi = {10.1017/pasa.2024.30},
archivePrefix = {arXiv},
       eprint = {2312.01875},
 primaryClass = {astro-ph.HE},
       adsurl = {https://ui.adsabs.harvard.edu/abs/2024PASA...41...36K}
}

@ARTICLE{2012MNRAS.420..361L,
       author = {{Liu}, K. and {Keane}, E.~F. and {Lee}, K.~J. and {Kramer}, M. and {Cordes}, J.~M. and {Purver}, M.~B.},
        title = "{Profile-shape stability and phase-jitter analyses of millisecond pulsars}",
      journal = {\mnras},
         year = 2012,
        month = feb,
       volume = {420},
       number = {1},
        pages = {361-368},
          doi = {10.1111/j.1365-2966.2011.20041.x},
archivePrefix = {arXiv},
       eprint = {1110.4759},
 primaryClass = {astro-ph.HE},
       adsurl = {https://ui.adsabs.harvard.edu/abs/2012MNRAS.420..361L}
}

@ARTICLE{2021MNRAS.502..407P,
       author = {{Parthasarathy}, A. and {Bailes}, M. and {Shannon}, R.~M. and {van Straten}, W. and {Os{\l}owski}, S. and {Johnston}, S. and {Spiewak}, R. and {Reardon}, D.~J. and {Kramer}, M. and {Venkatraman Krishnan}, V. and {Pennucci}, T.~T. and {Abbate}, F. and {Buchner}, S. and {Camilo}, F. and {Champion}, D.~J. and {Geyer}, M. and {Hugo}, B. and {Jameson}, A. and {Karastergiou}, A. and {Keith}, M.~J. and {Serylak}, M.},
        title = "{Measurements of pulse jitter and single-pulse variability in millisecond pulsars using MeerKAT}",
      journal = {\mnras},
         year = 2021,
        month = mar,
       volume = {502},
       number = {1},
        pages = {407-422},
          doi = {10.1093/mnras/stab037},
archivePrefix = {arXiv},
       eprint = {2101.08531},
 primaryClass = {astro-ph.HE},
       adsurl = {https://ui.adsabs.harvard.edu/abs/2021MNRAS.502..407P}
}

@ARTICLE{2024MNRAS.528.3658K,
       author = {{Kulkarni}, A.~D. and {Shannon}, R.~M. and {Reardon}, D.~J. and {Miles}, M.~T. and {Bailes}, M. and {Shamohammadi}, M.},
        title = "{An insight into chromatic behaviour of jitter in pulsars and its modelling: a case study of PSR J0437-4715}",
      journal = {\mnras},
         year = 2024,
        month = feb,
       volume = {528},
       number = {2},
        pages = {3658-3667},
          doi = {10.1093/mnras/stae041},
archivePrefix = {arXiv},
       eprint = {2401.03660},
 primaryClass = {astro-ph.HE},
       adsurl = {https://ui.adsabs.harvard.edu/abs/2024MNRAS.528.3658K}
}

@ARTICLE{1975ApJ...198..661H,
       author = {{Helfand}, D.~J. and {Manchester}, R.~N. and {Taylor}, J.~H.},
        title = "{Observations of pulsar radio emission. III. Stability of integrated profiles.}",
      journal = {\apj},
         year = 1975,
        month = jun,
       volume = {198},
        pages = {661-670},
          doi = {10.1086/153644},
       adsurl = {https://ui.adsabs.harvard.edu/abs/1975ApJ...198..661H}
}

@ARTICLE{2023ApJ...951L...6R,
       author = {{Reardon}, Daniel J. and {Zic}, Andrew and {Shannon}, Ryan M. and {Hobbs}, George B. and {Bailes}, Matthew and {Di Marco}, Valentina and {Kapur}, Agastya and {Rogers}, Axl F. and {Thrane}, Eric and {Askew}, Jacob and {Bhat}, N.~D. Ramesh and {Cameron}, Andrew and {Cury{\l}o}, Ma{\l}gorzata and {Coles}, William A. and {Dai}, Shi and {Goncharov}, Boris and {Kerr}, Matthew and {Kulkarni}, Atharva and {Levin}, Yuri and {Lower}, Marcus E. and {Manchester}, Richard N. and {Mandow}, Rami and {Miles}, Matthew T. and {Nathan}, Rowina S. and {Os{\l}owski}, Stefan and {Russell}, Christopher J. and {Spiewak}, Ren{\'e}e and {Zhang}, Songbo and {Zhu}, Xing-Jiang},
        title = "{Search for an Isotropic Gravitational-wave Background with the Parkes Pulsar Timing Array}",
      journal = {\apjl},
         year = 2023,
        month = jul,
       volume = {951},
       number = {1},
          eid = {L6},
        pages = {L6},
          doi = {10.3847/2041-8213/acdd02},
archivePrefix = {arXiv},
       eprint = {2306.16215},
 primaryClass = {astro-ph.HE},
       adsurl = {https://ui.adsabs.harvard.edu/abs/2023ApJ...951L...6R}
}

@ARTICLE{2023RAA....23g5024X,
       author = {{Xu}, Heng and {Chen}, Siyuan and {Guo}, Yanjun and {Jiang}, Jinchen and {Wang}, Bojun and {Xu}, Jiangwei and {Xue}, Zihan and {Caballero}, R. Nicolas and {Yuan}, Jianping and {Xu}, Yonghua and {Wang}, Jingbo and {Hao}, Longfei and {Luo}, Jingtao and {Lee}, Kejia and {Han}, Jinlin and {Jiang}, Peng and {Shen}, Zhiqiang and {Wang}, Min and {Wang}, Na and {Xu}, Renxin and {Wu}, Xiangping and {Manchester}, Richard and {Qian}, Lei and {Guan}, Xin and {Huang}, Menglin and {Sun}, Chun and {Zhu}, Yan},
        title = "{Searching for the Nano-Hertz Stochastic Gravitational Wave Background with the Chinese Pulsar Timing Array Data Release I}",
      journal = {Research in Astronomy and Astrophysics},
         year = 2023,
        month = jul,
       volume = {23},
       number = {7},
          eid = {075024},
        pages = {075024},
          doi = {10.1088/1674-4527/acdfa5},
archivePrefix = {arXiv},
       eprint = {2306.16216},
 primaryClass = {astro-ph.HE},
       adsurl = {https://ui.adsabs.harvard.edu/abs/2023RAA....23g5024X}
}

@ARTICLE{2023A&A...678A..48E,
       author = {{EPTA Collaboration} and {Antoniadis}, J. and {Babak}, S. and {Bak Nielsen}, A.-S. and {Bassa}, C.~G. and {Berthereau}, A. and {Bonetti}, M. and {Bortolas}, E. and {Brook}, P.~R. and {Burgay}, M. and {Caballero}, R.~N. and {Chalumeau}, A. and {Champion}, D.~J. and {Chanlaridis}, S. and {Chen}, S. and {Cognard}, I. and {Desvignes}, G. and {Falxa}, M. and {Ferdman}, R.~D. and {Franchini}, A. and {Gair}, J.~R. and {Goncharov}, B. and {Graikou}, E. and {Grie{\ss}meier}, J.-M. and {Guillemot}, L. and {Guo}, Y.~J. and {Hu}, H. and {Iraci}, F. and {Izquierdo-Villalba}, D. and {Jang}, J. and {Jawor}, J. and {Janssen}, G.~H. and {Jessner}, A. and {Karuppusamy}, R. and {Keane}, E.~F. and {Keith}, M.~J. and {Kramer}, M. and {Krishnakumar}, M.~A. and {Lackeos}, K. and {Lee}, K.~J. and {Liu}, K. and {Liu}, Y. and {Lyne}, A.~G. and {McKee}, J.~W. and {Main}, R.~A. and {Mickaliger}, M.~B. and {Ni{\c{t}}u}, I.~C. and {Parthasarathy}, A. and {Perera}, B.~B.~P. and {Perrodin}, D. and {Petiteau}, A. and {Porayko}, N.~K. and {Possenti}, A. and {Quelquejay Leclere}, H. and {Samajdar}, A. and {Sanidas}, S.~A. and {Sesana}, A. and {Shaifullah}, G. and {Speri}, L. and {Spiewak}, R. and {Stappers}, B.~W. and {Susarla}, S.~C. and {Theureau}, G. and {Tiburzi}, C. and {van der Wateren}, E. and {Vecchio}, A. and {Venkatraman Krishnan}, V. and {Verbiest}, J.~P.~W. and {Wang}, J. and {Wang}, L. and {Wu}, Z.},
        title = "{The second data release from the European Pulsar Timing Array. I. The dataset and timing analysis}",
      journal = {\aap},
         year = 2023,
        month = oct,
       volume = {678},
          eid = {A48},
        pages = {A48},
          doi = {10.1051/0004-6361/202346841},
archivePrefix = {arXiv},
       eprint = {2306.16224},
 primaryClass = {astro-ph.HE},
       adsurl = {https://ui.adsabs.harvard.edu/abs/2023A&A...678A..48E}
}

@ARTICLE{2023ApJ...951L...9A,
       author = {{Agazie}, Gabriella and {Alam}, Md Faisal and {Anumarlapudi}, Akash and {Archibald}, Anne M. and {Arzoumanian}, Zaven and {Baker}, Paul T. and {Blecha}, Laura and {Bonidie}, Victoria and {Brazier}, Adam and {Brook}, Paul R. and {Burke-Spolaor}, Sarah and {B{\'e}csy}, Bence and {Chapman}, Christopher and {Charisi}, Maria and {Chatterjee}, Shami and {Cohen}, Tyler and {Cordes}, James M. and {Cornish}, Neil J. and {Crawford}, Fronefield and {Cromartie}, H. Thankful and {Crowter}, Kathryn and {Decesar}, Megan E. and {Demorest}, Paul B. and {Dolch}, Timothy and {Drachler}, Brendan and {Ferrara}, Elizabeth C. and {Fiore}, William and {Fonseca}, Emmanuel and {Freedman}, Gabriel E. and {Garver-Daniels}, Nate and {Gentile}, Peter A. and {Glaser}, Joseph and {Good}, Deborah C. and {G{\"u}ltekin}, Kayhan and {Hazboun}, Jeffrey S. and {Jennings}, Ross J. and {Jessup}, Cody and {Johnson}, Aaron D. and {Jones}, Megan L. and {Kaiser}, Andrew R. and {Kaplan}, David L. and {Kelley}, Luke Zoltan and {Kerr}, Matthew and {Key}, Joey S. and {Kuske}, Anastasia and {Laal}, Nima and {Lam}, Michael T. and {Lamb}, William G. and {Lazio}, T. Joseph W. and {Lewandowska}, Natalia and {Lin}, Ye and {Liu}, Tingting and {Lorimer}, Duncan R. and {Luo}, Jing and {Lynch}, Ryan S. and {Ma}, Chung-Pei and {Madison}, Dustin R. and {Maraccini}, Kaleb and {McEwen}, Alexander and {McKee}, James W. and {McLaughlin}, Maura A. and {McMann}, Natasha and {Meyers}, Bradley W. and {Mingarelli}, Chiara M.~F. and {Mitridate}, Andrea and {Ng}, Cherry and {Nice}, David J. and {Ocker}, Stella Koch and {Olum}, Ken D. and {Panciu}, Elisa and {Pennucci}, Timothy T. and {Perera}, Benetge B.~P. and {Pol}, Nihan S. and {Radovan}, Henri A. and {Ransom}, Scott M. and {Ray}, Paul S. and {Romano}, Joseph D. and {Salo}, Laura and {Sardesai}, Shashwat C. and {Schmiedekamp}, Carl and {Schmiedekamp}, Ann and {Schmitz}, Kai and {Shapiro-Albert}, Brent J. and {Siemens}, Xavier and {Simon}, Joseph and {Siwek}, Magdalena S. and {Stairs}, Ingrid H. and {Stinebring}, Daniel R. and {Stovall}, Kevin and {Susobhanan}, Abhimanyu and {Swiggum}, Joseph K. and {Taylor}, Stephen R. and {Turner}, Jacob E. and {Unal}, Caner and {Vallisneri}, Michele and {Vigeland}, Sarah J. and {Wahl}, Haley M. and {Wang}, Qiaohong and {Witt}, Caitlin A. and {Young}, Olivia and {Nanograv Collaboration}},
        title = "{The NANOGrav 15 yr Data Set: Observations and Timing of 68 Millisecond Pulsars}",
      journal = {\apjl},
         year = 2023,
        month = jul,
       volume = {951},
       number = {1},
          eid = {L9},
        pages = {L9},
          doi = {10.3847/2041-8213/acda9a},
archivePrefix = {arXiv},
       eprint = {2306.16217},
 primaryClass = {astro-ph.HE},
       adsurl = {https://ui.adsabs.harvard.edu/abs/2023ApJ...951L...9A}
}

@ARTICLE{2025MNRAS.536.1489M,
       author = {{Miles}, Matthew T. and {Shannon}, Ryan M. and {Reardon}, Daniel J. and {Bailes}, Matthew and {Champion}, David J. and {Geyer}, Marisa and {Gitika}, Pratyasha and {Grunthal}, Kathrin and {Keith}, Michael J. and {Kramer}, Michael and {Kulkarni}, Atharva D. and {Nathan}, Rowina S. and {Parthasarathy}, Aditya and {Singha}, Jaikhomba and {Theureau}, Gilles and {Thrane}, Eric and {Abbate}, Federico and {Buchner}, Sarah and {Cameron}, Andrew D. and {Camilo}, Fernando and {Moreschi}, Beatrice E. and {Shaifullah}, Golam and {Shamohammadi}, Mohsen and {Possenti}, Andrea and {Krishnan}, Vivek Venkatraman},
        title = "{The MeerKAT Pulsar Timing Array: the first search for gravitational waves with the MeerKAT radio telescope}",
      journal = {\mnras},
         year = 2025,
        month = jan,
       volume = {536},
       number = {2},
        pages = {1489-1500},
          doi = {10.1093/mnras/stae2571},
archivePrefix = {arXiv},
       eprint = {2412.01153},
 primaryClass = {astro-ph.HE},
       adsurl = {https://ui.adsabs.harvard.edu/abs/2025MNRAS.536.1489M}
}

@ARTICLE{2024ApJ...974..295D,
       author = {{Dittmann}, Alexander J. and {Miller}, M. Coleman and {Lamb}, Frederick K. and {Holt}, Isiah M. and {Chirenti}, Cecilia and {Wolff}, Michael T. and {Bogdanov}, Slavko and {Guillot}, Sebastien and {Ho}, Wynn C.~G. and {Morsink}, Sharon M. and {Arzoumanian}, Zaven and {Gendreau}, Keith C.},
        title = "{A More Precise Measurement of the Radius of PSR J0740+6620 Using Updated NICER Data}",
      journal = {\apj},
         year = 2024,
        month = oct,
       volume = {974},
       number = {2},
          eid = {295},
        pages = {295},
          doi = {10.3847/1538-4357/ad5f1e},
archivePrefix = {arXiv},
       eprint = {2406.14467},
 primaryClass = {astro-ph.HE},
       adsurl = {https://ui.adsabs.harvard.edu/abs/2024ApJ...974..295D}
}

@ARTICLE{2025PhRvD.111l3021G,
       author = {{G{\"a}rtlein}, Christoph and {Sagun}, Violetta and {Ivanytskyi}, Oleksii and {Blaschke}, David and {Lopes}, Il{\'\i}dio},
        title = "{Fastest spinning millisecond pulsars: Indicators for quark matter in neutron stars?}",
      journal = {\prd},
         year = 2025,
        month = jun,
       volume = {111},
       number = {12},
          eid = {123021},
        pages = {123021},
          doi = {10.1103/PhysRevD.111.123021},
archivePrefix = {arXiv},
       eprint = {2412.07758},
 primaryClass = {nucl-th},
       adsurl = {https://ui.adsabs.harvard.edu/abs/2025PhRvD.111l3021G}
}

@ARTICLE{2025ApJ...995...60M,
       author = {{Mauviard}, Lucien and {Guillot}, Sebastien and {Salmi}, Tuomo and {Choudhury}, Devarshi and {Dorsman}, Bas and {Gonz{\'a}lez-Caniulef}, Denis and {Hoogkamer}, Mariska and {Huppenkothen}, Daniela and {Kazantsev}, Christine and {Kini}, Yves and {Olive}, Jean-Francois and {Stammler}, Pierre and {Watts}, Anna L. and {Mendes}, Melissa and {Rutherford}, Nathan and {Schwenk}, Achim and {Svensson}, Isak and {Bogdanov}, Slavko and {Kerr}, Matthew and {Ray}, Paul S. and {Guillemot}, Lucas and {Cognard}, Isma{\"e}l and {Theureau}, Gilles},
        title = "{A NICER View of the 1.4 M$_{{\ensuremath{\odot}}}$ Edge-on Pulsar PSR J0614-3329}",
      journal = {\apj},
         year = 2025,
        month = dec,
       volume = {995},
       number = {1},
          eid = {60},
        pages = {60},
          doi = {10.3847/1538-4357/ae145d},
archivePrefix = {arXiv},
       eprint = {2506.14883},
 primaryClass = {astro-ph.HE},
       adsurl = {https://ui.adsabs.harvard.edu/abs/2025ApJ...995...60M}
}

@ARTICLE{2021MNRAS.504.2094K,
       author = {{Kramer}, M. and {Stairs}, I.~H. and {Venkatraman Krishnan}, V. and {Freire}, P.~C.~C. and {Abbate}, F. and {Bailes}, M. and {Burgay}, M. and {Buchner}, S. and {Champion}, D.~J. and {Cognard}, I. and {Gautam}, T. and {Geyer}, M. and {Guillemot}, L. and {Hu}, H. and {Janssen}, G. and {Lower}, M.~E. and {Parthasarathy}, A. and {Possenti}, A. and {Ransom}, S. and {Reardon}, D.~J. and {Ridolfi}, A. and {Serylak}, M. and {Shannon}, R.~M. and {Spiewak}, R. and {Theureau}, G. and {van Straten}, W. and {Wex}, N. and {Oswald}, L.~S. and {Posselt}, B. and {Sobey}, C. and {Barr}, E.~D. and {Camilo}, F. and {Hugo}, B. and {Jameson}, A. and {Johnston}, S. and {Karastergiou}, A. and {Keith}, M. and {Os{\l}owski}, S.},
        title = "{The relativistic binary programme on MeerKAT: science objectives and first results}",
      journal = {\mnras},
         year = 2021,
        month = jun,
       volume = {504},
       number = {2},
        pages = {2094-2114},
          doi = {10.1093/mnras/stab375},
archivePrefix = {arXiv},
       eprint = {2102.05160},
 primaryClass = {astro-ph.HE},
       adsurl = {https://ui.adsabs.harvard.edu/abs/2021MNRAS.504.2094K}
}

@ARTICLE{2021PhRvX..11d1050K,
       author = {{Kramer}, M. and {Stairs}, I.~H. and {Manchester}, R.~N. and {Wex}, N. and {Deller}, A.~T. and {Coles}, W.~A. and {Ali}, M. and {Burgay}, M. and {Camilo}, F. and {Cognard}, I. and {Damour}, T. and {Desvignes}, G. and {Ferdman}, R.~D. and {Freire}, P.~C.~C. and {Grondin}, S. and {Guillemot}, L. and {Hobbs}, G.~B. and {Janssen}, G. and {Karuppusamy}, R. and {Lorimer}, D.~R. and {Lyne}, A.~G. and {McKee}, J.~W. and {McLaughlin}, M. and {M{\"u}nch}, L.~E. and {Perera}, B.~B.~P. and {Pol}, N. and {Possenti}, A. and {Sarkissian}, J. and {Stappers}, B.~W. and {Theureau}, G.},
        title = "{Strong-Field Gravity Tests with the Double Pulsar}",
      journal = {Physical Review X},
         year = 2021,
        month = oct,
       volume = {11},
       number = {4},
          eid = {041050},
        pages = {041050},
          doi = {10.1103/PhysRevX.11.041050},
archivePrefix = {arXiv},
       eprint = {2112.06795},
 primaryClass = {astro-ph.HE},
       adsurl = {https://ui.adsabs.harvard.edu/abs/2021PhRvX..11d1050K}
}

@ARTICLE{2022A&A...667A.149H,
       author = {{Hu}, H. and {Kramer}, M. and {Champion}, D.~J. and {Wex}, N. and {Parthasarathy}, A. and {Pennucci}, T.~T. and {Porayko}, N.~K. and {van Straten}, W. and {Venkatraman Krishnan}, V. and {Burgay}, M. and {Freire}, P.~C.~C. and {Manchester}, R.~N. and {Possenti}, A. and {Stairs}, I.~H. and {Bailes}, M. and {Buchner}, S. and {Cameron}, A.~D. and {Camilo}, F. and {Serylak}, M.},
        title = "{Gravitational signal propagation in the double pulsar studied with the MeerKAT telescope}",
      journal = {\aap},
         year = 2022,
        month = nov,
       volume = {667},
          eid = {A149},
        pages = {A149},
          doi = {10.1051/0004-6361/202244825},
archivePrefix = {arXiv},
       eprint = {2209.11798},
 primaryClass = {astro-ph.HE},
       adsurl = {https://ui.adsabs.harvard.edu/abs/2022A&A...667A.149H}
}

@ARTICLE{2022MNRAS.510.5908M,
       author = {{Miles}, M.~T. and {Shannon}, R.~M. and {Bailes}, M. and {Reardon}, D.~J. and {Buchner}, S. and {Middleton}, H. and {Spiewak}, R.},
        title = "{Mode changing in J1909 - 3744: the most precisely timed pulsar}",
      journal = {\mnras},
         year = 2022,
        month = mar,
       volume = {510},
       number = {4},
        pages = {5908-5915},
          doi = {10.1093/mnras/stab3549},
archivePrefix = {arXiv},
       eprint = {2112.00897},
 primaryClass = {astro-ph.HE},
       adsurl = {https://ui.adsabs.harvard.edu/abs/2022MNRAS.510.5908M}
}

@ARTICLE{2023MNRAS.523.4405N,
       author = {{Nathan}, Rowina S. and {Miles}, Matthew T. and {Ashton}, Gregory and {Lasky}, Paul D. and {Thrane}, Eric and {Reardon}, Daniel J. and {Shannon}, Ryan M. and {Cameron}, Andrew D.},
        title = "{Improving pulsar-timing solutions through dynamic pulse fitting}",
      journal = {\mnras},
         year = 2023,
        month = aug,
       volume = {523},
       number = {3},
        pages = {4405-4412},
          doi = {10.1093/mnras/stad1660},
archivePrefix = {arXiv},
       eprint = {2304.02793},
 primaryClass = {astro-ph.IM},
       adsurl = {https://ui.adsabs.harvard.edu/abs/2023MNRAS.523.4405N}
}

@ARTICLE{2018ApJ...868..122B,
       author = {{Brook}, P.~R. and {Karastergiou}, A. and {McLaughlin}, M.~A. and {Lam}, M.~T. and {Arzoumanian}, Z. and {Chatterjee}, S. and {Cordes}, J.~M. and {Crowter}, K. and {DeCesar}, M. and {Demorest}, P.~B. and {Dolch}, T. and {Ellis}, J.~A. and {Ferdman}, R.~D. and {Ferrara}, E. and {Fonseca}, E. and {Gentile}, P.~A. and {Jones}, G. and {Jones}, M.~L. and {Lazio}, T.~J.~W. and {Levin}, L. and {Lorimer}, D.~R. and {Lynch}, R.~S. and {Ng}, C. and {Nice}, D.~J. and {Pennucci}, T.~T. and {Ransom}, S.~M. and {Ray}, P.~S. and {Spiewak}, R. and {Stairs}, I.~H. and {Stinebring}, D.~R. and {Stovall}, K. and {Swiggum}, J.~K. and {Zhu}, W.~W.},
        title = "{The NANOGrav 11-year Data Set: Pulse Profile Variability}",
      journal = {\apj},
         year = 2018,
        month = dec,
       volume = {868},
       number = {2},
          eid = {122},
        pages = {122},
          doi = {10.3847/1538-4357/aae9e3},
archivePrefix = {arXiv},
       eprint = {1810.08269},
 primaryClass = {astro-ph.HE},
       adsurl = {https://ui.adsabs.harvard.edu/abs/2018ApJ...868..122B}
}

@ARTICLE{2021MNRAS.500.1178P,
       author = {{Padmanabh}, Prajwal V. and {Barr}, Ewan D. and {Champion}, David J. and {Karuppusamy}, Ramesh and {Kramer}, Michael and {Jessner}, Axel and {Lazarus}, Patrick},
        title = "{Revisiting profile instability of PSR J1022+1001}",
      journal = {\mnras},
         year = 2021,
        month = jan,
       volume = {500},
       number = {1},
        pages = {1178-1187},
          doi = {10.1093/mnras/staa3174},
archivePrefix = {arXiv},
       eprint = {2010.04206},
 primaryClass = {astro-ph.IM},
       adsurl = {https://ui.adsabs.harvard.edu/abs/2021MNRAS.500.1178P}
}

@ARTICLE{2025ApJ...984..139F,
       author = {{Fiore}, William and {McLaughlin}, Maura A. and {Agazie}, Gabriella and {Anumarlapudi}, Akash and {Archibald}, Anne M. and {Arzoumanian}, Zaven and {Baker}, Paul T. and {Brook}, Paul R. and {Cromartie}, H. Thankful and {Crowter}, Kathryn and {DeCesar}, Megan E. and {Demorest}, Paul B. and {Dey}, Lankeswar and {Dolch}, Timothy and {Ferrara}, Elizabeth C. and {Fonseca}, Emmanuel and {Freedman}, Gabriel E. and {Garver-Daniels}, Nate and {Gentile}, Peter A. and {Glaser}, Joseph and {Good}, Deborah C. and {Hazboun}, Jeffrey S. and {Jennings}, Ross J. and {Jones}, Megan L. and {Kaplan}, David L. and {Kerr}, Matthew and {Lam}, Michael T. and {Lorimer}, Duncan R. and {Luo}, Jing and {Lynch}, Ryan S. and {McEwen}, Alexander and {McMann}, Natasha and {Meyers}, Bradley W. and {Ng}, Cherry and {Nice}, David J. and {Pennucci}, Timothy T. and {Perera}, Benetge B.~P. and {Pol}, Nihan S. and {Radovan}, Henri A. and {Ransom}, Scott M. and {Ray}, Paul S. and {Schmiedekamp}, Ann and {Schmiedekamp}, Carl and {Shapiro-Albert}, Brent J. and {Stairs}, Ingrid H. and {Stovall}, Kevin and {Susobhanan}, Abhimanyu and {Swiggum}, Joseph K. and {Wahl}, Haley M.},
        title = "{Pulse Profile Variability of PSR J1022+1001 in NANOGrav Data}",
      journal = {\apj},
         year = 2025,
        month = may,
       volume = {984},
       number = {2},
          eid = {139},
        pages = {139},
          doi = {10.3847/1538-4357/adc255},
archivePrefix = {arXiv},
       eprint = {2412.05452},
 primaryClass = {astro-ph.HE},
       adsurl = {https://ui.adsabs.harvard.edu/abs/2025ApJ...984..139F}
}

@ARTICLE{2014MNRAS.443.1463S,
       author = {{Shannon}, R.~M. and {Os{\l}owski}, S. and {Dai}, S. and {Bailes}, M. and {Hobbs}, G. and {Manchester}, R.~N. and {van Straten}, W. and {Raithel}, C.~A. and {Ravi}, V. and {Toomey}, L. and {Bhat}, N.~D.~R. and {Burke-Spolaor}, S. and {Coles}, W.~A. and {Keith}, M.~J. and {Kerr}, M. and {Levin}, Y. and {Sarkissian}, J.~M. and {Wang}, J.-B. and {Wen}, L. and {Zhu}, X.-J.},
        title = "{Limitations in timing precision due to single-pulse shape variability in millisecond pulsars}",
      journal = {\mnras},
         year = 2014,
        month = sep,
       volume = {443},
       number = {2},
        pages = {1463-1481},
          doi = {10.1093/mnras/stu1213},
archivePrefix = {arXiv},
       eprint = {1406.4716},
 primaryClass = {astro-ph.SR},
       adsurl = {https://ui.adsabs.harvard.edu/abs/2014MNRAS.443.1463S}
}

@ARTICLE{2016ApJ...828L...1S,
       author = {{Shannon}, R.~M. and {Lentati}, L.~T. and {Kerr}, M. and {Bailes}, M. and {Bhat}, N.~D.~R. and {Coles}, W.~A. and {Dai}, S. and {Dempsey}, J. and {Hobbs}, G. and {Keith}, M.~J. and {Lasky}, P.~D. and {Levin}, Y. and {Manchester}, R.~N. and {Os{\l}owski}, S. and {Ravi}, V. and {Reardon}, D.~J. and {Rosado}, P.~A. and {Spiewak}, R. and {van Straten}, W. and {Toomey}, L. and {Wang}, J.-B. and {Wen}, L. and {You}, X.-P. and {Zhu}, X.-J.},
        title = "{The Disturbance of a Millisecond Pulsar Magnetosphere}",
      journal = {\apjl},
         year = 2016,
        month = sep,
       volume = {828},
       number = {1},
          eid = {L1},
        pages = {L1},
          doi = {10.3847/2041-8205/828/1/L1},
archivePrefix = {arXiv},
       eprint = {1608.02163},
 primaryClass = {astro-ph.HE},
       adsurl = {https://ui.adsabs.harvard.edu/abs/2016ApJ...828L...1S}
}

@ARTICLE{2025PASA...42..142M,
       author = {{Mandow}, Rami F. and {Zic}, Andrew and {Dawson}, J.~R. and {Wang}, Shuangqiang and {Cury{\l}o}, Ma{\l}gorzata and {Dai}, Shi and {Di Marco}, Valentina and {Hobbs}, George and {Gupta}, Vivek and {Kapur}, Agastya and {Kerr}, Matthew and {Lower}, Marcus E. and {Mishra}, Saurav and {Reardon}, Daniel and {Russell}, Christopher and {Shannon}, Ryan M. and {Zhang}, Lei and {Zhu}, Xingjiang},
        title = "{Ultra-wideband polarimetry of the April 2021 profile change event in PSR J1713+0747}",
      journal = {\pasa},
         year = 2025,
        month = oct,
       volume = {42},
          eid = {e142},
        pages = {e142},
          doi = {10.1017/pasa.2025.10099},
archivePrefix = {arXiv},
       eprint = {2509.18972},
 primaryClass = {astro-ph.HE},
       adsurl = {https://ui.adsabs.harvard.edu/abs/2025PASA...42..142M}
}

@ARTICLE{1997ApJ...475L..33A,
       author = {{Ables}, J.~G. and {McConnell}, D. and {Deshpande}, A.~A. and {Vivekanand}, M.},
        title = "{Coherent Radiation Patterns Suggested by Single-Pulse Observations of a Millisecond Pulsar}",
      journal = {\apjl},
         year = 1997,
        month = jan,
       volume = {475},
       number = {1},
        pages = {L33-L36},
          doi = {10.1086/310455},
       adsurl = {https://ui.adsabs.harvard.edu/abs/1997ApJ...475L..33A}
}

@ARTICLE{2000ApJ...543..979V,
       author = {{Vivekanand}, M.},
        title = "{Timescales of Radio Emission in Pulsar J0437-4715 at 327MHz}",
      journal = {\apj},
         year = 2000,
        month = nov,
       volume = {543},
       number = {2},
        pages = {979-986},
          doi = {10.1086/317141},
archivePrefix = {arXiv},
       eprint = {astro-ph/0006248},
 primaryClass = {astro-ph},
       adsurl = {https://ui.adsabs.harvard.edu/abs/2000ApJ...543..979V}
}

@ARTICLE{2001MNRAS.326L..33V,
       author = {{Vivekanand}, M.},
        title = "{Radio profile instability in the millisecond pulsar PSR J0437-4715 resulting from spiky emission}",
      journal = {\mnras},
         year = 2001,
        month = sep,
       volume = {326},
       number = {2},
        pages = {L33-L36},
          doi = {10.1046/j.1365-8711.2001.04811.x},
archivePrefix = {arXiv},
       eprint = {astro-ph/0205179},
 primaryClass = {astro-ph},
       adsurl = {https://ui.adsabs.harvard.edu/abs/2001MNRAS.326L..33V}
}

@ARTICLE{2021MNRAS.502..478G,
       author = {{Goncharov}, Boris and {Reardon}, D.~J. and {Shannon}, R.~M. and {Zhu}, Xing-Jiang and {Thrane}, Eric and {Bailes}, M. and {Bhat}, N.~D.~R. and {Dai}, S. and {Hobbs}, G. and {Kerr}, M. and {Manchester}, R.~N. and {Os{\l}owski}, S. and {Parthasarathy}, A. and {Russell}, C.~J. and {Spiewak}, R. and {Thyagarajan}, N. and {Wang}, J.~B.},
        title = "{Identifying and mitigating noise sources in precision pulsar timing data sets}",
      journal = {\mnras},
         year = 2021,
        month = mar,
       volume = {502},
       number = {1},
        pages = {478-493},
          doi = {10.1093/mnras/staa3411},
archivePrefix = {arXiv},
       eprint = {2010.06109},
 primaryClass = {astro-ph.HE},
       adsurl = {https://ui.adsabs.harvard.edu/abs/2021MNRAS.502..478G}
}

@ARTICLE{2023MNRAS.520.4961O,
       author = {{Oswald}, L.~S. and {Johnston}, S. and {Karastergiou}, A. and {Dai}, S. and {Kerr}, M. and {Lower}, M.~E. and {Manchester}, R.~N. and {Shannon}, R.~M. and {Sobey}, C. and {Weltevrede}, P.},
        title = "{Pulsar polarization: a broad-band population view with the Parkes Ultra-Wideband receiver}",
      journal = {\mnras},
         year = 2023,
        month = apr,
       volume = {520},
       number = {4},
        pages = {4961-4980},
          doi = {10.1093/mnras/stad070},
archivePrefix = {arXiv},
       eprint = {2301.05628},
 primaryClass = {astro-ph.HE},
       adsurl = {https://ui.adsabs.harvard.edu/abs/2023MNRAS.520.4961O}
}

@ARTICLE{2024MNRAS.530.4839J,
       author = {{Johnston}, Simon and {Mitra}, Dipanjan and {Keith}, Michael J. and {Oswald}, Lucy S. and {Karastergiou}, Aris},
        title = "{The Thousand-Pulsar-Array programme on MeerKAT ─ XIV. On the high linearly polarized pulsar signals}",
      journal = {\mnras},
         year = 2024,
        month = jun,
       volume = {530},
       number = {4},
        pages = {4839-4849},
          doi = {10.1093/mnras/stae1175},
archivePrefix = {arXiv},
       eprint = {2404.10254},
 primaryClass = {astro-ph.HE},
       adsurl = {https://ui.adsabs.harvard.edu/abs/2024MNRAS.530.4839J}
}

@ARTICLE{2024MNRAS.532.3558K,
       author = {{Karastergiou}, A. and {Johnston}, S. and {Posselt}, B. and {Oswald}, L.~S. and {Kramer}, M. and {Weltevrede}, P.},
        title = "{The Thousand-Pulsar-Array programme on MeerKAT ─ XV. A comparison of the radio emission properties of slow and millisecond pulsars}",
      journal = {\mnras},
         year = 2024,
        month = aug,
       volume = {532},
       number = {3},
        pages = {3558-3566},
          doi = {10.1093/mnras/stae1694},
archivePrefix = {arXiv},
       eprint = {2407.06836},
 primaryClass = {astro-ph.HE},
       adsurl = {https://ui.adsabs.harvard.edu/abs/2024MNRAS.532.3558K}
}

@ARTICLE{2025A&A...695A.173X,
       author = {{Xu}, Jiangwei and {Jiang}, Jinchen and {Xu}, Heng and {Wang}, Bojun and {Xue}, Zihan and {Chen}, Siyuan and {Guo}, Yanjun and {Caballero}, R. Nicolas and {Lee}, Kejia and {Yuan}, Jianping and {Xu}, Yonghua and {Wang}, Jingbo and {Hao}, Longfei and {Li}, Zhixuan and {Huang}, Yuxiang and {Xu}, Zezhong and {Luo}, Jintao and {Han}, Jinlin and {Jiang}, Peng and {Shen}, Zhiqiang and {Wang}, Min and {Wang}, Na and {Xu}, Renxin and {Wu}, Xiangping and {Qian}, Lei and {Yue}, Youling and {Guan}, Xin and {Huang}, Menglin and {Sun}, Chun and {Zhu}, Yan},
        title = "{The Chinese pulsar timing array data release I: Polarimetry for 56 millisecond pulsars}",
      journal = {\aap},
         year = 2025,
        month = mar,
       volume = {695},
          eid = {A173},
        pages = {A173},
          doi = {10.1051/0004-6361/202452960},
archivePrefix = {arXiv},
       eprint = {2502.20820},
 primaryClass = {astro-ph.HE},
       adsurl = {https://ui.adsabs.harvard.edu/abs/2025A&A...695A.173X}
}

@ARTICLE{1969Natur.221..443R,
       author = {{Radhakrishnan}, V. and {Cooke}, D.~J. and {Komesaroff}, M.~M. and {Morris}, D.},
        title = "{Evidence in Support of a Rotational Model for the Pulsar PSR 0833-45}",
      journal = {\nat},
         year = 1969,
        month = feb,
       volume = {221},
       number = {5179},
        pages = {443-446},
          doi = {10.1038/221443a0},
       adsurl = {https://ui.adsabs.harvard.edu/abs/1969Natur.221..443R}
}

@ARTICLE{2020PASA...37...12H,
       author = {{Hobbs}, George and {Manchester}, Richard N. and {Dunning}, Alex and {Jameson}, Andrew and {Roberts}, Paul and {George}, Daniel and {Green}, J.~A. and {Tuthill}, John and {Toomey}, Lawrence and {Kaczmarek}, Jane F. and {Mader}, Stacy and {Marquarding}, Malte and {Ahmed}, Azeem and {Amy}, Shaun W. and {Bailes}, Matthew and {Beresford}, Ron and {Bhat}, N.~D.~R. and {Bock}, Douglas C.-J. and {Bourne}, Michael and {Bowen}, Mark and {Brothers}, Michael and {Cameron}, Andrew D. and {Carretti}, Ettore and {Carter}, Nick and {Castillo}, Santy and {Chekkala}, Raji and {Cheng}, Wan and {Chung}, Yoon and {Craig}, Daniel A. and {Dai}, Shi and {Dawson}, Joanne and {Dempsey}, James and {Doherty}, Paul and {Dong}, Bin and {Edwards}, Philip and {Ergesh}, Tuohutinuer and {Gao}, Xuyang and {Han}, JinLin and {Hayman}, Douglas and {Indermuehle}, Balthasar and {Jeganathan}, Kanapathippillai and {Johnston}, Simon and {Kanoniuk}, Henry and {Kesteven}, Michael and {Kramer}, Michael and {Leach}, Mark and {Mcintyre}, Vince and {Moss}, Vanessa and {Os{\l}owski}, Stefan and {Phillips}, Chris and {Pope}, Nathan and {Preisig}, Brett and {Price}, Daniel and {Reeves}, Ken and {Reilly}, Les and {Reynolds}, John and {Robishaw}, Tim and {Roush}, Peter and {Ruckley}, Tim and {Sadler}, Elaine and {Sarkissian}, John and {Severs}, Sean and {Shannon}, Ryan and {Smart}, Ken and {Smith}, Malcolm and {Smith}, Stephanie and {Sobey}, Charlotte and {Staveley-Smith}, Lister and {Tzioumis}, Anastasios and {van Straten}, Willem and {Wang}, Nina and {Wen}, Linqing and {Whiting}, Matthew},
        title = "{An ultra-wide bandwidth (704 to 4 032 MHz) receiver for the Parkes radio telescope}",
      journal = {\pasa},
         year = 2020,
        month = apr,
       volume = {37},
          eid = {e012},
        pages = {e012},
          doi = {10.1017/pasa.2020.2},
archivePrefix = {arXiv},
       eprint = {1911.00656},
 primaryClass = {astro-ph.IM},
       adsurl = {https://ui.adsabs.harvard.edu/abs/2020PASA...37...12H}
}

@ARTICLE{2023PASA...40...49Z,
       author = {{Zic}, Andrew and {Reardon}, Daniel J. and {Kapur}, Agastya and {Hobbs}, George and {Mandow}, Rami and {Cury{\l}o}, Ma{\l}gorzata and {Shannon}, Ryan M. and {Askew}, Jacob and {Bailes}, Matthew and {Bhat}, N.~D. Ramesh and {Cameron}, Andrew and {Chen}, Zu-Cheng and {Dai}, Shi and {Di Marco}, Valentina and {Feng}, Yi and {Kerr}, Matthew and {Kulkarni}, Atharva and {Lower}, Marcus E. and {Luo}, Rui and {Manchester}, Richard N. and {Miles}, Matthew T. and {Nathan}, Rowina S. and {Os{\l}owski}, Stefan and {Rogers}, Axl F. and {Russell}, Christopher J. and {Sarkissian}, John M. and {Shamohammadi}, Mohsen and {Spiewak}, Ren{\'e}e and {Thyagarajan}, Nithyanandan and {Toomey}, Lawrence and {Wang}, Shuangqiang and {Zhang}, Lei and {Zhang}, Songbo and {Zhu}, Xing-Jiang},
        title = "{The Parkes Pulsar Timing Array third data release}",
      journal = {\pasa},
         year = 2023,
        month = dec,
       volume = {40},
          eid = {e049},
        pages = {e049},
          doi = {10.1017/pasa.2023.36},
archivePrefix = {arXiv},
       eprint = {2306.16230},
 primaryClass = {astro-ph.HE},
       adsurl = {https://ui.adsabs.harvard.edu/abs/2023PASA...40...49Z}
}

@ARTICLE{2004PASA...21..302H,
       author = {{Hotan}, A.~W. and {van Straten}, W. and {Manchester}, R.~N.},
        title = "{PSRCHIVE and PSRFITS: An Open Approach to Radio Pulsar Data Storage and Analysis}",
      journal = {\pasa},
         year = 2004,
        month = jan,
       volume = {21},
       number = {3},
        pages = {302-309},
          doi = {10.1071/AS04022},
archivePrefix = {arXiv},
       eprint = {astro-ph/0404549},
 primaryClass = {astro-ph},
       adsurl = {https://ui.adsabs.harvard.edu/abs/2004PASA...21..302H}
}

@ARTICLE{2011PASA...28....1V,
       author = {{van Straten}, W. and {Bailes}, M.},
        title = "{DSPSR: Digital Signal Processing Software for Pulsar Astronomy}",
      journal = {\pasa},
         year = 2011,
        month = jan,
       volume = {28},
       number = {1},
        pages = {1-14},
          doi = {10.1071/AS10021},
archivePrefix = {arXiv},
       eprint = {1008.3973},
 primaryClass = {astro-ph.IM},
       adsurl = {https://ui.adsabs.harvard.edu/abs/2011PASA...28....1V}
}

@ARTICLE{2006MNRAS.372.1549E,
       author = {{Edwards}, R.~T. and {Hobbs}, G.~B. and {Manchester}, R.~N.},
        title = "{TEMPO2, a new pulsar timing package - II. The timing model and precision estimates}",
      journal = {\mnras},
         year = 2006,
        month = nov,
       volume = {372},
       number = {4},
        pages = {1549-1574},
          doi = {10.1111/j.1365-2966.2006.10870.x},
archivePrefix = {arXiv},
       eprint = {astro-ph/0607664},
 primaryClass = {astro-ph},
       adsurl = {https://ui.adsabs.harvard.edu/abs/2006MNRAS.372.1549E}
}

@ARTICLE{2002ApJ...581..495B,
       author = {{Bogdanov}, Slavko and {Pruszy{\'n}ska}, Ma{\l}gorzata and {Lewandowski}, Wojciech and {Wolszczan}, Alex},
        title = "{Interstellar Scintillation Velocities of the Relativistic Binary PSR B1534+12 and Three Other Millisecond Pulsars}",
      journal = {\apj},
         year = 2002,
        month = dec,
       volume = {581},
       number = {1},
        pages = {495-500},
          doi = {10.1086/344169},
archivePrefix = {arXiv},
       eprint = {astro-ph/0204270},
 primaryClass = {astro-ph},
       adsurl = {https://ui.adsabs.harvard.edu/abs/2002ApJ...581..495B}
}

@ARTICLE{1992RSPTA.341..117T,
       author = {{Taylor}, J.~H.},
        title = "{Pulsar Timing and Relativistic Gravity}",
      journal = {Philosophical Transactions of the Royal Society of London Series A},
         year = 1992,
        month = oct,
       volume = {341},
       number = {1660},
        pages = {117-134},
          doi = {10.1098/rsta.1992.0088},
       adsurl = {https://ui.adsabs.harvard.edu/abs/1992RSPTA.341..117T}
}

@article{scikit-learn,
  title={Scikit-learn: Machine Learning in {P}ython},
  author={Pedregosa, F. and Varoquaux, G. and Gramfort, A. and Michel, V.
          and Thirion, B. and Grisel, O. and Blondel, M. and Prettenhofer, P.
          and Weiss, R. and Dubourg, V. and Vanderplas, J. and Passos, A. and
          Cournapeau, D. and Brucher, M. and Perrot, M. and Duchesnay, E.},
  journal={Journal of Machine Learning Research},
  volume={12},
  pages={2825--2830},
  year={2011}
}

@ARTICLE{1989ApJ...347.1030W,
       author = {{Weisberg}, J.~M. and {Romani}, R.~W. and {Taylor}, J.~H.},
        title = "{Evidence for Geodetic Spin Precession in the Binary Pulsar PSR 1913+16}",
      journal = {\apj},
         year = 1989,
        month = dec,
       volume = {347},
        pages = {1030},
          doi = {10.1086/168193},
       adsurl = {https://ui.adsabs.harvard.edu/abs/1989ApJ...347.1030W}
}

@ARTICLE{2000Natur.406..484S,
       author = {{Stairs}, I.~H. and {Lyne}, A.~G. and {Shemar}, S.~L.},
        title = "{Evidence for free precession in a pulsar}",
      journal = {\nat},
         year = 2000,
        month = aug,
       volume = {406},
       number = {6795},
        pages = {484-486},
          doi = {10.1038/35020010},
       adsurl = {https://ui.adsabs.harvard.edu/abs/2000Natur.406..484S}
}

@ARTICLE{2011MNRAS.411.1917W,
       author = {{Weltevrede}, Patrick and {Johnston}, Simon and {Espinoza}, Crist{\'o}bal M.},
        title = "{The glitch-induced identity changes of PSR J1119-6127}",
      journal = {\mnras},
         year = 2011,
        month = mar,
       volume = {411},
       number = {3},
        pages = {1917-1934},
          doi = {10.1111/j.1365-2966.2010.17821.x},
archivePrefix = {arXiv},
       eprint = {1010.0857},
 primaryClass = {astro-ph.SR},
       adsurl = {https://ui.adsabs.harvard.edu/abs/2011MNRAS.411.1917W}
}

@ARTICLE{2011MNRAS.410..499G,
       author = {{Graham Smith}, F. and {Lyne}, A.~G. and {Jordan}, C.},
        title = "{The 1997 event in the Crab pulsar revisited}",
      journal = {\mnras},
         year = 2011,
        month = jan,
       volume = {410},
       number = {1},
        pages = {499-503},
          doi = {10.1111/j.1365-2966.2010.17459.x},
archivePrefix = {arXiv},
       eprint = {1008.4494},
 primaryClass = {astro-ph.GA},
       adsurl = {https://ui.adsabs.harvard.edu/abs/2011MNRAS.410..499G}
}

@ARTICLE{2013MNRAS.434...69L,
       author = {{Lewandowski}, Wojciech and {Dembska}, Marta and {Kijak}, Jaroslaw and {Kowali{\'n}ska}, Magdalena},
        title = "{Pulse broadening analysis for several new pulsars and anomalous scattering}",
      journal = {\mnras},
         year = 2013,
        month = sep,
       volume = {434},
       number = {1},
        pages = {69-83},
          doi = {10.1093/mnras/stt989},
archivePrefix = {arXiv},
       eprint = {1306.0738},
 primaryClass = {astro-ph.HE},
       adsurl = {https://ui.adsabs.harvard.edu/abs/2013MNRAS.434...69L}
}

@ARTICLE{2013ApJ...766....5S,
       author = {{Shannon}, R.~M. and {Cordes}, J.~M. and {Metcalfe}, T.~S. and {Lazio}, T.~J.~W. and {Cognard}, I. and {Desvignes}, G. and {Janssen}, G.~H. and {Jessner}, A. and {Kramer}, M. and {Lazaridis}, K. and {Purver}, M.~B. and {Stappers}, B.~W. and {Theureau}, G.},
        title = "{An Asteroid Belt Interpretation for the Timing Variations of the Millisecond Pulsar B1937+21}",
      journal = {\apj},
         year = 2013,
        month = mar,
       volume = {766},
       number = {1},
          eid = {5},
        pages = {5},
          doi = {10.1088/0004-637X/766/1/5},
archivePrefix = {arXiv},
       eprint = {1301.6429},
 primaryClass = {astro-ph.SR},
       adsurl = {https://ui.adsabs.harvard.edu/abs/2013ApJ...766....5S}
}

@ARTICLE{2014ApJ...780L..31B,
       author = {{Brook}, P.~R. and {Karastergiou}, A. and {Buchner}, S. and {Roberts}, S.~J. and {Keith}, M.~J. and {Johnston}, S. and {Shannon}, R.~M.},
        title = "{Evidence of an Asteroid Encountering a Pulsar}",
      journal = {\apjl},
         year = 2014,
        month = jan,
       volume = {780},
       number = {2},
          eid = {L31},
        pages = {L31},
          doi = {10.1088/2041-8205/780/2/L31},
archivePrefix = {arXiv},
       eprint = {1311.3541},
 primaryClass = {astro-ph.HE},
       adsurl = {https://ui.adsabs.harvard.edu/abs/2014ApJ...780L..31B}
}

@ARTICLE{1988MNRAS.234..477L,
       author = {{Lyne}, A.~G. and {Manchester}, R.~N.},
        title = "{The shape of pulsar radio beams.}",
      journal = {\mnras},
         year = 1988,
        month = oct,
       volume = {234},
        pages = {477-508},
          doi = {10.1093/mnras/234.3.477},
       adsurl = {https://ui.adsabs.harvard.edu/abs/1988MNRAS.234..477L}
}

@ARTICLE{1993ApJ...405..285R,
       author = {{Rankin}, Joanna M.},
        title = "{Toward an Empirical Theory of Pulsar Emission. VI. The Geometry of the Conal Emission Region}",
      journal = {\apj},
         year = 1993,
        month = mar,
       volume = {405},
        pages = {285},
          doi = {10.1086/172361},
       adsurl = {https://ui.adsabs.harvard.edu/abs/1993ApJ...405..285R}
}

@ARTICLE{2007MNRAS.380.1678K,
       author = {{Karastergiou}, Aris and {Johnston}, Simon},
        title = "{An empirical model for the beams of radio pulsars}",
      journal = {\mnras},
         year = 2007,
        month = oct,
       volume = {380},
       number = {4},
        pages = {1678-1684},
          doi = {10.1111/j.1365-2966.2007.12237.x},
archivePrefix = {arXiv},
       eprint = {0707.2547},
 primaryClass = {astro-ph},
       adsurl = {https://ui.adsabs.harvard.edu/abs/2007MNRAS.380.1678K}
}

@ARTICLE{2015A&A...576A..62N,
       author = {{Noutsos}, A. and {Sobey}, C. and {Kondratiev}, V.~I. and {Weltevrede}, P. and {Verbiest}, J.~P.~W. and {Karastergiou}, A. and {Kramer}, M. and {Kuniyoshi}, M. and {Alexov}, A. and {Breton}, R.~P. and {Bilous}, A.~V. and {Cooper}, S. and {Falcke}, H. and {Grie{\ss}meier}, J.-M. and {Hassall}, T.~E. and {Hessels}, J.~W.~T. and {Keane}, E.~F. and {Os{\l}owski}, S. and {Pilia}, M. and {Serylak}, M. and {Stappers}, B.~W. and {ter Veen}, S. and {van Leeuwen}, J. and {Zagkouris}, K. and {Anderson}, K. and {B{\"a}hren}, L. and {Bell}, M. and {Broderick}, J. and {Carbone}, D. and {Cendes}, Y. and {Coenen}, T. and {Corbel}, S. and {Eisl{\"o}ffel}, J. and {Fender}, R. and {Garsden}, H. and {Jonker}, P. and {Law}, C. and {Markoff}, S. and {Masters}, J. and {Miller-Jones}, J. and {Molenaar}, G. and {Osten}, R. and {Pietka}, M. and {Rol}, E. and {Rowlinson}, A. and {Scheers}, B. and {Spreeuw}, H. and {Staley}, T. and {Stewart}, A. and {Swinbank}, J. and {Wijers}, R. and {Wijnands}, R. and {Wise}, M. and {Zarka}, P. and {van der Horst}, A.},
        title = "{Pulsar polarisation below 200 MHz: Average profiles and propagation effects}",
      journal = {\aap},
         year = 2015,
        month = apr,
       volume = {576},
          eid = {A62},
        pages = {A62},
          doi = {10.1051/0004-6361/201425186},
archivePrefix = {arXiv},
       eprint = {1501.03312},
 primaryClass = {astro-ph.GA},
       adsurl = {https://ui.adsabs.harvard.edu/abs/2015A&A...576A..62N}
}

@ARTICLE{1984ApJS...55..247S,
       author = {{Stinebring}, D.~R. and {Cordes}, J.~M. and {Rankin}, J.~M. and {Weisberg}, J.~M. and {Boriakoff}, V.},
        title = "{Pulsar polarization fluctuations. I. 1404 MHz statistical summaries.}",
      journal = {\apjs},
         year = 1984,
        month = jun,
       volume = {55},
        pages = {247-277},
          doi = {10.1086/190954},
       adsurl = {https://ui.adsabs.harvard.edu/abs/1984ApJS...55..247S}
}

@ARTICLE{1997ApJ...486.1019N,
       author = {{Navarro}, J. and {Manchester}, R.~N. and {Sandhu}, J.~S. and {Kulkarni}, S.~R. and {Bailes}, M.},
        title = "{Mean Pulse Shape and Polarization of PSR J0437-4715}",
      journal = {\apj},
         year = 1997,
        month = sep,
       volume = {486},
       number = {2},
        pages = {1019-1025},
          doi = {10.1086/304563},
       adsurl = {https://ui.adsabs.harvard.edu/abs/1997ApJ...486.1019N}
}

@ARTICLE{1983ApJ...274..333R,
       author = {{Rankin}, J.~M.},
        title = "{Toward an empirical theory of pulsar emission. I. Morphological taxonomy.}",
      journal = {\apj},
         year = 1983,
        month = nov,
       volume = {274},
        pages = {333-358},
          doi = {10.1086/161450},
       adsurl = {https://ui.adsabs.harvard.edu/abs/1983ApJ...274..333R}
}

@ARTICLE{1990ApJ...352..247R,
       author = {{Rankin}, Joanna M.},
        title = "{Toward an Empirical Theory of Pulsar Emission. IV. Geometry of the Core Emission Region}",
      journal = {\apj},
         year = 1990,
        month = mar,
       volume = {352},
        pages = {247},
          doi = {10.1086/168530},
       adsurl = {https://ui.adsabs.harvard.edu/abs/1990ApJ...352..247R}
}

@ARTICLE{1997MNRAS.285..561G,
       author = {{Gil}, Janusz and {Krawczyk}, Agnieszka},
        title = "{PSR J0437-4715: a challenge for pulsar modelling}",
      journal = {\mnras},
         year = 1997,
        month = mar,
       volume = {285},
       number = {3},
        pages = {561-566},
          doi = {10.1093/mnras/285.3.561},
       adsurl = {https://ui.adsabs.harvard.edu/abs/1997MNRAS.285..561G}
}

@ARTICLE{1998ApJ...501..823V,
       author = {{Vivekanand}, M. and {Ables}, J.~G. and {McConnell}, D.},
        title = "{Radio Emission of PSR J0437-4715 at 327 MHz}",
      journal = {\apj},
         year = 1998,
        month = jul,
       volume = {501},
       number = {2},
        pages = {823-829},
          doi = {10.1086/305847},
       adsurl = {https://ui.adsabs.harvard.edu/abs/1998ApJ...501..823V}
}

@ARTICLE{2023ApJ...951L...7R,
       author = {{Reardon}, Daniel J. and {Zic}, Andrew and {Shannon}, Ryan M. and {Di Marco}, Valentina and {Hobbs}, George B. and {Kapur}, Agastya and {Lower}, Marcus E. and {Mandow}, Rami and {Middleton}, Hannah and {Miles}, Matthew T. and {Rogers}, Axl F. and {Askew}, Jacob and {Bailes}, Matthew and {Bhat}, N.~D. Ramesh and {Cameron}, Andrew and {Kerr}, Matthew and {Kulkarni}, Atharva and {Manchester}, Richard N. and {Nathan}, Rowina S. and {Russell}, Christopher J. and {Os{\l}owski}, Stefan and {Zhu}, Xing-Jiang},
        title = "{The Gravitational-wave Background Null Hypothesis: Characterizing Noise in Millisecond Pulsar Arrival Times with the Parkes Pulsar Timing Array}",
      journal = {\apjl},
         year = 2023,
        month = jul,
       volume = {951},
       number = {1},
          eid = {L7},
        pages = {L7},
          doi = {10.3847/2041-8213/acdd03},
archivePrefix = {arXiv},
       eprint = {2306.16229},
 primaryClass = {astro-ph.HE},
       adsurl = {https://ui.adsabs.harvard.edu/abs/2023ApJ...951L...7R}
}

@ARTICLE{2025OJAp....854638J,
       author = {{Joshi}, Bhal Chandra and {Karastergiou}, Aris and {Burgay}, Marta},
        title = "{Pulsar Science with the SKA Observatory}",
      journal = {The Open Journal of Astrophysics},
         year = 2025,
        month = dec,
       volume = {8},
        pages = {54638},
          doi = {10.33232/001c.154638},
archivePrefix = {arXiv},
       eprint = {2512.16152},
 primaryClass = {astro-ph.HE},
       adsurl = {https://ui.adsabs.harvard.edu/abs/2025OJAp....854638J}
}

@INPROCEEDINGS{2018ASPC..517..751C,
       author = {{Chatterjee}, S.},
        title = "{Science with Pulsar Timing Arrays and the ngVLA}",
    booktitle = {Science with a Next Generation Very Large Array},
         year = 2018,
       editor = {{Murphy}, Eric},
       series = {Astronomical Society of the Pacific Conference Series},
       volume = {517},
        month = dec,
        pages = {751},
       adsurl = {https://ui.adsabs.harvard.edu/abs/2018ASPC..517..751C}
}

@ARTICLE{2019RAA....19...20H,
       author = {{Hobbs}, George and {Dai}, Shi and {Manchester}, Richard N. and {Shannon}, Ryan M. and {Kerr}, Matthew and {Lee}, Ke-Jia and {Xu}, Ren-Xin},
        title = "{The role of FAST in pulsar timing arrays}",
      journal = {Research in Astronomy and Astrophysics},
         year = 2019,
        month = feb,
       volume = {19},
       number = {2},
          eid = {020},
        pages = {020},
          doi = {10.1088/1674-4527/19/2/20},
       adsurl = {https://ui.adsabs.harvard.edu/abs/2019RAA....19...20H}
}

@ARTICLE{2025arXiv251209220C,
       author = {{Cury{\l}o}, Ma{\l}gorzata and {Zic}, Andrew and {Wang}, Shuangqiang and {Thrane}, Eric and {Lasky}, Paul D. and {Cardinal Tremblay}, Jacob and {Chen}, Zu-Cheng and {Dai}, Shi and {Di Marco}, Valentina and {Hobbs}, George and {Kapur}, Agastya and {Ling}, Wenhua and {Lower}, Marcus E. and {Mandow}, Rami F. and {Mishra}, Saurav and {Reardon}, Daniel J. and {Russell}, Christopher J. and {Shannon}, Ryan M. and {Zhu}, Xing-Jiang},
        title = "{Frequency- and phase-resolved polarimetry of millisecond pulsars and its application to timing}",
      journal = {arXiv e-prints},
         year = 2025,
        month = dec,
          eid = {arXiv:2512.09220},
        pages = {arXiv:2512.09220},
          doi = {10.48550/arXiv.2512.09220},
archivePrefix = {arXiv},
       eprint = {2512.09220},
 primaryClass = {astro-ph.HE},
       adsurl = {https://ui.adsabs.harvard.edu/abs/2025arXiv251209220C}
}

@ARTICLE{2000ApJ...532.1240B,
       author = {{Britton}, M.~C.},
        title = "{Radio Astronomical Polarimetry and the Lorentz Group}",
      journal = {\apj},
         year = 2000,
        month = apr,
       volume = {532},
       number = {2},
        pages = {1240-1244},
          doi = {10.1086/308595},
archivePrefix = {arXiv},
       eprint = {astro-ph/9911101},
 primaryClass = {astro-ph},
       adsurl = {https://ui.adsabs.harvard.edu/abs/2000ApJ...532.1240B}
}

@ARTICLE{2011MNRAS.418.1258O,
       author = {{Os{\l}owski}, S. and {van Straten}, W. and {Hobbs}, G.~B. and {Bailes}, M. and {Demorest}, P.},
        title = "{High signal-to-noise ratio observations and the ultimate limits of precision pulsar timing}",
      journal = {\mnras},
         year = 2011,
        month = dec,
       volume = {418},
       number = {2},
        pages = {1258-1271},
          doi = {10.1111/j.1365-2966.2011.19578.x},
archivePrefix = {arXiv},
       eprint = {1108.0812},
 primaryClass = {astro-ph.GA},
       adsurl = {https://ui.adsabs.harvard.edu/abs/2011MNRAS.418.1258O}
}

@ARTICLE{2026ApJ..1005...58J,
       author = {{Jacobson-Bell}, Ben and {Cordes}, James M. and {Chatterjee}, Shami and {Niedbalski}, Sashabaw and {Agazie}, Gabriella and {Anumarlapudi}, Akash and {Archibald}, Anne M. and {Arzoumanian}, Zaven and {Baier}, Jeremy G. and {Baker}, Paul T. and {Brook}, Paul R. and {Cromartie}, H. Thankful and {Crowter}, Kathryn and {DeCesar}, Megan E. and {Demorest}, Paul B. and {Dey}, Lankeswar and {Dolch}, Timothy and {Ferrara}, Elizabeth C. and {Fiore}, William and {Fonseca}, Emmanuel and {Freedman}, Gabriel E. and {Garver-Daniels}, Nate and {Gentile}, Peter A. and {Glaser}, Joseph and {Good}, Deborah C. and {Hazboun}, Jeffrey S. and {Jennings}, Ross J. and {Jones}, Megan L. and {Kaplan}, David L. and {Kerr}, Matthew and {Lam}, Michael T. and {Larsen}, Bjorn and {Lorimer}, Duncan R. and {Lowes}, Georgia A. and {Luo}, Jing and {Lynch}, Ryan S. and {Martsen}, Ashley and {McEwen}, Alexander and {McLaughlin}, Maura A. and {McMann}, Natasha and {Meyers}, Bradley W. and {Meyers}, Patrick M. and {Ng}, Cherry and {Ng}, Mason and {Nice}, David J. and {Nichols}, Shania and {Oliver}, Daniel J. and {Pennucci}, Timothy T. and {Perera}, Benetge B.~P. and {Pol}, Nihan S. and {Radovan}, Henri A. and {Ransom}, Scott M. and {Ray}, Paul S. and {Saffer}, Alexander and {Schmiedekamp}, Ann and {Schmiedekamp}, Carl and {Shapiro-Albert}, Brent J. and {Stairs}, Ingrid H. and {Stovall}, Kevin and {Susobhanan}, Abhimanyu and {Swiggum}, Joseph K. and {Thompson}, Mercedes S. and {Tresnjic}, Amir and {Wahl}, Haley M.},
        title = "{The NANOGrav 15 yr and 20 yr Datasets: Timing Events and Pulse Shape Changes}",
      journal = {\apj},
         year = 2026,
        month = jul,
       volume = {1005},
       number = {1},
          eid = {58},
        pages = {58},
          doi = {10.3847/1538-4357/ae6db6},
       adsurl = {https://ui.adsabs.harvard.edu/abs/2026ApJ..1005...58J}
}

@ARTICLE{2026arXiv260712038N,
       author = {{Nichols}, Shania A. and {Lam}, Michael T. and {Agazie}, Gabriella and {Ashok}, Anjana and {Baier}, Jeremy G. and {Cromartie}, H. Thankful and {Crowter}, Kathryn and {DeCesar}, Megan E. and {Demorest}, Paul B. and {Dey}, Lankeswar and {Fiore}, William and {Fonseca}, Emmanuel and {Glaser}, Joseph and {Good}, Deborah C. and {Hazboun}, Jeffrey S. and {Jennings}, Ross J. and {Kaplan}, David L. and {Larsen}, Bjorn and {Lowes}, Georgia A. and {Lynch}, Ryan S. and {Martsen}, Ashley and {Meyers}, Bradley W. and {Meyers}, Patrick M. and {Ng}, Mason and {Oliver}, Daniel J. and {Perera}, Benetge B.~P. and {Radovan}, Henri A. and {Ransom}, Scott M. and {Saffer}, Alexander and {Stairs}, Ingrid H. and {Thompson}, Mercedes S. and {Tresnjic}, Amir and {Verbiest}, Joris P.~W. and {Wright}, David},
        title = "{Mitigating the Timing Impact of Anomalous Pulse Profile Shape Variability in PSR J1713+0747 with Gaussian Component Modeling}",
      journal = {arXiv e-prints},
         year = 2026,
        month = jul,
          eid = {arXiv:2607.12038},
        pages = {arXiv:2607.12038},
archivePrefix = {arXiv},
       eprint = {2607.12038},
 primaryClass = {astro-ph.HE},
       adsurl = {https://ui.adsabs.harvard.edu/abs/2026arXiv260712038N}
}

@ARTICLE{2025OJAp....854243S,
       author = {{Shannon}, Ryan M. and {Bhat}, N.~D. Ramesh and {Chalumeau}, Aurelien and {Chen}, Siyuan and {Cromartie}, H. Thankful and {Gopukumar}, A. and {Grunthal}, Kathrin and {Hazboun}, Jeffrey S. and {Iraci}, Francesco and {Joshi}, Bhal Chandra and {Kato}, Ryo and {Keith}, MIchael J. and {Lee}, Kejia and {Liu}, Kuo and {Middleton}, Hannah and {Miles}, Matthew T. and {Mingarelli}, Chiara M.~F. and {Parthasarathy}, Adtiya and {Reardon}, Daniel J. and {Shaifullah}, Golam M. and {Takahashi}, Keitaro and {Tiburzi}, Caterina and {Truant}, Riccardo and {Xue}, Xiao and {Zic}, Andrew},
        title = "{The SKAO Pulsar Timing Array}",
      journal = {The Open Journal of Astrophysics},
         year = 2025,
        month = dec,
       volume = {8},
        pages = {54243},
          doi = {10.33232/001c.154243},
archivePrefix = {arXiv},
       eprint = {2607.03059},
 primaryClass = {astro-ph.IM},
       adsurl = {https://ui.adsabs.harvard.edu/abs/2025OJAp....854243S}
}

@ARTICLE{2025OJAp....854244O,
       author = {{Oswald}, Lucy S. and {Basu}, Avishek and {Chakraborty}, Manoneeta and {Joshi}, Bhal Chandra and {Lewandowska}, Natalia and {Liu}, Kuo and {Lower}, Marcus and {Phillipov}, Alexander and {Song}, Xiaoxi and {Tarafdar}, Pratik and {van Leeuwen}, Joeri and {Watts}, Anna and {Weltevrede}, Patrick and {Wright}, Geoff and {Benacek}, Jan and {Beri}, Aru and {Cao}, Shunshun and {Esposito}, Paolo and {Jankowski}, Fabian and {Jiang}, Jinchen and {Karastergiou}, Aris and {Lee}, Kejia and {Rea}, Nanda and {Vohl}, Dany},
        title = "{Understanding pulsar magnetospheres with the SKAO}",
      journal = {The Open Journal of Astrophysics},
         year = 2025,
        month = dec,
       volume = {8},
        pages = {54244},
          doi = {10.33232/001c.154244},
archivePrefix = {arXiv},
       eprint = {2512.16157},
 primaryClass = {astro-ph.HE},
       adsurl = {https://ui.adsabs.harvard.edu/abs/2025OJAp....854244O}
}

@ARTICLE{2017A&A...606A..41M,
       author = {{M{\"u}ller}, Peter and {Krause}, Marita and {Beck}, Rainer and {Schmidt}, Philip},
        title = "{The NOD3 software package: A graphical user interface-supported reduction package for single-dish radio continuum and polarisation observations}",
      journal = {\aap},
         year = 2017,
        month = oct,
       volume = {606},
          eid = {A41},
        pages = {A41},
          doi = {10.1051/0004-6361/201731257},
archivePrefix = {arXiv},
       eprint = {1707.05573},
 primaryClass = {astro-ph.IM},
       adsurl = {https://ui.adsabs.harvard.edu/abs/2017A&A...606A..41M}
}

@ARTICLE{2017ApJ...846..104K,
       author = {{Krishnakumar}, M.~A. and {Joshi}, Bhal Chandra and {Manoharan}, P.~K.},
        title = "{Multi-frequency Scatter Broadening Evolution of Pulsars. I}",
      journal = {\apj},
         year = 2017,
        month = sep,
       volume = {846},
       number = {2},
          eid = {104},
        pages = {104},
          doi = {10.3847/1538-4357/aa7af2},
archivePrefix = {arXiv},
       eprint = {1706.05799},
 primaryClass = {astro-ph.HE},
       adsurl = {https://ui.adsabs.harvard.edu/abs/2017ApJ...846..104K}
}

@ARTICLE{2016ApJ...833L..10D,
       author = {{De}, Kishalay and {Gupta}, Yashwant and {Sharma}, Prateek},
        title = "{Detection of Polarized Quasi-periodic Microstructure Emission in Millisecond Pulsars}",
      journal = {\apjl},
         year = 2016,
        month = dec,
       volume = {833},
       number = {1},
          eid = {L10},
        pages = {L10},
          doi = {10.3847/2041-8213/833/1/L10},
archivePrefix = {arXiv},
       eprint = {1611.07330},
 primaryClass = {astro-ph.HE},
       adsurl = {https://ui.adsabs.harvard.edu/abs/2016ApJ...833L..10D}
}

@ARTICLE{2024ApJ...961...48W,
       author = {{Wang}, Xiao-Wei and {Yan}, Zhen and {Shen}, Zhi-Qiang and {Tong}, Hao and {Zhou}, Xia and {Zhao}, Rong-Bing and {Wu}, Ya-Jun and {Huang}, Zhi-Peng and {Wang}, Rui and {Liu}, Jie},
        title = "{Observations of Nine Millisecond Pulsars at 8600 MHz Using the TMRT}",
      journal = {\apj},
         year = 2024,
        month = jan,
       volume = {961},
       number = {1},
          eid = {48},
        pages = {48},
          doi = {10.3847/1538-4357/ad0724},
       adsurl = {https://ui.adsabs.harvard.edu/abs/2024ApJ...961...48W}
}

@ARTICLE{2024ApJ...964..179J,
       author = {{Jennings}, Ross J. and {Cordes}, James M. and {Chatterjee}, Shami and {McLaughlin}, Maura A. and {Demorest}, Paul B. and {Arzoumanian}, Zaven and {Baker}, Paul T. and {Blumer}, Harsha and {Brook}, Paul R. and {Cohen}, Tyler and {Crawford}, Fronefield and {Cromartie}, H. Thankful and {DeCesar}, Megan E. and {Dolch}, Timothy and {Ferrara}, Elizabeth C. and {Fonseca}, Emmanuel and {Good}, Deborah C. and {Hazboun}, Jeffrey S. and {Jones}, Megan L. and {Kaplan}, David L. and {Lam}, Michael T. and {Lazio}, T. Joseph W. and {Lorimer}, Duncan R. and {Luo}, Jing and {Lynch}, Ryan S. and {McKee}, James W. and {Madison}, Dustin R. and {Meyers}, Bradley W. and {Mingarelli}, Chiara M.~F. and {Nice}, David J. and {Pennucci}, Timothy T. and {Perera}, Benetge B.~P. and {Pol}, Nihan S. and {Ransom}, Scott M. and {Ray}, Paul S. and {Shapiro-Albert}, Brent J. and {Siemens}, Xavier and {Stairs}, Ingrid H. and {Stinebring}, Daniel R. and {Swiggum}, Joseph K. and {Tan}, Chia Min and {Taylor}, Stephen R. and {Vigeland}, Sarah J. and {Witt}, Caitlin A.},
        title = "{An Unusual Pulse Shape Change Event in PSR J1713+0747 Observed with the Green Bank Telescope and CHIME}",
      journal = {\apj},
         year = 2024,
        month = apr,
       volume = {964},
       number = {2},
          eid = {179},
        pages = {179},
          doi = {10.3847/1538-4357/ad2930},
archivePrefix = {arXiv},
       eprint = {2210.12266},
 primaryClass = {astro-ph.HE},
       adsurl = {https://ui.adsabs.harvard.edu/abs/2024ApJ...964..179J}
}

@ARTICLE{1990ApJ...361..300F,
       author = {{Foster}, R.~S. and {Backer}, D.~C.},
        title = "{Constructing a Pulsar Timing Array}",
      journal = {\apj},
         year = 1990,
        month = sep,
       volume = {361},
        pages = {300},
          doi = {10.1086/169195},
       adsurl = {https://ui.adsabs.harvard.edu/abs/1990ApJ...361..300F}
}

@ARTICLE{2001ApJ...555...31G,
       author = {{Gangadhara}, R.~T. and {Gupta}, Y.},
        title = "{Understanding the Radio Emission Geometry of PSR B0329+54}",
      journal = {\apj},
         year = 2001,
        month = jul,
       volume = {555},
       number = {1},
        pages = {31-39},
          doi = {10.1086/321439},
archivePrefix = {arXiv},
       eprint = {astro-ph/0103109},
 primaryClass = {astro-ph},
       adsurl = {https://ui.adsabs.harvard.edu/abs/2001ApJ...555...31G}
}

@ARTICLE{2003ApJ...584..418G,
       author = {{Gupta}, Y. and {Gangadhara}, R.~T.},
        title = "{Understanding the Radio Emission Geometry of Multiple-Component Radio Pulsars from Retardation and Aberration Effects}",
      journal = {\apj},
         year = 2003,
        month = feb,
       volume = {584},
       number = {1},
        pages = {418-426},
          doi = {10.1086/345682},
archivePrefix = {arXiv},
       eprint = {astro-ph/0210411},
 primaryClass = {astro-ph},
       adsurl = {https://ui.adsabs.harvard.edu/abs/2003ApJ...584..418G}
}

@ARTICLE{1999ApJ...526..957K,
       author = {{Kramer}, Michael and {Lange}, Christoph and {Lorimer}, Duncan R. and {Backer}, Donald C. and {Xilouris}, Kiriaki M. and {Jessner}, Axel and {Wielebinski}, Richard},
        title = "{The Characteristics of Millisecond Pulsar Emission. III. From Low to High Frequencies}",
      journal = {\apj},
         year = 1999,
        month = dec,
       volume = {526},
       number = {2},
        pages = {957-975},
          doi = {10.1086/308042},
archivePrefix = {arXiv},
       eprint = {astro-ph/9906442},
 primaryClass = {astro-ph},
       adsurl = {https://ui.adsabs.harvard.edu/abs/1999ApJ...526..957K}
}

@ARTICLE{1978ApJ...222.1006C,
       author = {{Cordes}, J.~M.},
        title = "{Observational limits on the location of pulsar emission regions.}",
      journal = {\apj},
         year = 1978,
        month = jun,
       volume = {222},
        pages = {1006-1011},
          doi = {10.1086/156218},
       adsurl = {https://ui.adsabs.harvard.edu/abs/1978ApJ...222.1006C}
}

@ARTICLE{1998MNRAS.299..855K,
       author = {{Kijak}, Jaroslaw and {Gil}, Janusz},
        title = "{Radio emission regions in pulsars}",
      journal = {\mnras},
         year = 1998,
        month = sep,
       volume = {299},
       number = {3},
        pages = {855-861},
          doi = {10.1046/j.1365-8711.1998.01832.x},
       adsurl = {https://ui.adsabs.harvard.edu/abs/1998MNRAS.299..855K}
}

@ARTICLE{1998ApJ...501..286X,
       author = {{Xilouris}, Kiriaki M. and {Kramer}, Michael and {Jessner}, Axel and {von Hoensbroech}, Alexis and {Lorimer}, Duncan R. and {Wielebinski}, Richard and {Wolszczan}, Alexander and {Camilo}, Fernando},
        title = "{The Characteristics of Millisecond Pulsar Emission. II. Polarimetry}",
      journal = {\apj},
         year = 1998,
        month = jul,
       volume = {501},
       number = {1},
        pages = {286-306},
          doi = {10.1086/305791},
archivePrefix = {arXiv},
       eprint = {astro-ph/9801178},
 primaryClass = {astro-ph},
       adsurl = {https://ui.adsabs.harvard.edu/abs/1998ApJ...501..286X}
}

@ARTICLE{2026MNRAS.549ag835C,
       author = {{Cury{\l}o}, Ma{\l}gorzata and {Zic}, Andrew and {Wang}, Shuangqiang and {Thrane}, Eric and {Lasky}, Paul D. and {Tremblay}, Jacob Cardinal and {Chen}, Zu-Cheng and {Dai}, Shi and {Marco}, Valentina Di and {Hobbs}, George and {Kapur}, Agastya and {Ling}, Wenhua and {Lower}, Marcus E. and {Mandow}, Rami F. and {Mishra}, Saurav and {Reardon}, Daniel J. and {Russell}, Christopher J. and {Shannon}, Ryan M. and {Zhu}, Xing-Jiang},
        title = "{Frequency- and phase-resolved polarimetry of millisecond pulsars and its application to timing}",
      journal = {\mnras},
         year = 2026,
        month = jul,
       volume = {549},
       number = {4},
          eid = {stag835},
        pages = {stag835},
          doi = {10.1093/mnras/stag835},
archivePrefix = {arXiv},
       eprint = {2512.09220},
 primaryClass = {astro-ph.HE},
       adsurl = {https://ui.adsabs.harvard.edu/abs/2026MNRAS.549ag835C}
}

@ARTICLE{2017ApJ...845...23R,
       author = {{Rankin}, Joanna M. and {Archibald}, Anne and {Hessels}, Jason and {van Leeuwen}, Joeri and {Mitra}, Dipanjan and {Ransom}, Scott and {Stairs}, Ingrid and {van Straten}, Willem and {Weisberg}, Joel M.},
        title = "{Toward an Empirical Theory of Pulsar Emission. XII. Exploring the Physical Conditions in Millisecond Pulsar Emission Regions}",
      journal = {\apj},
         year = 2017,
        month = aug,
       volume = {845},
       number = {1},
          eid = {23},
        pages = {23},
          doi = {10.3847/1538-4357/aa7b73},
archivePrefix = {arXiv},
       eprint = {1710.11465},
 primaryClass = {astro-ph.HE},
       adsurl = {https://ui.adsabs.harvard.edu/abs/2017ApJ...845...23R}
}

@ARTICLE{2017MNRAS.471L.131D,
       author = {{Dyks}, J.},
        title = "{The geometry of a radio pulsar beam}",
      journal = {\mnras},
         year = 2017,
        month = oct,
       volume = {471},
       number = {1},
        pages = {L131-L134},
          doi = {10.1093/mnrasl/slx120},
archivePrefix = {arXiv},
       eprint = {1705.05133},
 primaryClass = {astro-ph.HE},
       adsurl = {https://ui.adsabs.harvard.edu/abs/2017MNRAS.471L.131D}
}

@ARTICLE{2015MNRAS.448..771W,
       author = {{Wang}, P.~F. and {Wang}, C. and {Han}, J.~L.},
        title = "{On the frequency dependence of pulsar linear polarization}",
      journal = {\mnras},
         year = 2015,
        month = mar,
       volume = {448},
       number = {1},
        pages = {771-780},
          doi = {10.1093/mnras/stu2765},
archivePrefix = {arXiv},
       eprint = {1501.00066},
 primaryClass = {astro-ph.HE},
       adsurl = {https://ui.adsabs.harvard.edu/abs/2015MNRAS.448..771W}
}

@ARTICLE{2015MNRAS.449.3223D,
       author = {{Dai}, S. and {Hobbs}, G. and {Manchester}, R.~N. and {Kerr}, M. and {Shannon}, R.~M. and {van Straten}, W. and {Mata}, A. and {Bailes}, M. and {Bhat}, N.~D.~R. and {Burke-Spolaor}, S. and {Coles}, W.~A. and {Johnston}, S. and {Keith}, M.~J. and {Levin}, Y. and {Os{\l}owski}, S. and {Reardon}, D. and {Ravi}, V. and {Sarkissian}, J.~M. and {Tiburzi}, C. and {Toomey}, L. and {Wang}, H.~G. and {Wang}, J.-B. and {Wen}, L. and {Xu}, R.~X. and {Yan}, W.~M. and {Zhu}, X.-J.},
        title = "{A study of multifrequency polarization pulse profiles of millisecond pulsars}",
      journal = {\mnras},
         year = 2015,
        month = may,
       volume = {449},
       number = {3},
        pages = {3223-3262},
          doi = {10.1093/mnras/stv508},
archivePrefix = {arXiv},
       eprint = {1503.01841},
 primaryClass = {astro-ph.GA},
       adsurl = {https://ui.adsabs.harvard.edu/abs/2015MNRAS.449.3223D}
}

@ARTICLE{2008ApJ...683L..41B,
       author = {{Beloborodov}, Andrei M.},
        title = "{Polar-Cap Accelerator and Radio Emission from Pulsars}",
      journal = {\apjl},
         year = 2008,
        month = aug,
       volume = {683},
       number = {1},
        pages = {L41},
          doi = {10.1086/590079},
archivePrefix = {arXiv},
       eprint = {0710.0920},
 primaryClass = {astro-ph},
       adsurl = {https://ui.adsabs.harvard.edu/abs/2008ApJ...683L..41B}
}

@ARTICLE{2010MNRAS.408L..41T,
       author = {{Timokhin}, A.~N.},
        title = "{A model for nulling and mode changing in pulsars}",
      journal = {\mnras},
         year = 2010,
        month = oct,
       volume = {408},
       number = {1},
        pages = {L41-L45},
          doi = {10.1111/j.1745-3933.2010.00924.x},
archivePrefix = {arXiv},
       eprint = {0912.2995},
 primaryClass = {astro-ph.HE},
       adsurl = {https://ui.adsabs.harvard.edu/abs/2010MNRAS.408L..41T}
}

@ARTICLE{2010MNRAS.408.2092T,
       author = {{Timokhin}, A.~N.},
        title = "{Time-dependent pair cascades in magnetospheres of neutron stars - I. Dynamics of the polar cap cascade with no particle supply from the neutron star surface}",
      journal = {\mnras},
         year = 2010,
        month = nov,
       volume = {408},
       number = {4},
        pages = {2092-2114},
          doi = {10.1111/j.1365-2966.2010.17286.x},
archivePrefix = {arXiv},
       eprint = {1006.2384},
 primaryClass = {astro-ph.HE},
       adsurl = {https://ui.adsabs.harvard.edu/abs/2010MNRAS.408.2092T}
}

@ARTICLE{2017MNRAS.469.2049Y,
       author = {{Yuen}, R. and {Melrose}, D.~B.},
        title = "{A model for abrupt changes in pulsar pulse profile}",
      journal = {\mnras},
         year = 2017,
        month = aug,
       volume = {469},
       number = {2},
        pages = {2049-2058},
          doi = {10.1093/mnras/stx1023},
       adsurl = {https://ui.adsabs.harvard.edu/abs/2017MNRAS.469.2049Y}
}

@ARTICLE{2026MNRAS.547f2258K,
       author = {{Kramer}, Michael and {Johnston}, Simon},
        title = "{Radio emission from beyond the light cylinder in millisecond pulsars}",
      journal = {\mnras},
         year = 2026,
        month = apr,
       volume = {547},
       number = {4},
          eid = {staf2258},
        pages = {staf2258},
          doi = {10.1093/mnras/staf2258},
archivePrefix = {arXiv},
       eprint = {2510.05778},
 primaryClass = {astro-ph.HE},
       adsurl = {https://ui.adsabs.harvard.edu/abs/2026MNRAS.547f2258K}
}

@ARTICLE{2023MNRAS.519.3976M,
       author = {{Miles}, M.~T. and {Shannon}, R.~M. and {Bailes}, M. and {Reardon}, D.~J. and {Keith}, M.~J. and {Cameron}, A.~D. and {Parthasarathy}, A. and {Shamohammadi}, M. and {Spiewak}, R. and {van Straten}, W. and {Buchner}, S. and {Camilo}, F. and {Geyer}, M. and {Karastergiou}, A. and {Kramer}, M. and {Serylak}, M. and {Theureau}, G. and {Venkatraman Krishnan}, V.},
        title = "{The MeerKAT Pulsar Timing Array: first data release}",
      journal = {\mnras},
         year = 2023,
        month = mar,
       volume = {519},
       number = {3},
        pages = {3976-3991},
          doi = {10.1093/mnras/stac3644},
archivePrefix = {arXiv},
       eprint = {2212.04648},
 primaryClass = {astro-ph.HE},
       adsurl = {https://ui.adsabs.harvard.edu/abs/2023MNRAS.519.3976M}
}

@ARTICLE{2023MNRAS.524.5904L,
       author = {{Lower}, M.~E. and {Johnston}, S. and {Karastergiou}, A. and {Brook}, P.~R. and {Bailes}, M. and {Buchner}, S. and {Deller}, A.~T. and {Dunn}, L. and {Flynn}, C. and {Kerr}, M. and {Manchester}, R.~N. and {Mandlik}, A. and {Oswald}, L.~S. and {Parthasarathy}, A. and {Shannon}, R.~M. and {Sobey}, C. and {Weltevrede}, P.},
        title = "{Rotational and radio emission properties of PSR J0738-4042 over half a century}",
      journal = {\mnras},
         year = 2023,
        month = oct,
       volume = {524},
       number = {4},
        pages = {5904-5917},
          doi = {10.1093/mnras/stad2243},
archivePrefix = {arXiv},
       eprint = {2307.11953},
 primaryClass = {astro-ph.HE},
       adsurl = {https://ui.adsabs.harvard.edu/abs/2023MNRAS.524.5904L}
}

@ARTICLE{2025MNRAS.538.3104L,
       author = {{Lower}, M.~E. and {Karastergiou}, A. and {Johnston}, S. and {Brook}, P.~R. and {Dai}, S. and {Kerr}, M. and {Manchester}, R.~N. and {Oswald}, L.~S. and {Shannon}, R.~M. and {Sobey}, C. and {Weltevrede}, P.},
        title = "{The ubiquity of variable radio emission and spin-down rates in pulsars}",
      journal = {\mnras},
         year = 2025,
        month = apr,
       volume = {538},
       number = {4},
        pages = {3104-3129},
          doi = {10.1093/mnras/staf427},
archivePrefix = {arXiv},
       eprint = {2501.03500},
 primaryClass = {astro-ph.HE},
       adsurl = {https://ui.adsabs.harvard.edu/abs/2025MNRAS.538.3104L}
}

@ARTICLE{2017MNRAS.466.3706L,
       author = {{Lentati}, L. and {Kerr}, M. and {Dai}, S. and {Hobson}, M.~P. and {Shannon}, R.~M. and {Hobbs}, G. and {Bailes}, M. and {Bhat}, N.~D. Ramesh and {Burke-Spolaor}, S. and {Coles}, W. and {Dempsey}, J. and {Lasky}, P.~D. and {Levin}, Y. and {Manchester}, R.~N. and {Os{\l}owski}, S. and {Ravi}, V. and {Reardon}, D.~J. and {Rosado}, P.~A. and {Spiewak}, R. and {van Straten}, W. and {Toomey}, L. and {Wang}, J. and {Wen}, L. and {You}, X. and {Zhu}, X.},
        title = "{Wide-band profile domain pulsar timing analysis}",
      journal = {\mnras},
         year = 2017,
        month = apr,
       volume = {466},
       number = {3},
        pages = {3706-3727},
          doi = {10.1093/mnras/stw3359},
archivePrefix = {arXiv},
       eprint = {1612.05258},
 primaryClass = {astro-ph.IM},
       adsurl = {https://ui.adsabs.harvard.edu/abs/2017MNRAS.466.3706L}
}

@ARTICLE{2023ApJ...958L...9B,
       author = {{Bransgrove}, Ashley and {Beloborodov}, Andrei M. and {Levin}, Yuri},
        title = "{Radio Emission and Electric Gaps in Pulsar Magnetospheres}",
      journal = {\apjl},
         year = 2023,
        month = nov,
       volume = {958},
       number = {1},
          eid = {L9},
        pages = {L9},
          doi = {10.3847/2041-8213/ad0556},
archivePrefix = {arXiv},
       eprint = {2209.11362},
 primaryClass = {astro-ph.HE},
       adsurl = {https://ui.adsabs.harvard.edu/abs/2023ApJ...958L...9B}
}

@ARTICLE{2016MNRAS.456.1374B,
       author = {{Brook}, P.~R. and {Karastergiou}, A. and {Johnston}, S. and {Kerr}, M. and {Shannon}, R.~M. and {Roberts}, S.~J.},
        title = "{Emission-rotation correlation in pulsars: new discoveries with optimal techniques}",
      journal = {\mnras},
         year = 2016,
        month = feb,
       volume = {456},
       number = {2},
        pages = {1374-1393},
          doi = {10.1093/mnras/stv2715},
archivePrefix = {arXiv},
       eprint = {1511.05481},
 primaryClass = {astro-ph.HE},
       adsurl = {https://ui.adsabs.harvard.edu/abs/2016MNRAS.456.1374B}
}

@ARTICLE{2022MNRAS.513.5861S,
       author = {{Shaw}, B. and {Stappers}, B.~W. and {Weltevrede}, P. and {Brook}, P.~R. and {Karastergiou}, A. and {Jordan}, C.~A. and {Keith}, M.~J. and {Kramer}, M. and {Lyne}, A.~G.},
        title = "{Long-term rotational and emission variability of 17 radio pulsars}",
      journal = {\mnras},
         year = 2022,
        month = jul,
       volume = {513},
       number = {4},
        pages = {5861-5880},
          doi = {10.1093/mnras/stac1156},
archivePrefix = {arXiv},
       eprint = {2204.10767},
 primaryClass = {astro-ph.HE},
       adsurl = {https://ui.adsabs.harvard.edu/abs/2022MNRAS.513.5861S}
}

@ARTICLE{2024MNRAS.528.7458B,
       author = {{Basu}, A. and {Weltevrede}, P. and {Keith}, M.~J. and {Johnston}, S. and {Karastergiou}, A. and {Oswald}, L.~S. and {Posselt}, B. and {Song}, X. and {Cameron}, A.~D.},
        title = "{The Thousand-Pulsar-Array programme on MeerKAT - XII. Discovery of long-term pulse profile evolution in seven young pulsars}",
      journal = {\mnras},
         year = 2024,
        month = mar,
       volume = {528},
       number = {4},
        pages = {7458-7476},
          doi = {10.1093/mnras/stae483},
archivePrefix = {arXiv},
       eprint = {2402.09065},
 primaryClass = {astro-ph.HE},
       adsurl = {https://ui.adsabs.harvard.edu/abs/2024MNRAS.528.7458B}
}

@ARTICLE{2010Sci...329..408L,
       author = {{Lyne}, Andrew and {Hobbs}, George and {Kramer}, Michael and {Stairs}, Ingrid and {Stappers}, Ben},
        title = "{Switched Magnetospheric Regulation of Pulsar Spin-Down}",
      journal = {Science},
         year = 2010,
        month = jul,
       volume = {329},
       number = {5990},
        pages = {408},
          doi = {10.1126/science.1186683},
archivePrefix = {arXiv},
       eprint = {1006.5184},
 primaryClass = {astro-ph.GA},
       adsurl = {https://ui.adsabs.harvard.edu/abs/2010Sci...329..408L}
}

@ARTICLE{2011MNRAS.415..251K,
       author = {{Karastergiou}, A. and {Roberts}, S.~J. and {Johnston}, S. and {Lee}, H. and {Weltevrede}, P. and {Kramer}, M.},
        title = "{A transient component in the pulse profile of PSR J0738-4042}",
      journal = {\mnras},
         year = 2011,
        month = jul,
       volume = {415},
       number = {1},
        pages = {251-256},
          doi = {10.1111/j.1365-2966.2011.18697.x},
archivePrefix = {arXiv},
       eprint = {1103.2247},
 primaryClass = {astro-ph.HE},
       adsurl = {https://ui.adsabs.harvard.edu/abs/2011MNRAS.415..251K}
}




\appendix

\label{app:model_select}
\begin{table*}
\centering
\caption{Model-selection comparison between exponential and power-law recovery models for the prior and primary profile-change events.}
\label{tab:model_selection_recovery}
\begin{tabular}{ccccccccc}
\hline
Model & Event & Recovery model & $\chi^2_{\rm r}$ & AIC & BIC & $\Delta$AIC & $\Delta$BIC & Preferred model \\
\hline
\hline
M01 & Prior   & Exponential & 1.35 & -18275 & -18186 &      &      &  \\
M02 & Prior   & Power law   & 1.43 & -18269 & -18176 & +6   & +10  & M01 (exponential) \\
\hline
M03 & Primary & Exponential & 1.20 & -12749 & -12653 &      &      &  \\
M04 & Primary & Power law   & 1.40 & -12867 & -12767 & -118 & -114 & M04 (power law) \\
\hline
\end{tabular}
\end{table*}

\begin{table*}
\centering
\caption{Model-selection tests for the additional components and frequency dependence adopted for the primary-event power-law recovery model. All models are compared with the adopted cubic-$\tau$ power-law model (M04).}
\label{tab:model_selection_additional}
\begin{tabular}{cccccccc}
\hline
Model & Model description & $\chi^2_{\rm r}$ & AIC & BIC & $\Delta$AIC & $\Delta$BIC & Preferred model \\
\hline
\hline
\textbf{M04} & \textbf{Power law, cubic $\tau(\nu)$} & \textbf{1.40}  & \textbf{-12867} & \textbf{-12767}  &  &  &   \\
M04a & Power law, no $m_{\rm post}$          & 1.78 & -12705 & -12609 & +162 & +158 & M04 \\
M04b & Power law, linear $\tau(\nu)$         & 1.53 & -12854 & -12763 & +13  & +4   & M04 \\
M04c & Power law, quadratic $\tau(\nu)$      & 1.51 & -12858 & -12763 & +9   & +4   & M04 \\
M04d & Power law, quartic $\tau(\nu)$        & 1.55 & -12842 & -12737 & +25  & +30  & M04 \\
\hline
\end{tabular}
\end{table*}

\label{app:global_mcmc_params}
\begin{table*}
\centering
\caption{Global parameter values from the MCMC modelling of the prior and primary profile change events. These parameters are fitted simultaneously across all eight frequency sub-bands and describe the frequency dependence of the recovery time-scales, the post-event linear trend ($m_{\rm post}$), and the power-law index ($\alpha_{\rm post}$). The fixed onset epochs of the prior and primary events, $t_{0,\rm pre}$ and $t_{0,\rm post}$, are also listed. Reported uncertainties correspond to the 16th and 84th percentiles of the posterior distributions.}
\label{tab:MCMC_global_params}
\begin{tabular}{ccc}
\hline
Parameter & Value & Units \\
\hline
\hline
$\tau_{0,\rm pre}$ & $322\pm49$ & d \\
$\tau_{\rm slope,pre}$ & $24\pm44$ & d \\
$\tau_{0,\rm post}$ & $59\pm23$ & d \\
$\tau_{\rm slope,post}$ & $45\pm50$ & d \\
$\tau_{\rm quad,post}$ & $-50\pm35$ & d \\
$\tau_{\rm cubic,post}$ & $11.6\pm7.9$ & d \\
$m_{\rm post}$ & $1.17\pm0.11\times10^{-7}$ & d$^{-1}$ \\
$\alpha_{\rm post}$ & $1.03\pm0.04$ & -- \\
\hline
$t_{0,\rm pre}$ & $58082.00$ & MJD \\
$t_{0,\rm post}$ & $59670.00$ & MJD \\
\hline
\end{tabular}
\end{table*}

\label{app:banded_mcmc_params}
\begin{table*}
\centering
\caption{Frequency-dependent parameter values from the MCMC modelling of the prior and primary profile change events. For each sub-band, we report the fitted event amplitudes ($A_{\rm pre}$ and $A_{\rm post}$), recovery half-lives ($t_{1/2,\rm pre}$ and $t_{1/2,\rm post}$), and constant offsets ($C_{\rm pre}$ and $C_{\rm post}$). The recovery half-lives are derived from the fitted exponential and power-law models to provide a common measure of the recovery time-scale for the two events. Reported uncertainties correspond to the 16th and 84th percentiles of the posterior distributions.}
\label{tab:MCMC_subband_params}
\begin{tabular}{cccccccc}
\hline
Band & $\nu$ (MHz) & $A_{\rm pre}$ & $A_{\rm post}$ & $t_{1/2,\rm pre}$ (d) & $t_{1/2,\rm post}$ (d) & $C_{\rm pre}$ & $C_{\rm post}$ \\
\hline
\hline
sbA & 736 & $-1.44\pm0.23\times10^{-3}$ & $-1.16\pm0.17\times10^{-3}$ & $232\pm20$ & $66\pm6$ & $1.86\pm0.28\times10^{-4}$ & $2.86\pm0.39\times10^{-4}$ \\
sbB & 816 & $-1.39\pm0.16\times10^{-3}$ & $-1.56\pm0.08\times10^{-3}$ & $233\pm19$ & $66\pm5$ & $1.62\pm0.17\times10^{-4}$ & $2.30\pm0.21\times10^{-4}$ \\
sbC & 912 & $-1.37\pm0.14\times10^{-3}$ & $-1.62\pm0.07\times10^{-3}$ & $234\pm17$ & $66\pm4$ & $1.29\pm0.15\times10^{-4}$ & $2.81\pm0.22\times10^{-4}$ \\
sbD & 1104 & $-1.08\pm0.08\times10^{-3}$ & $-1.31\pm0.03\times10^{-3}$ & $237\pm15$ & $65\pm3$ & $1.34\pm0.08\times10^{-4}$ & $2.16\pm0.11\times10^{-4}$ \\
sbE & 1376 & $-7.05\pm0.46\times10^{-4}$ & $-1.03\pm0.02\times10^{-3}$ & $240\pm13$ & $61\pm3$ & $9.19\pm0.47\times10^{-5}$ & $1.64\pm0.08\times10^{-4}$ \\
sbF & 1776 & $-5.63\pm0.41\times10^{-4}$ & $-8.05\pm0.19\times10^{-4}$ & $244\pm15$ & $55\pm3$ & $5.77\pm0.45\times10^{-5}$ & $1.33\pm0.07\times10^{-4}$ \\
sbG & 2320 & $-4.75\pm0.46\times10^{-4}$ & $-5.70\pm0.18\times10^{-4}$ & $251\pm23$ & $46\pm4$ & $6.04\pm0.53\times10^{-5}$ & $7.54\pm0.66\times10^{-5}$ \\
sbH & 3312 & $-2.80\pm0.58\times10^{-4}$ & $-3.81\pm0.30\times10^{-4}$ & $263\pm43$ & $39\pm8$ & $3.87\pm0.64\times10^{-5}$ & $4.46\pm0.78\times10^{-5}$ \\
\hline
\end{tabular}
\end{table*}


\bsp	
\label{lastpage}
\end{document}